\documentclass[conference]{IEEEtran}
\IEEEoverridecommandlockouts

\usepackage{cite}
\usepackage{amsmath,amssymb,amsfonts,mathrsfs, url}
\usepackage{algorithmic}
\usepackage{graphicx}
\usepackage{dblfloatfix}
\usepackage{textcomp}
\usepackage{balance}
\usepackage{tikz}
\usetikzlibrary{calc}
\usetikzlibrary{positioning, fit, backgrounds}
\usepackage{xcolor}
\usepackage[firstpage=true]{background}
\def\BibTeX{{\rm B\kern-.05em{\sc i\kern-.025em b}\kern-.08em
    T\kern-.1667em\lower.7ex\hbox{E}\kern-.125emX}}

\begin{document}
\SetBgContents{\parbox{\textwidth}{\footnotesize \copyright 2025 IEEE. Published in 2025 International Conference on Visual Communications and Image Processing (VCIP), scheduled for 01-04 December 2025 in Klagenfurt, Austria. Personal use of this material is permitted. However, permission to reprint/republish this material for advertising or promotional purposes or for creating new collective works for resale or redistribution to servers or lists, or to reuse any copyrighted component of this work in other works, must be obtained from the IEEE. DOI: \url{10.1109/VCIP67698.2025.11396903}}}
\SetBgScale{1}
\SetBgAngle{0}
\SetBgPosition{current page.north}
\SetBgVshift{-1cm}
\SetBgColor{black}
\SetBgOpacity{1}

\title{Towards Object Segmentation Mask Selection Using Specular Reflections\\
\thanks{The authors gratefully acknowledge that this work has been supported by the Bayerische Forschungsstiftung (BFS, Bavarian Research Foundation) under project number AZ-1547-22.}
}

\author{\IEEEauthorblockN{Katja Kossira, Yunxuan Zhu, Jürgen Seiler, and André Kaup}
\IEEEauthorblockA{\textit{Multimedia Communications and Signal Processing} \\
\textit{Friedrich-Alexander University Erlangen-Nürnberg (FAU)}\\
Cauerstr.7, 91058 Erlangen, Germany\\
\{katja.kossira, yunxuan.zhu, juergen.seiler, andre.kaup\} @fau.de
}
}

\maketitle

\begin{abstract}
Specular reflections pose a significant challenge for object segmentation, as their sharp intensity transitions often mislead both conventional algorithms and deep learning based methods. However, as the specular reflection must lie on the surface of the object, this fact can be exploited to improve the segmentation masks. By identifying the largest region containing the reflection as the object, we derive a more accurate object mask without requiring specialized training data or model adaption.
We evaluate our method on both synthetic and real world images and compare it against established and state-of-the-art techniques including Otsu thresholding, YOLO, and SAM2. Compared to the best performing baseline SAM2, our approach achieves up to 26.7\% improvement in IoU, 22.3\% in DSC, and 9.7\% in pixel accuracy. Qualitative evaluations on real world images further confirm the robustness and generalizability of the proposed approach. 
\end{abstract}


\section{Introduction}
Accurate object segmentation is a fundamental task in computer vision, enabling the precise identification of object boundaries within digital images. This capability is essential for a wide range of applications, including industrial automation \cite{AutoApp}, robotic manipulation \cite{RobotApp}, medical diagnostics \cite{MedApp}, and environmental monitoring \cite{EnvApp}. 
In such contexts, the ability to isolate specific objects such as plastic parts on an assembly line is a prerequisite for reliable downstream processing, whether it involves classification, measurement, or interaction \cite{AutoApp2}. 

Over the years, a diverse set of object segmentation methods has been developed. Classical image processing techniques remain relevant in controlled environments. For example, Otsu's thresholding method \cite{Otsu} automatically determines an optimal grayscale threshold to separate foreground from background. This method is fast and interpretable, but it often fails in complex scenes where lighting, texture, or occlusion interfere with signal quality. To address such challenges, in recent years the field has shifted towards deep learning-based segmentation, which enables more robust and generalizable solutions. A widely adopted model is Mask R-CNN \cite{MaskRCNN}, which extends object detection models to pixel-level instance segmentation by adding a dedicated mask prediction branch. Other architectures like DeepLabv3+ \cite{DeepLabv3+} incorporate dilated convolutions to capture multi-scale contextual information, while HRNet \cite{HRNet} maintains high-resolution representations throughout the network, producing particularly accurate and sharp segmentation masks. Among the most influential models is YOLO (You Only Look Once) \cite{YOLO}, which was originally designed for real-time object detection. Recent iterations extend this functionality to instance segmentation, offering fast and reasonably accurate segmentation capabilities. Due to their speed and lightweight architecture, YOLO-based models are widely deployed in industrial environments where real-time processing is critical. More recently, foundation models have emerged in computer vision. A leading example is the Segment Anything Model (SAM) \cite{SAM} and its successor SAM2 \cite{SAM2}. These models enable zero-shot segmentation, allowing them to predict object masks in previously unseen domains without task-specific training. SAM2 improves upon SAM by offering enhanced prompt handling, refined multi-mask generation, greater generalization capabilities, and a significantly faster runtime. Crucially, SAM2 can produce multiple candidate masks for a given object, reflecting ambiguity or uncertainty in its boundary, particularly in complex scenes.
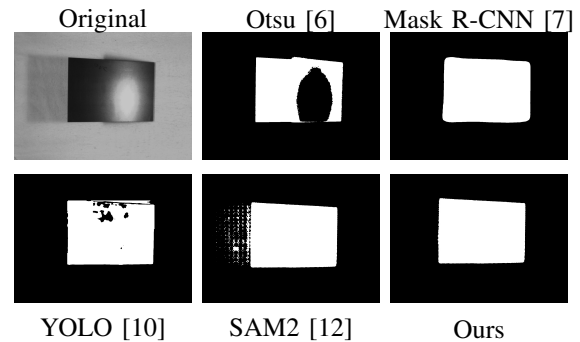
\begin{figure}
	\centering
	\begin{tikzpicture}
		\node(Original)[]{\includegraphics[width=0.13\textwidth]{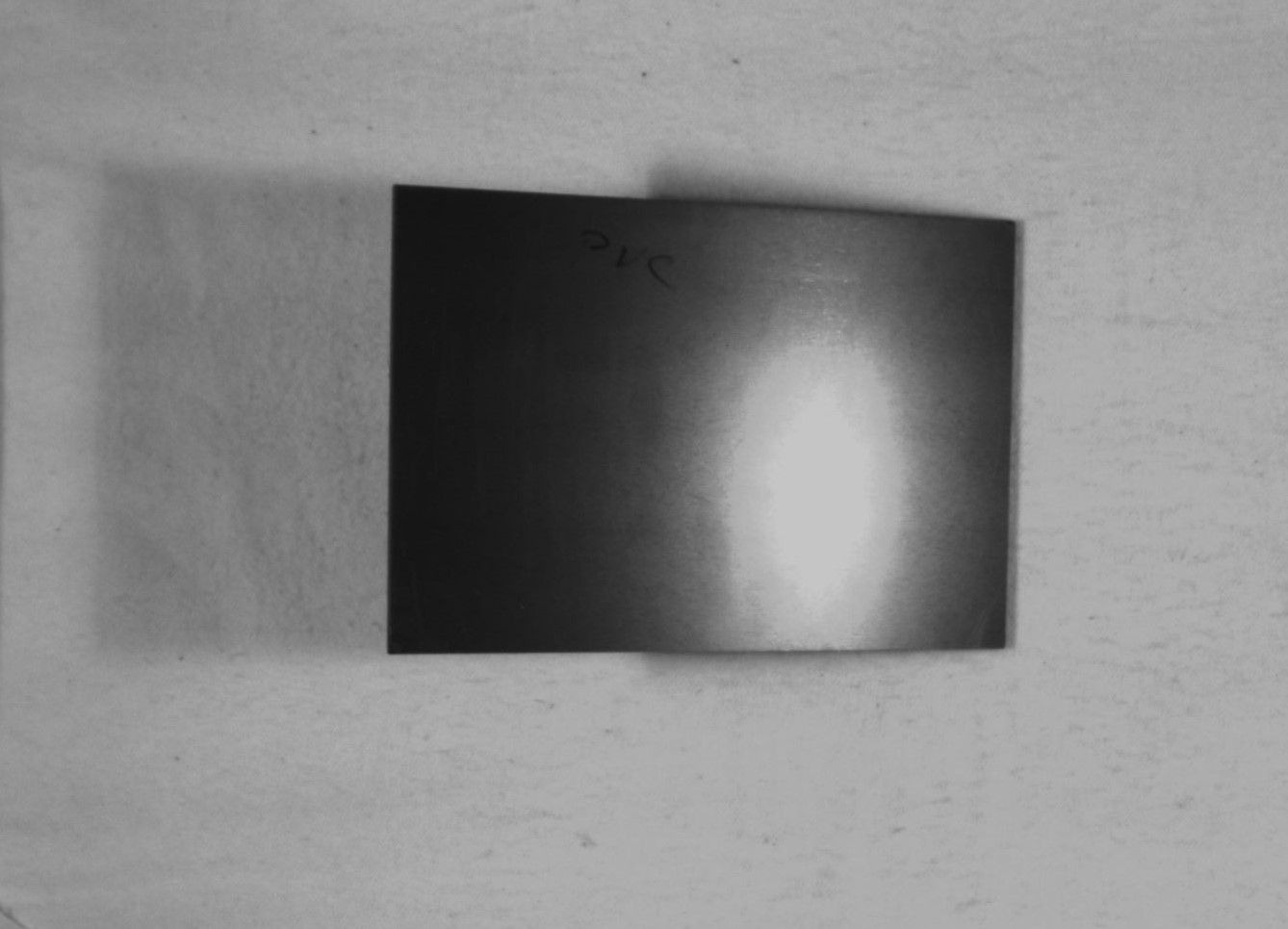}};
		\node(ori)[above of=Original]{Original};
		\node(Otsu)[right of=Original, xshift=1.5cm]{\includegraphics[width=0.13\textwidth]{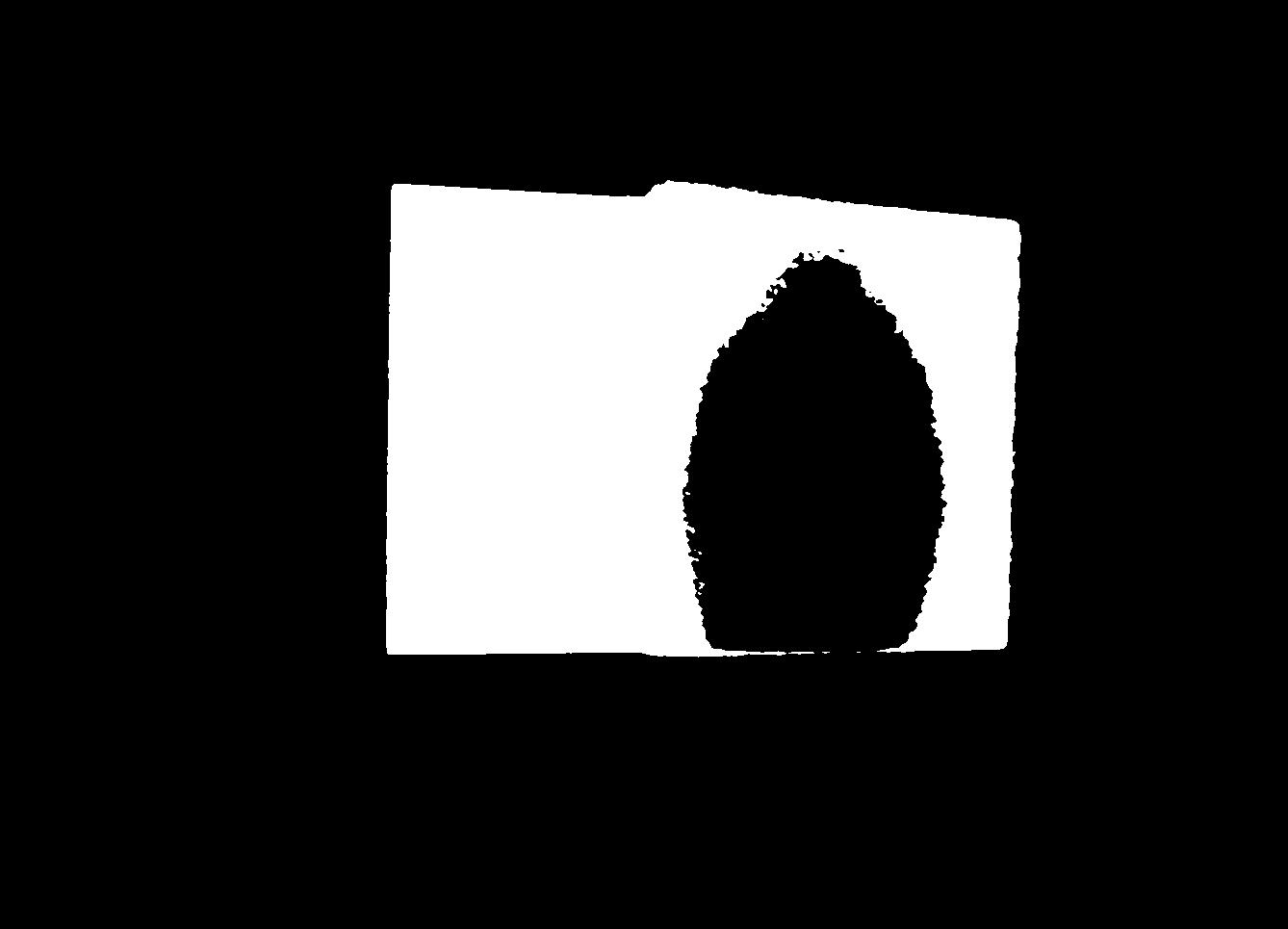}};
		\node(otsu)[above of=Otsu]{Otsu \cite{Otsu}};
		\node(MRCNN)[right of=Otsu, xshift=1.5cm]{\includegraphics[width=0.13\textwidth]{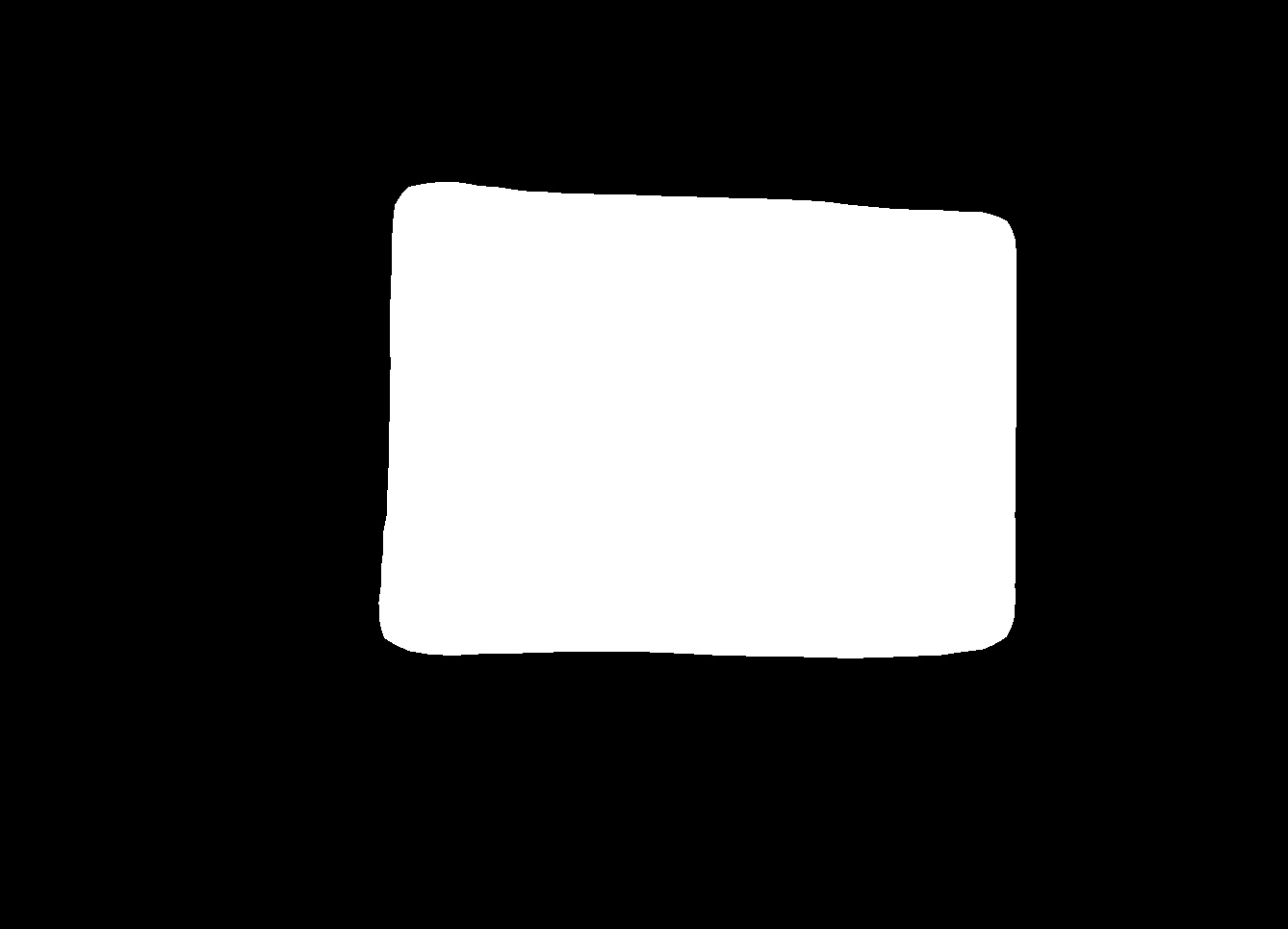}};
		\node(mrcnn)[above of=MRCNN]{Mask R-CNN \cite{MaskRCNN}};
		\node(YOLOv11)[below of=Original, yshift=-0.9cm]{\includegraphics[width=0.13\textwidth]{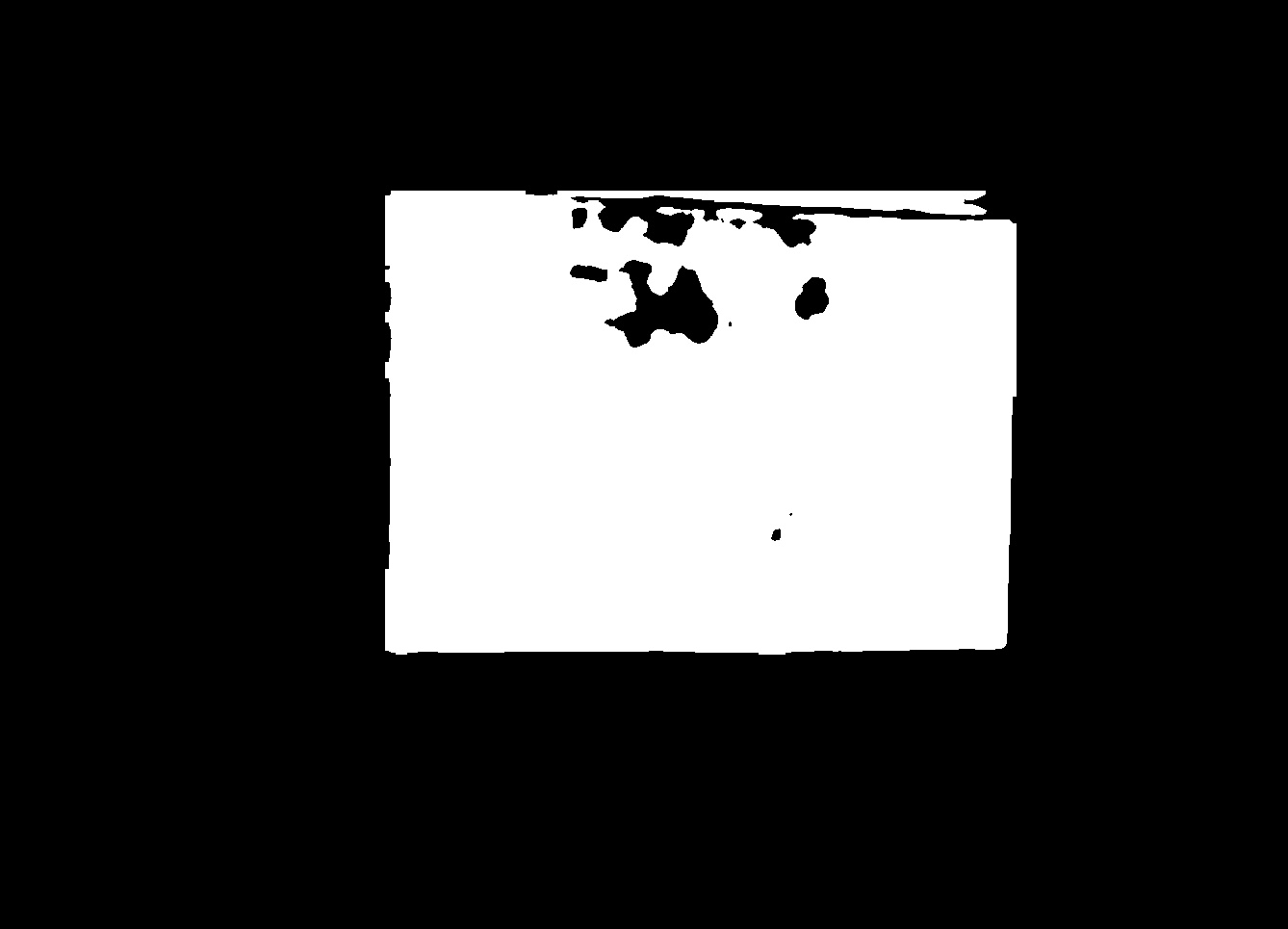}};
		\node(yolo)[below of=YOLOv11, yshift=-0.2cm]{YOLO \cite{YOLO}};
		\node(SAM2)[right of=YOLOv11, xshift=1.5cm]{\includegraphics[width=0.13\textwidth]{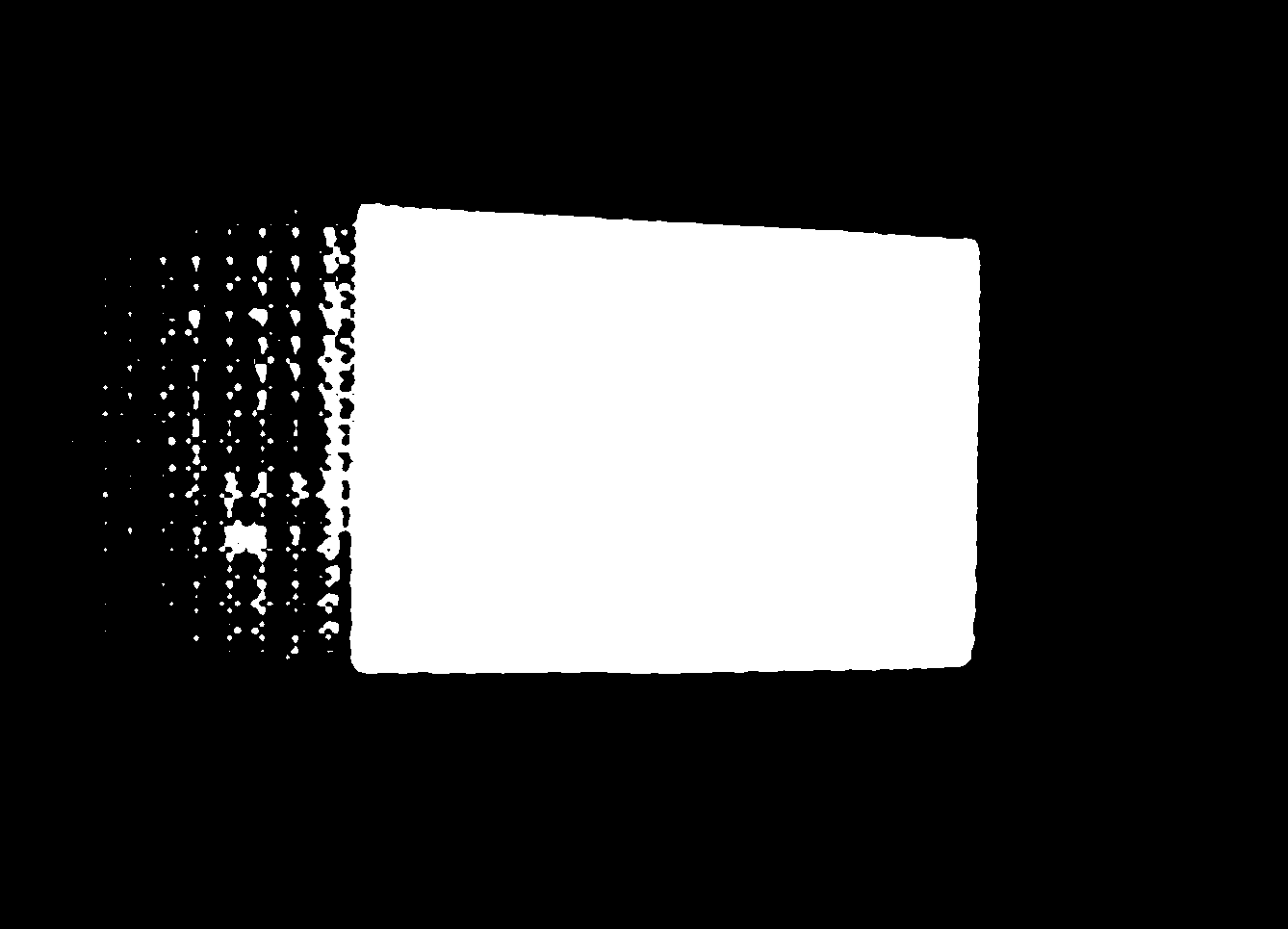}};
		\node(sam2)[below of=SAM2, yshift=-0.2cm]{SAM2 \cite{SAM2}};
		\node(Ours)[right of=SAM2, xshift=1.5cm]{\includegraphics[width=0.13\textwidth]{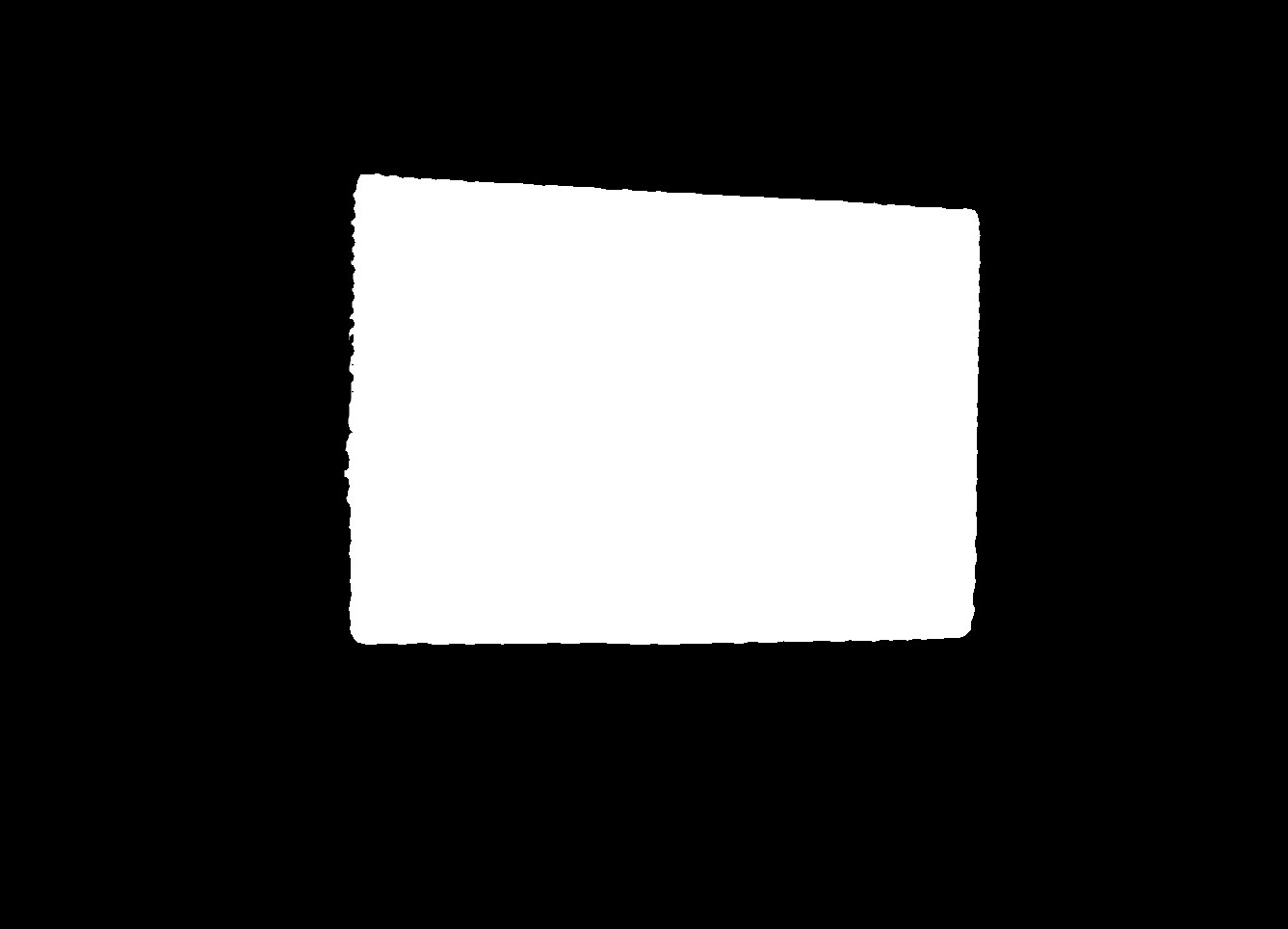}};
		\node(ours)[below of=Ours, yshift=-0.2cm]{Ours};
	\end{tikzpicture}
	\vspace{-0.3cm}
	\caption{Original image and the object masks generated using Otsu \cite{Otsu}, Mask R-CNN \cite{MaskRCNN}, YOLO \cite{YOLO}, SAM2 \cite{SAM2} as well as our proposed method. The white regions represent the calculated masks.}
	\label{fig:ComparisonDifferentMethods}
	\vspace{-0.6cm}
\end{figure}
\begin{figure*}[ht!]
	\centering
	\begin{tikzpicture}
		\node(Image)[]{$\mathcal{I}$};
		\node(SRD)[text width=1.8cm, minimum height=2.5cm, align=center, right of=Image, xshift=1.8cm, draw]{Specular reflection detection \cite{SRD1, SRD2, SRD3}};
		\node(sam2)[draw, minimum height=2.2cm, text width=1.4cm, right of=SRD, align=center, xshift=3cm]{SAM2 \cite{SAM2}};
		\node(m2)[right of=sam2, align=center, xshift=1.5cm]{$\mathcal{M}_2$};
		\node(m21)[right of=m2, xshift=-0.2cm]{\includegraphics[width=0.05\textwidth]{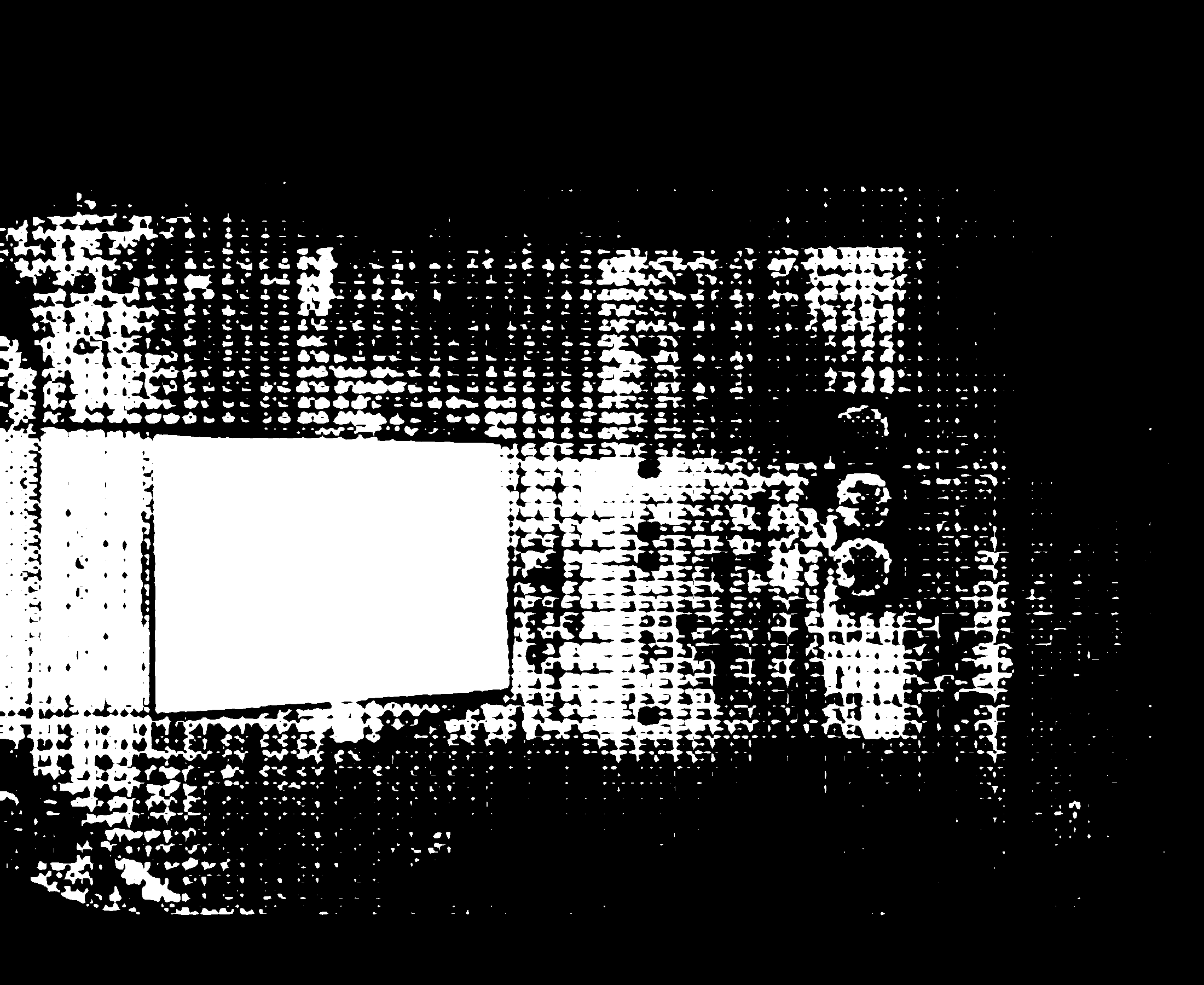}};
		\node(m1)[above of=m2, align=center, yshift=-0.1cm]{$\mathcal{M}_1$};
		\node(m11)[right of=m1,xshift=-0.2cm]{\includegraphics[width=0.05\textwidth]{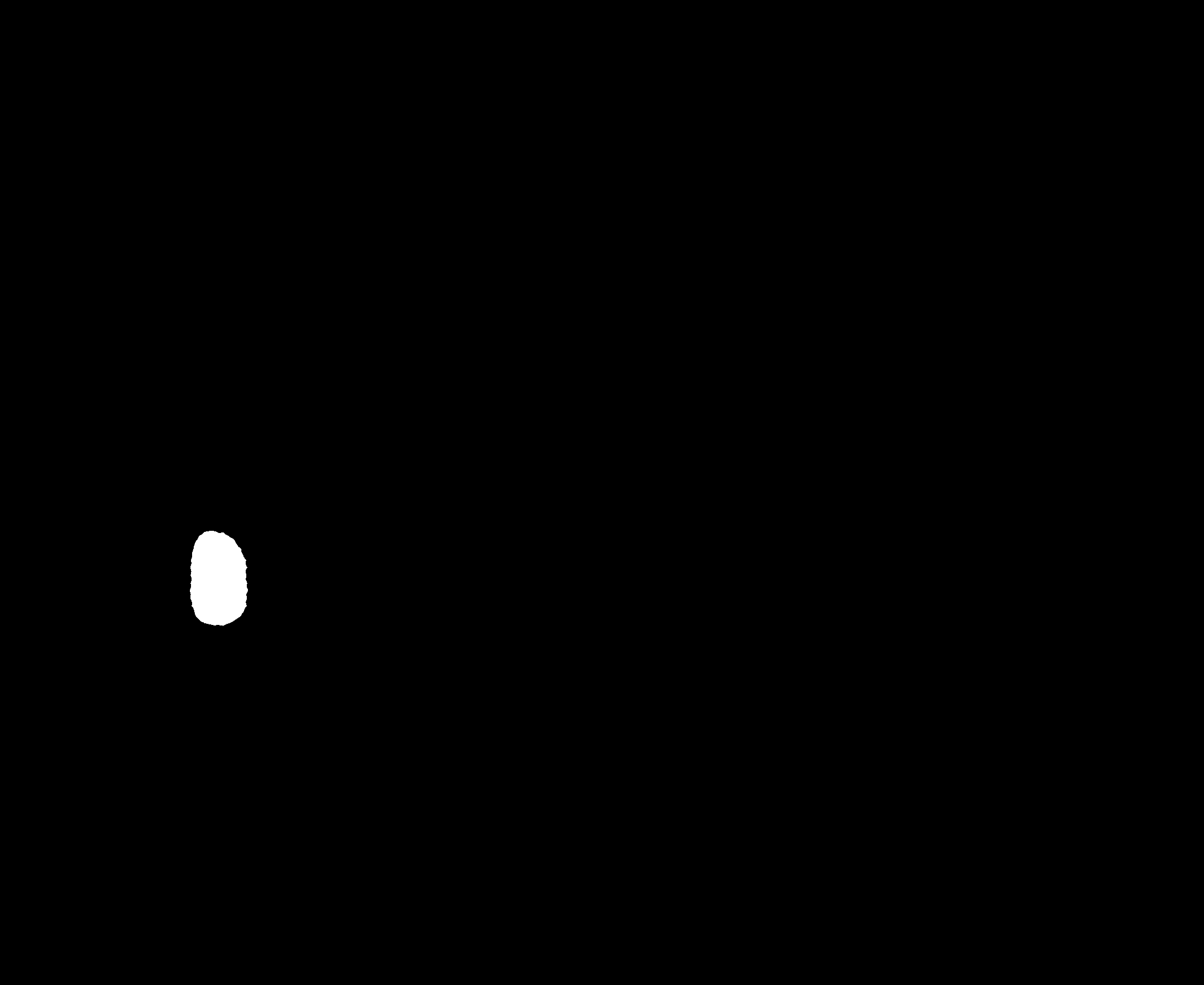}};
		\node(m3)[below of=m2, align=center, yshift=0.1cm]{$\mathcal{M}_3$};
		\node(m31)[right of=m3, xshift=-0.2cm]{\includegraphics[width=0.05\textwidth]{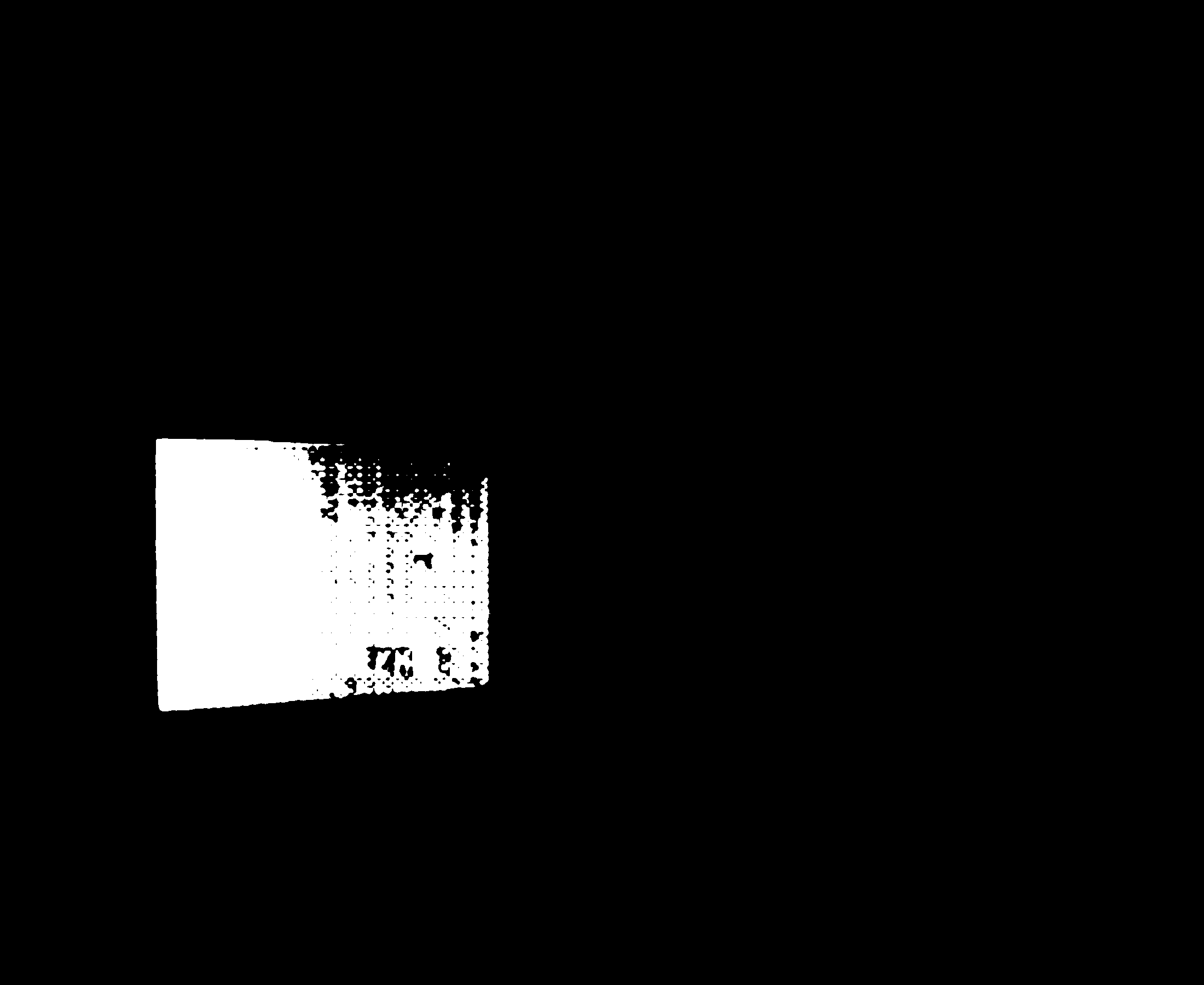}};
		
		\node(MS)[right of=m21, align=center, draw, dashed, color=blue, fill=blue!10, minimum width=5cm, minimum height=5.2cm, xshift = 3.7cm, yshift=-1.3cm] {};
		\node(MSLabel)[below of=MS,yshift=-2cm] {\color{blue}\textbf{Mask selector}};
		\node(step1)[below of=MS, fill=white, yshift=2.6cm, text width=3.6cm, align=center]{Test maximum white pixel ratio};
		\node(step2) [below of=step1, xshift=1.2cm, fill=white, text width=1.2cm, yshift=-1.8cm, align=center]{Discard mask};
		\node(step3)[below of=step1, xshift=-1.2cm, fill=white, yshift=-1.8cm, text width=1.6cm, align=center]{Find $\arg\max R_i$};
		\draw[->](step1.south) |- ++(0,-1.1) 
		-| (step2.north);
		\node(no)[align=center, above of=step2, xshift=0.3cm, yshift=0.5cm]{$R_i > R_\text{max}$};
		\draw[->](step1.south) |- ++(0,-1.1) 
		-| (step3.north);
		\node(yes)[align=center, above of=step3, xshift=-0.3cm, yshift=0.5cm]{$R_i \leq R_\text{max}$};

		\node(PP)[left of=MS, align=center, draw, dashed, color=red, fill=red!10, minimum width=3.4cm, minimum height=3.1cm, xshift=-5.5cm, yshift=-2cm] {};
		\node(PPLabel)[below of=PP,yshift=-0.8cm] {\color{red}\textbf{Post processing}};
		\node(step1)[below of=PP, yshift=2.2cm, fill=white]{CCA};
		\node(step2) [below of=step1, yshift=0.2cm, fill=white]{Invert mask};
		\node(step3)[below of=step2, yshift=0.2cm, fill=white]{CCA};
		\node(step4)[below of=step3, yshift=0.2cm, fill=white]{Invert mask};
		\draw[->](step1) -- (step2);
		\draw[->](step2) -- (step3);
		\draw[->](step3) -- (step4);
		
		\node(M)[left of=PP, align=center, xshift=-3.5cm]{$\mathcal{M}$};
		
		\draw[->](Image)--(SRD) node[midway, above]{\includegraphics[width=0.07\textwidth]{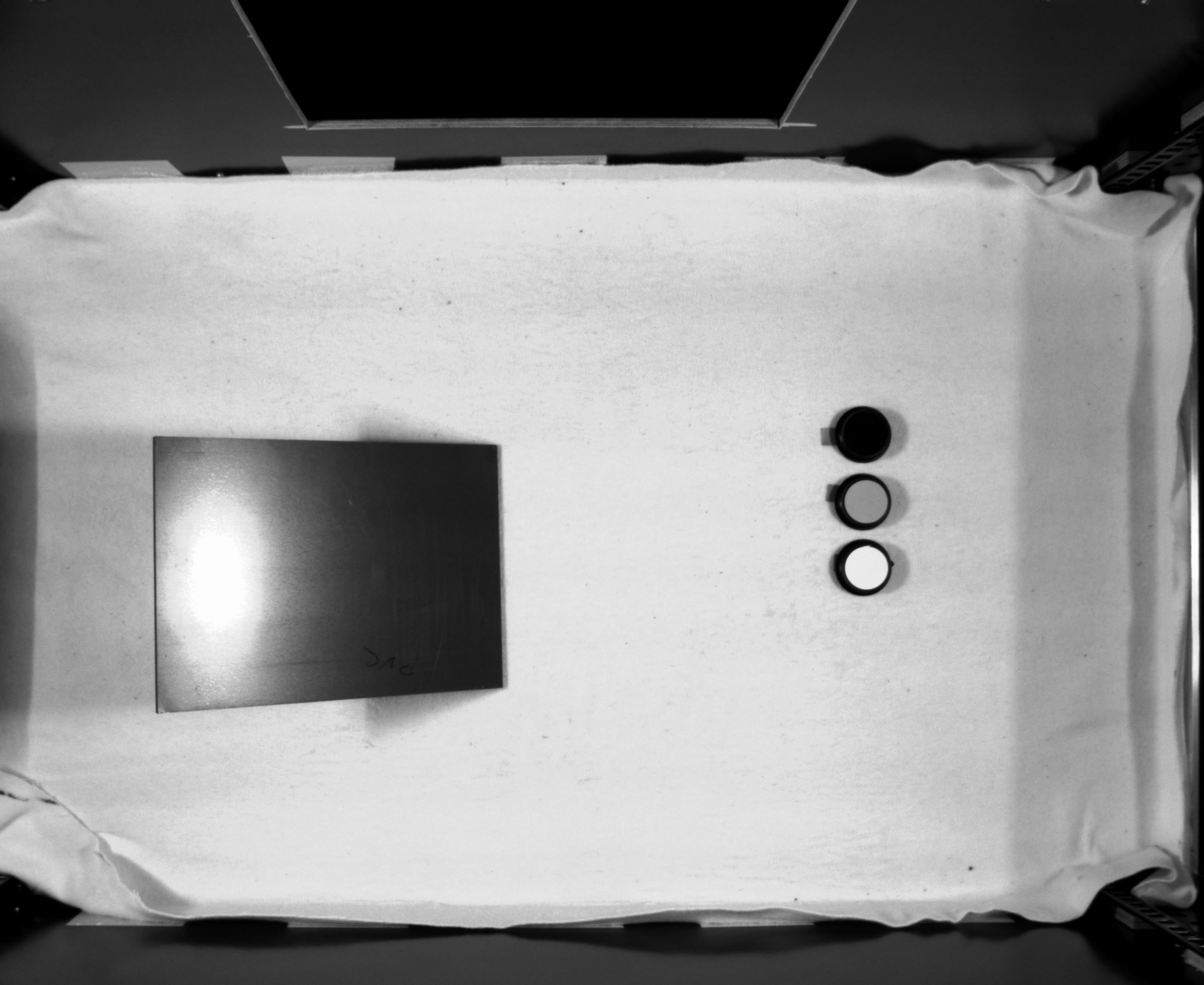}};
		\draw[->](SRD)--(sam2) node[midway, above]{$\Omega, [c_x,c_y] \in \Omega$} ;
		\draw[->](sam2)--(m2);
		\draw[->](sam2.east)++(0,0.9)--(m1);
		\draw[->](sam2.east)++(0,-0.9)--(m3);
		\draw[->](m21)--($(MS.west) + (0,1.3)$);
		\draw[->](m11.east)-| ++(0.6,-1.0) 
		|-($(MS.west) + (0,1.3)$);
		\draw[->](m31.east)-| ++(0.6,1.0) 
		|-($(MS.west) + (0,1.3)$);
		\node[circle, fill=black, inner sep=1.5pt, right of=m21, xshift=0.2cm] {};		
		\draw[->](MS.west)++(0,-2)--(PP.east) node[midway, above]{\includegraphics[width=0.07\textwidth]{Images/scene4/output_mask_2.png}}node[midway, below]{$\mathcal{M}_{\text{selected}}$};
		\draw[->](PP)--(M)node[midway, above]{\includegraphics[width=0.07\textwidth]{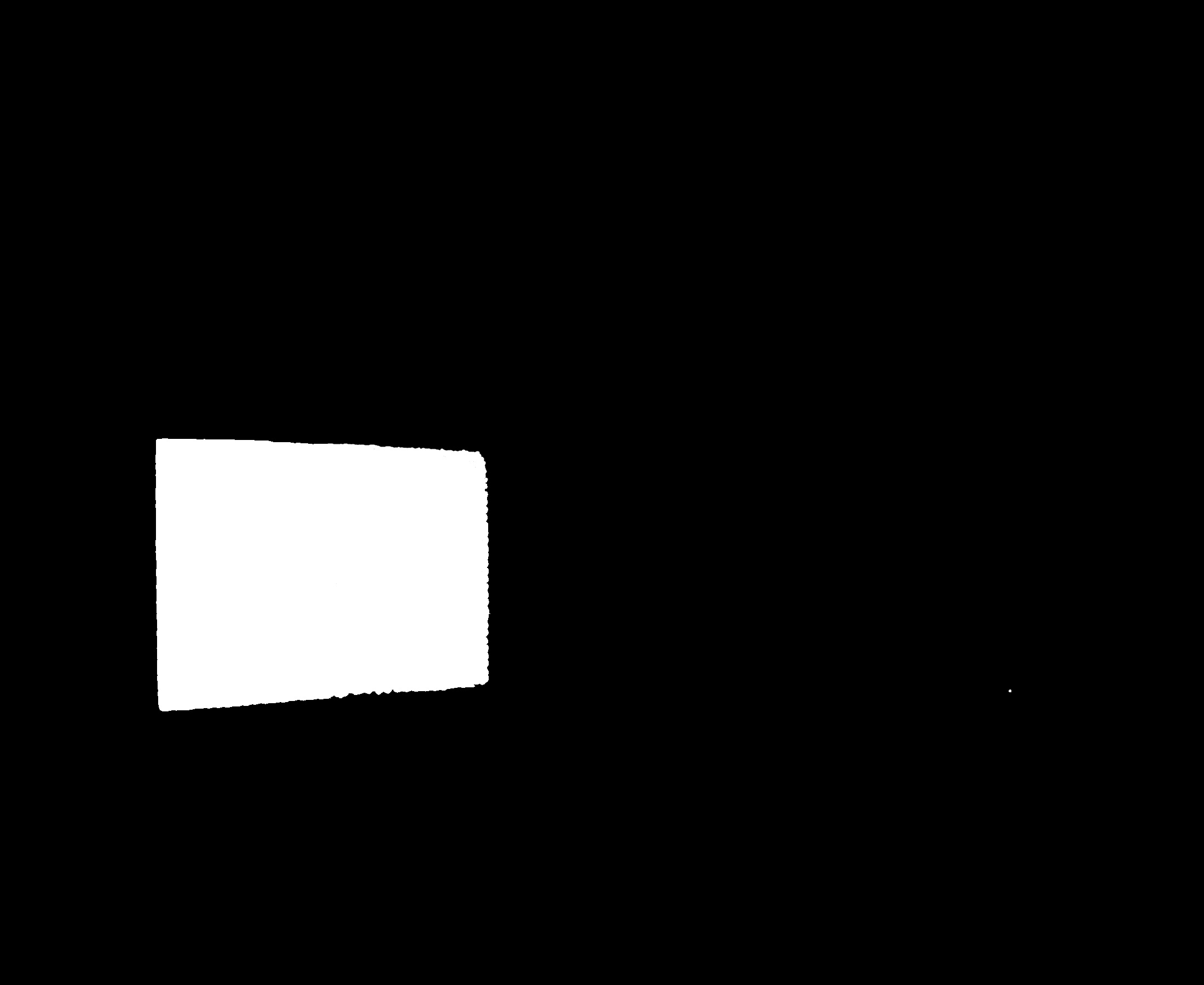}};
	\end{tikzpicture}
	\vspace{-0.4cm}
	\caption{Overview of the proposed pipeline for object mask selection using specular reflections. Starting from the input image $\mathcal{I}$, specular reflection detection identifies candidate regions, which are then processed by SAM2 \cite{SAM2} to generate multiple segmentation masks $\mathcal{M}_1$, $\mathcal{M}_2$, $\mathcal{M}_3$. The mask selector evaluates each candidate using the white-pixel ratio $R_i$ and discards masks above a predefined threshold $R_{\text{max}}$. Among the remaining candidates, the mask with the highest $R_i$ is selected as $\mathcal{M}_\text{selected}$. Finally, the post-processing step refines this mask through connected component analysis and mask inversion to obtain the final segmentation result $\mathcal{M}$.}
	\label{fig:Pipeline}
	\vspace{-0.4cm}
\end{figure*}

Fig. \ref{fig:ComparisonDifferentMethods} shows the segmentation results for a flat, textureless object with a specular reflection using Otsu \cite{Otsu}, Mask R-CNN \cite{MaskRCNN}, YOLO \cite{YOLO}, SAM2 \cite{SAM2} and our proposed solution. It is obvious that classical methods like Otsu misinterpret the reflections as additional object, leading to distorted masks. Even the more modern deep learning methods like Mask R-CNN or YOLO produce unsharp mask boundaries and often show inconsistencies within the predicted masks. In contrast, foundation models like SAM2 show improved segmentation by leveraging large-scale training data and richer contextual reasoning. While reflective regions are included in the object masks, shadows are occasionally partially segmented, resulting in incomplete or fragmented mask artifacts. Furthermore, with multiple masks being available, the best mask still has to be determined. 

In robotic assembly, a misaligned grasp point caused by an inaccurate segmentation can result in mechanical errors or system shutdowns \cite{AutoApp2}. In automated recycling, over- or undersegmentation of reflective packaging can lead to misclassification and reduced material purity \cite{Recycling}. In industrial inspection systems, reflections misidentified as surface defects may cause false alarms and unnecessary rejection of products \cite{FoodApp}. Therefore, this work focuses on developing and evaluating an improved strategy for automatically selecting the most accurate mask from a set of candidates, with attention to the exclusion of specular reflections in the segmentation of reflective objects. Importantly, our approach does not rely on models that have been specifically trained to handle specular reflections. Instead, it identifies the largest connected region containing the reflection, thereby increasing the robustness of generic segmentation pipelines. By addressing this gap, we aim to improve the reliability and accuracy of vision-based automation systems in demanding environments.


\section{Proposed Method}
Our novel method \underline{Re}flection-aware \underline{Po}stprocessed \underline{Seg}mentation (RePoSeg) is based on a concatenation of a specular reflection detection, the SAM2 model outputs, a mask selection and post processing. The proposed pipeline and concept is shown in Fig. \ref{fig:Pipeline}.

\begin{figure}[t!]
	\begin{tikzpicture}
		\node(i1)[]{\includegraphics[width=0.115\textwidth]{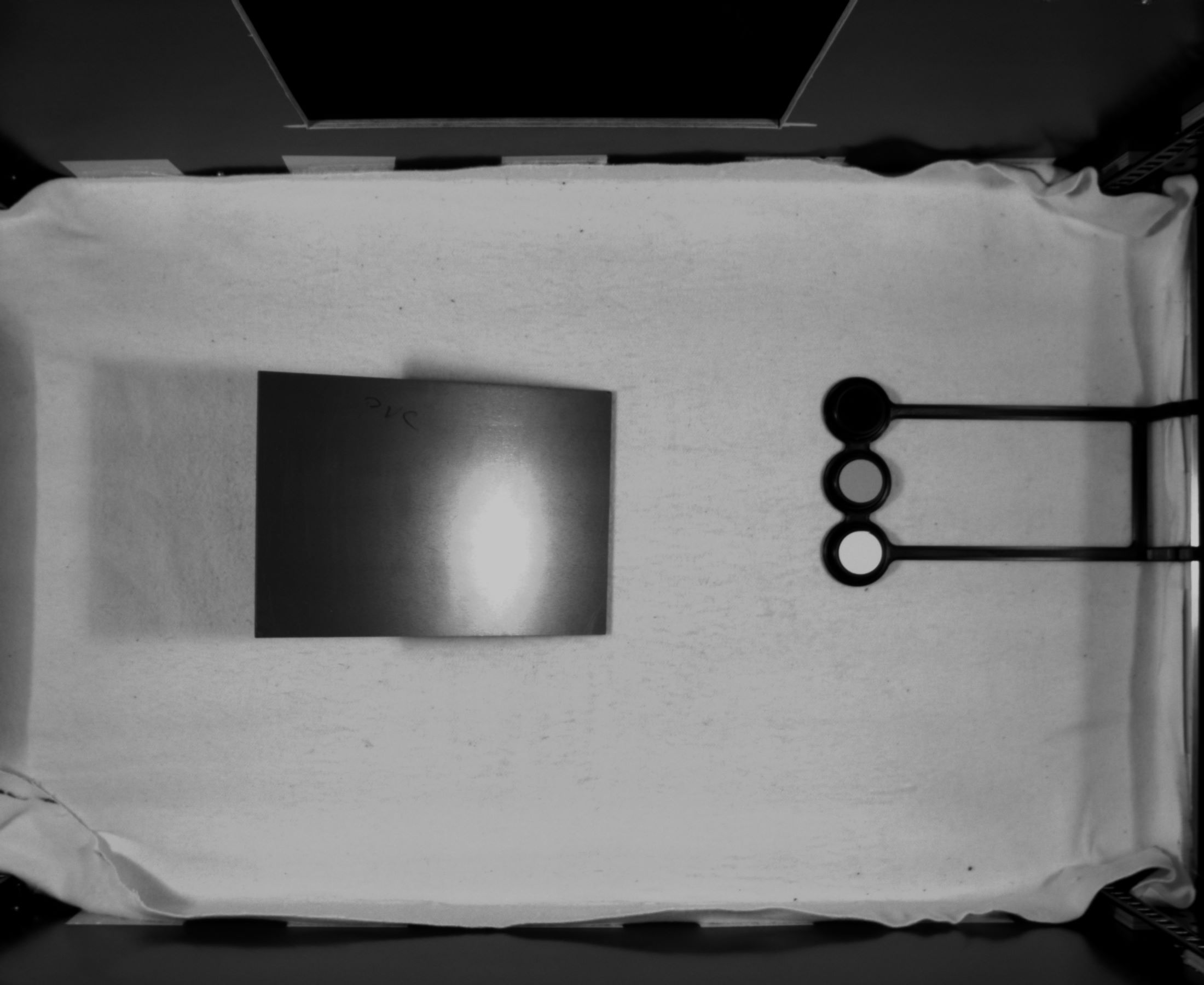}};
		\node(i2)[right of=i1, xshift=1.2cm]{\includegraphics[width=0.115\textwidth]{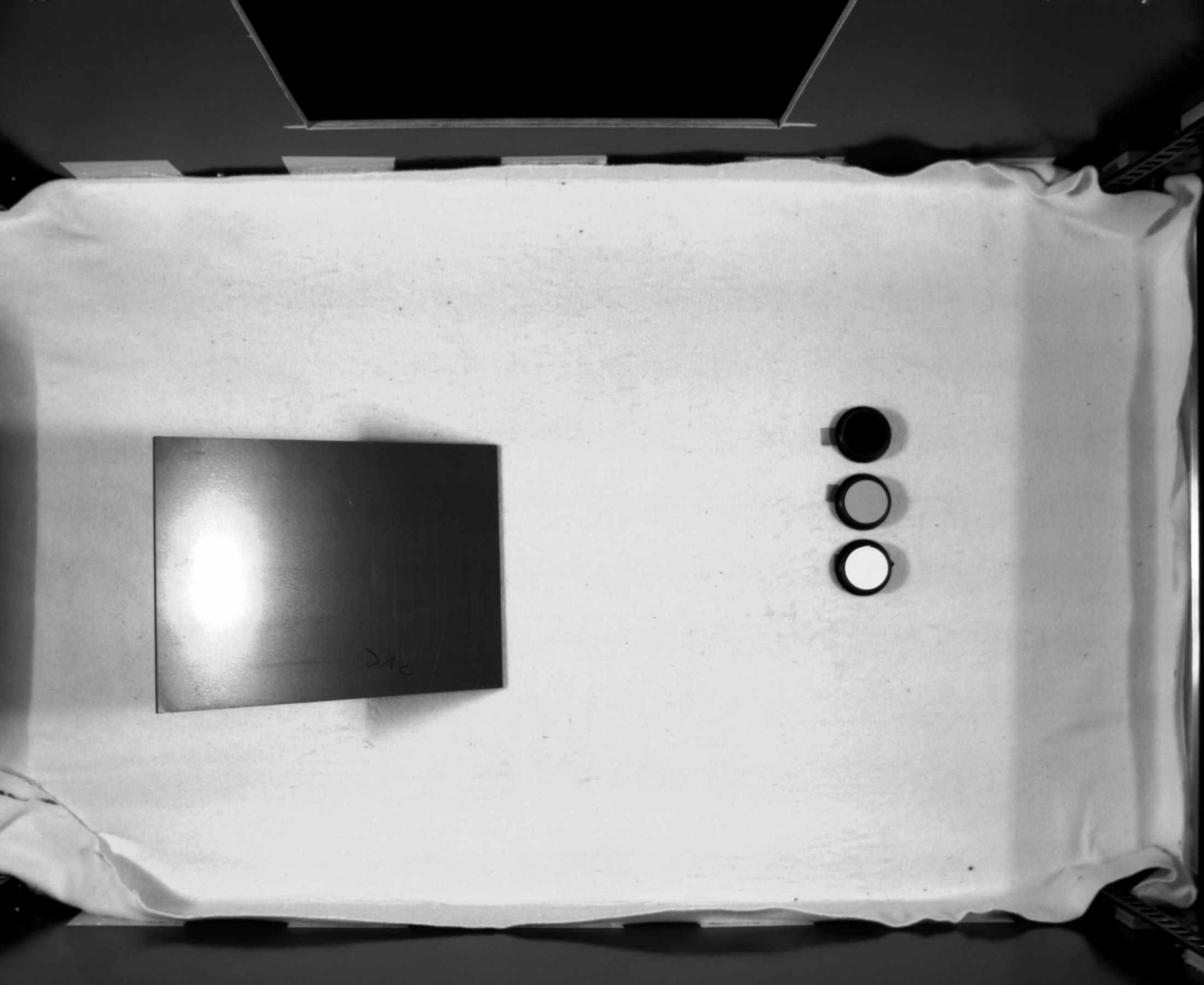}};
		\node(i3)[right of=i2, xshift=1.2cm]{\includegraphics[width=0.115\textwidth]{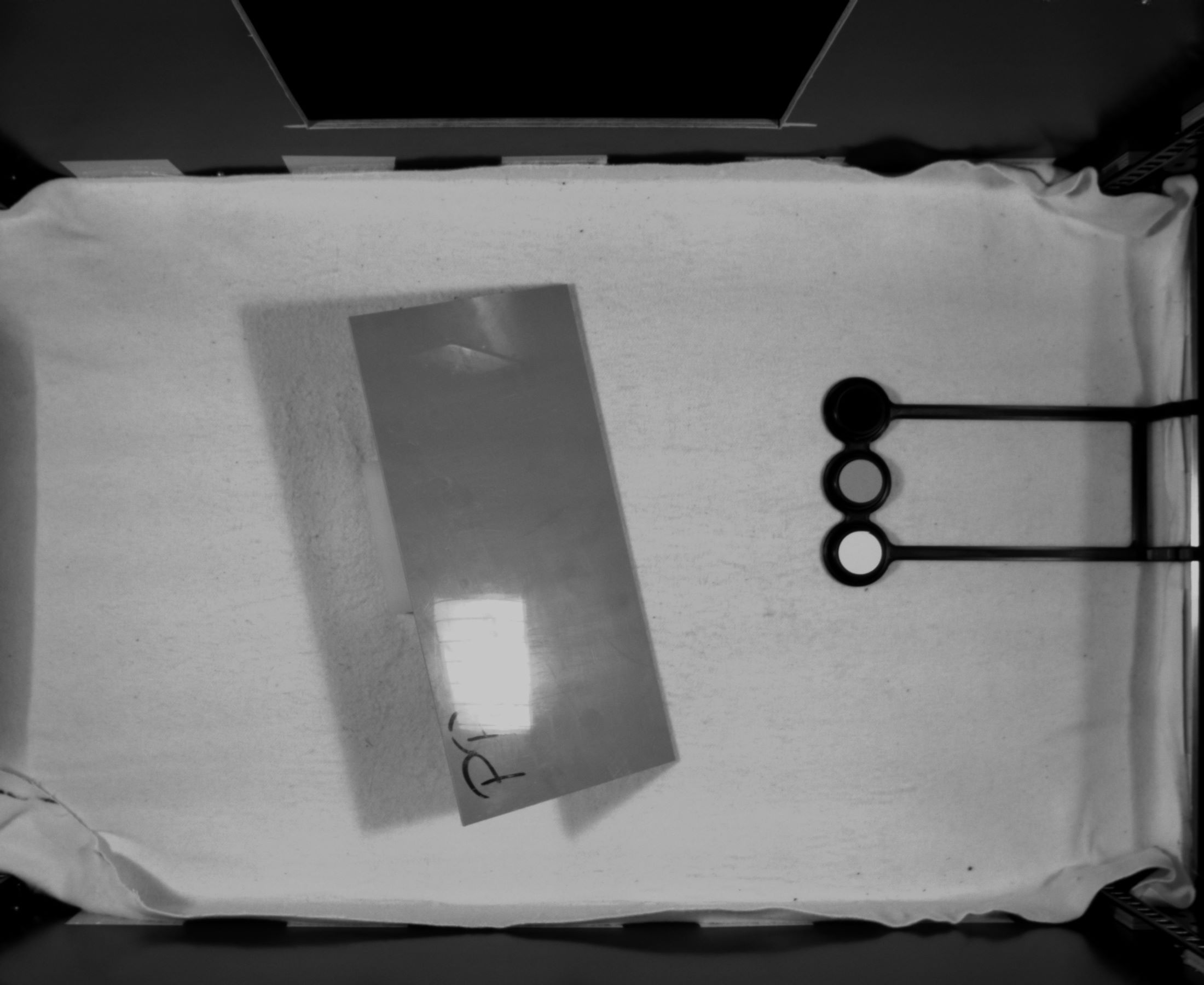}};
		\node(i4)[right of=i3, xshift=1.2cm]{\includegraphics[width=0.115\textwidth]{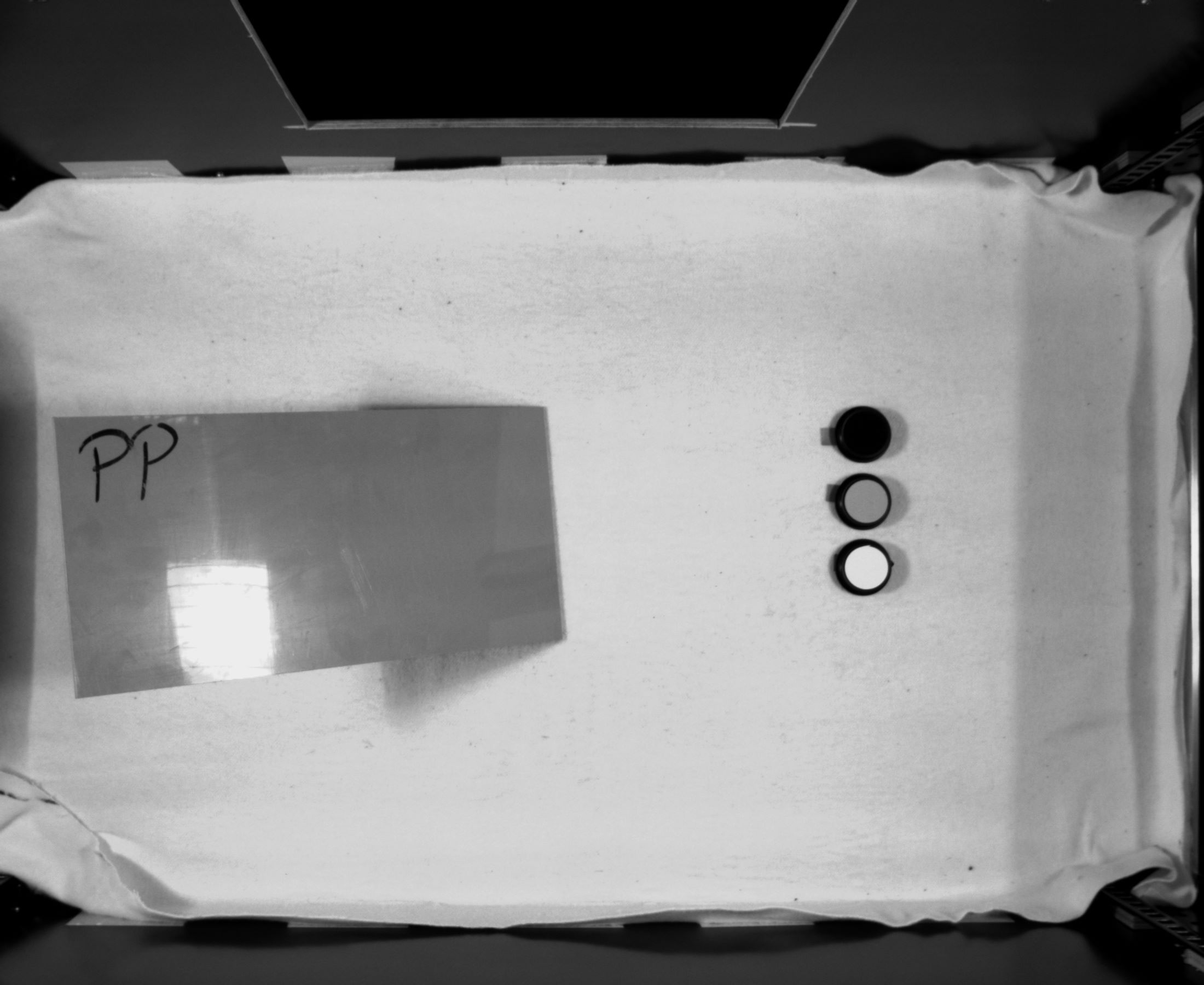}};
		\node(m1)[below of=i1, yshift=-0.8cm]{\includegraphics[width=0.115\textwidth]{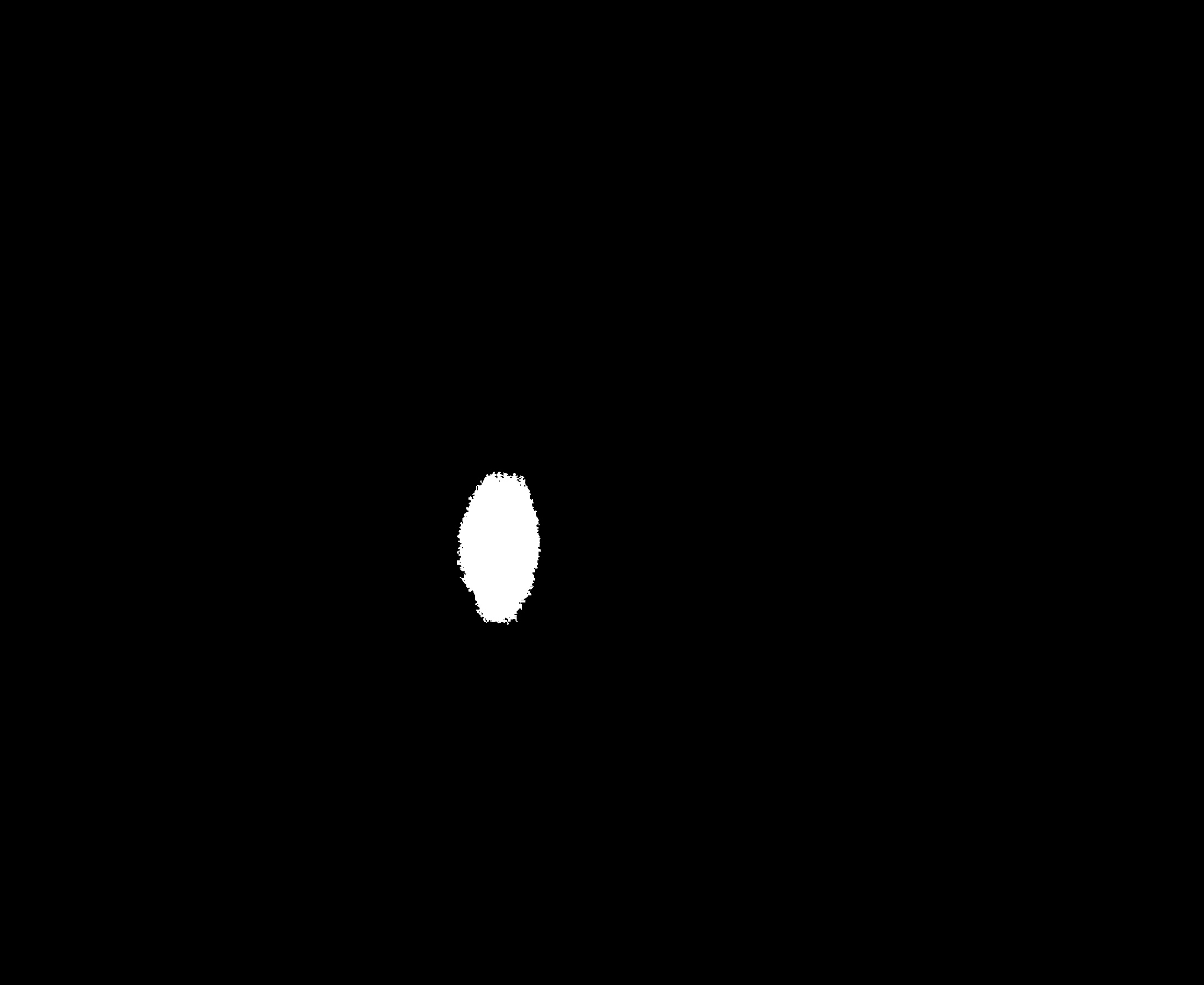}};
		\node(m2)[below of=i2, yshift=-0.8cm]{\includegraphics[width=0.115\textwidth]{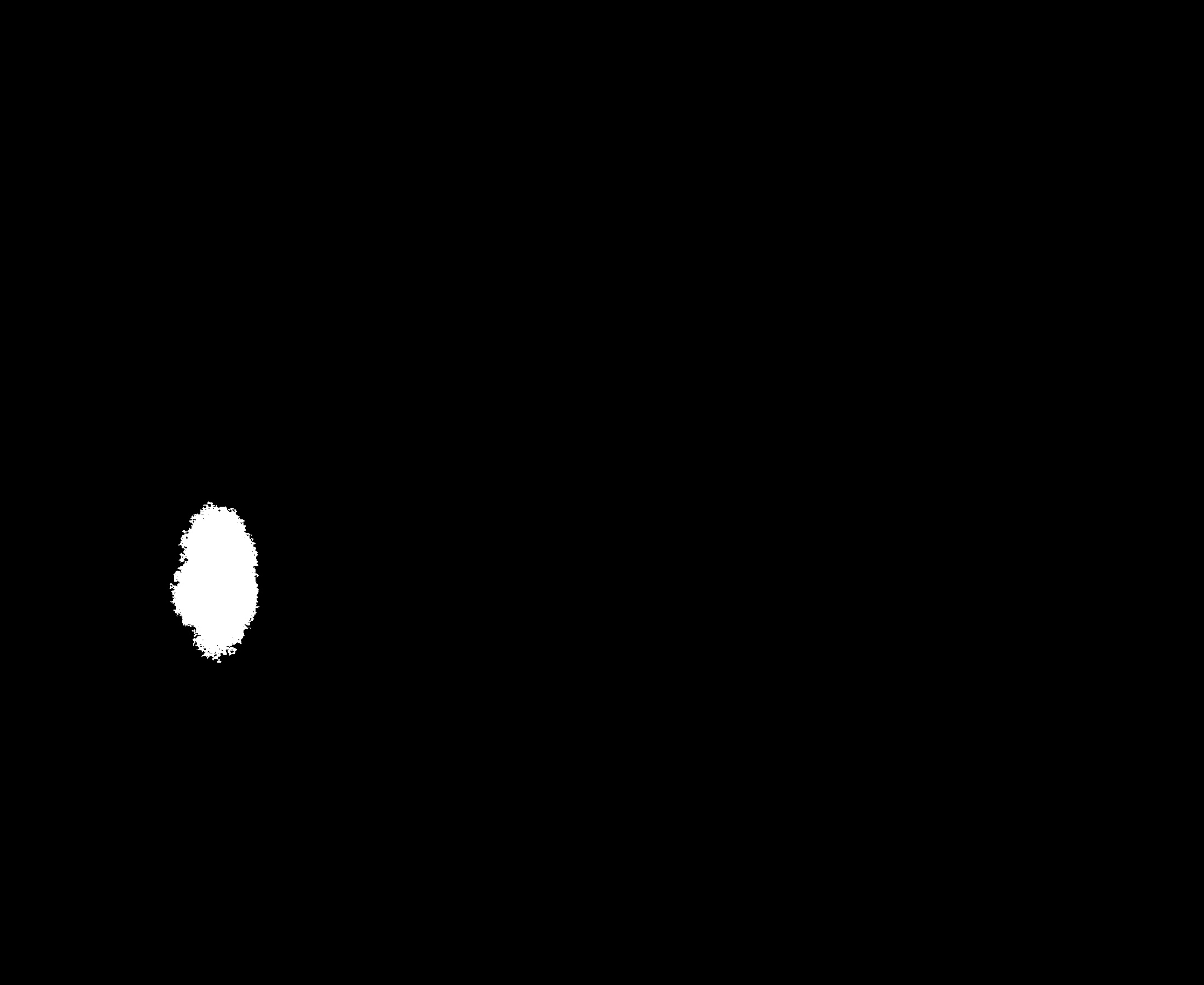}};
		\node(m3)[below of=i3, yshift=-0.8cm]{\includegraphics[width=0.115\textwidth]{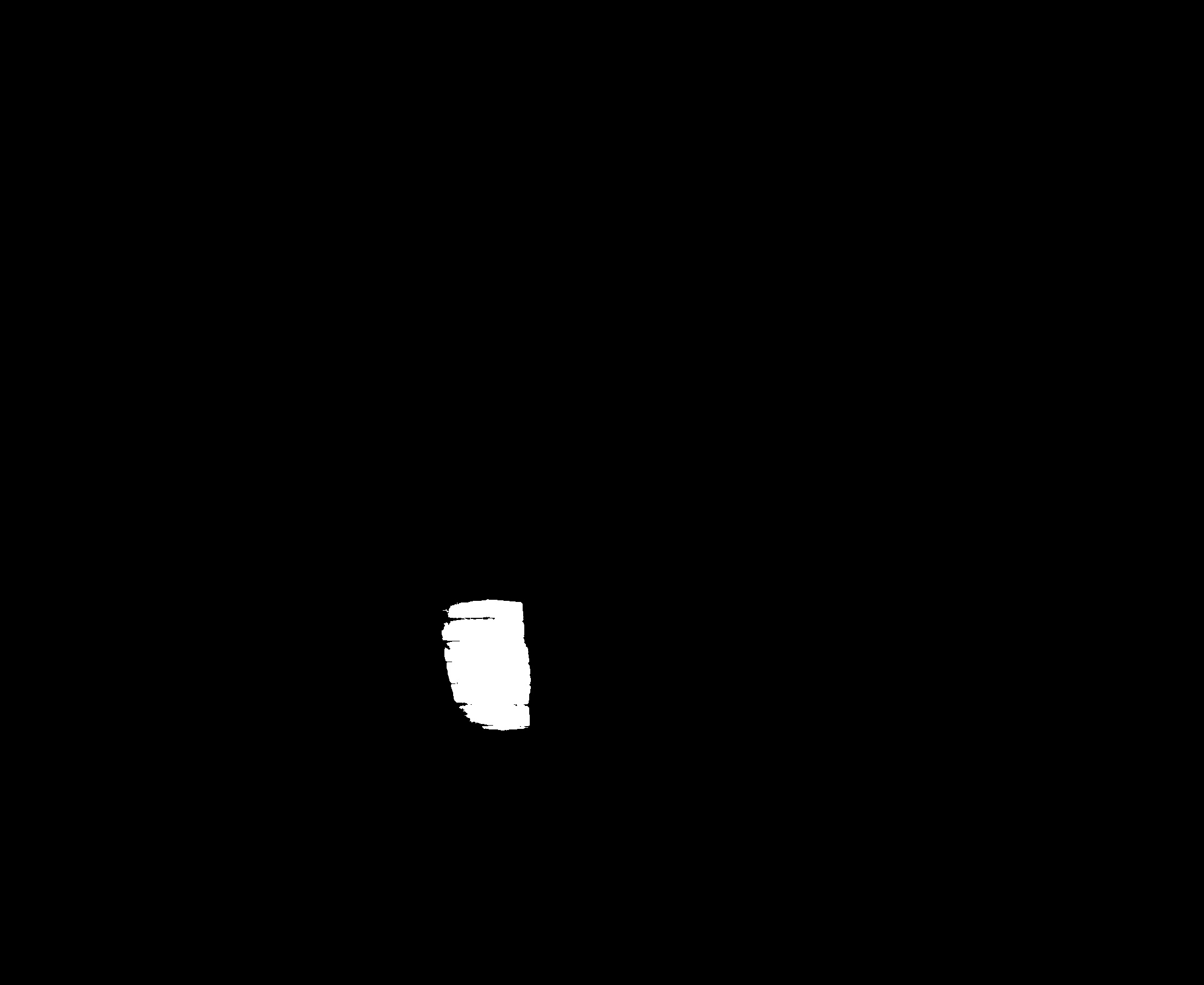}};
		\node(m4)[below of=i4, yshift=-0.8cm]{\includegraphics[width=0.115\textwidth]{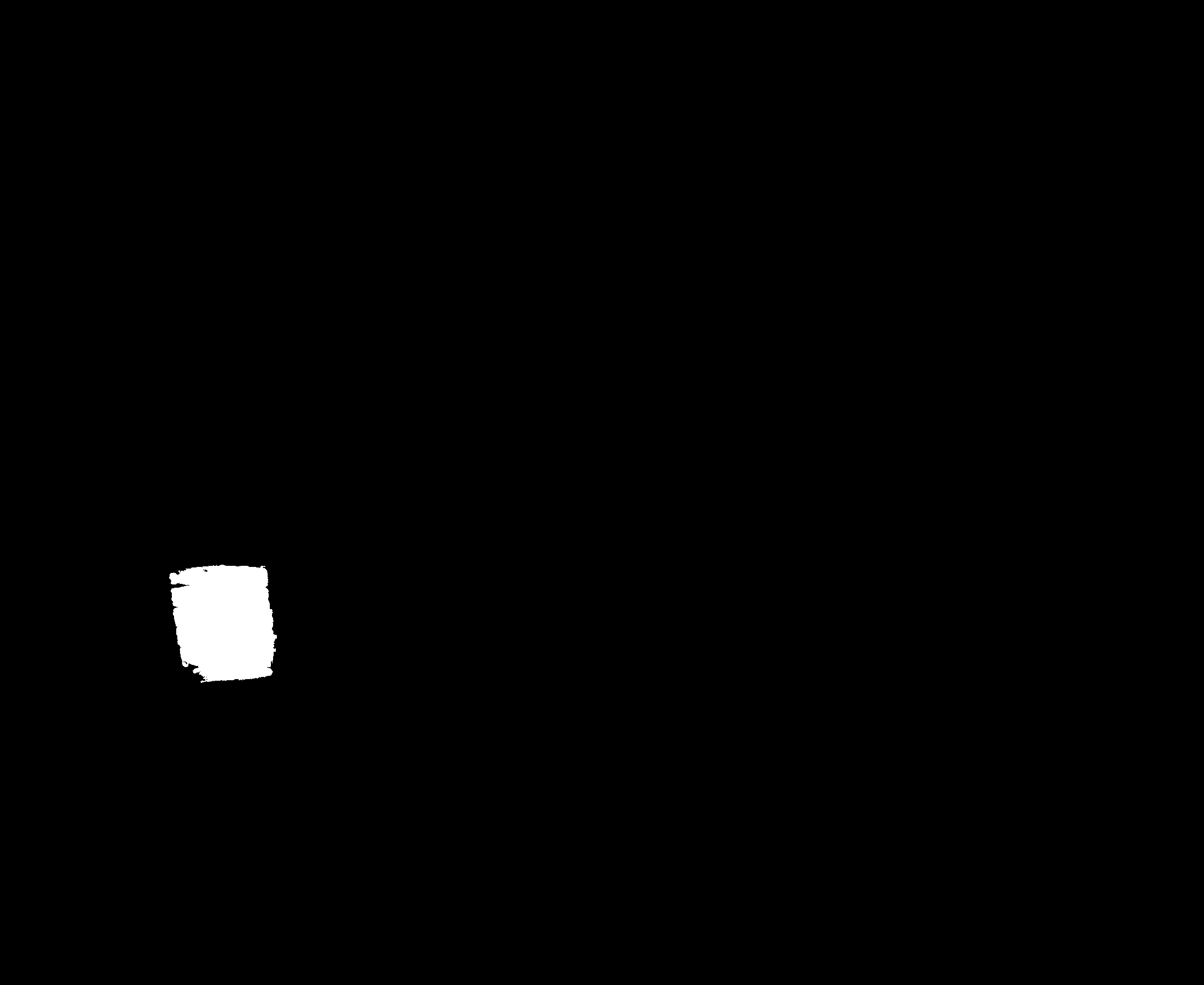}};
	\end{tikzpicture}
	\vspace{-0.7cm}
	\caption{Examples of the specular reflection detection based on \cite{SRD1}. The upper row shows the original image, while the bottom row displays the corresponding mask $\Omega$, which captures only the core specular highlights and ignores the surrounding intensity falloff.}
	\label{fig:ReflDet}
	\vspace{-0.3cm}
\end{figure}

Specular reflections manifest as intense highlights on surfaces, resulting from the coherent reflection of incident light. They occur predominantly on smooth surfaces like glass, plastics, or polished materials. As they hide the actual surface information of the object, the first step is to detect their location in the image $\mathcal{I}$. For this, already established methods such as Y-channel histogram exploration \cite{SRD1}, adaptive thresholding histogram exploration \cite{SRD2}, or neural network based approaches like \cite{SRD3} can be employed. Fig. \ref{fig:ReflDet} shows some results of the specular reflection detection based on \cite{SRD1}.
The white area of each specular mask is denoted as $\Omega$. Its center of mass $C = [c_x, c_y]$ for positions $[x,y] \in \Omega$ is generally computed as
\begin{equation}
	c_x = \frac{1}{|\Omega|}\sum_{(x,y \in \Omega)} x ~~~, ~~~ c_y = \frac{1}{|\Omega|}\sum_{(x,y \in \Omega)} y ~~.
\end{equation}

\begin{figure}[t!]
	\centering
	\begin{tikzpicture}
		\node(Or1)[]{\includegraphics[width=0.09\textwidth]{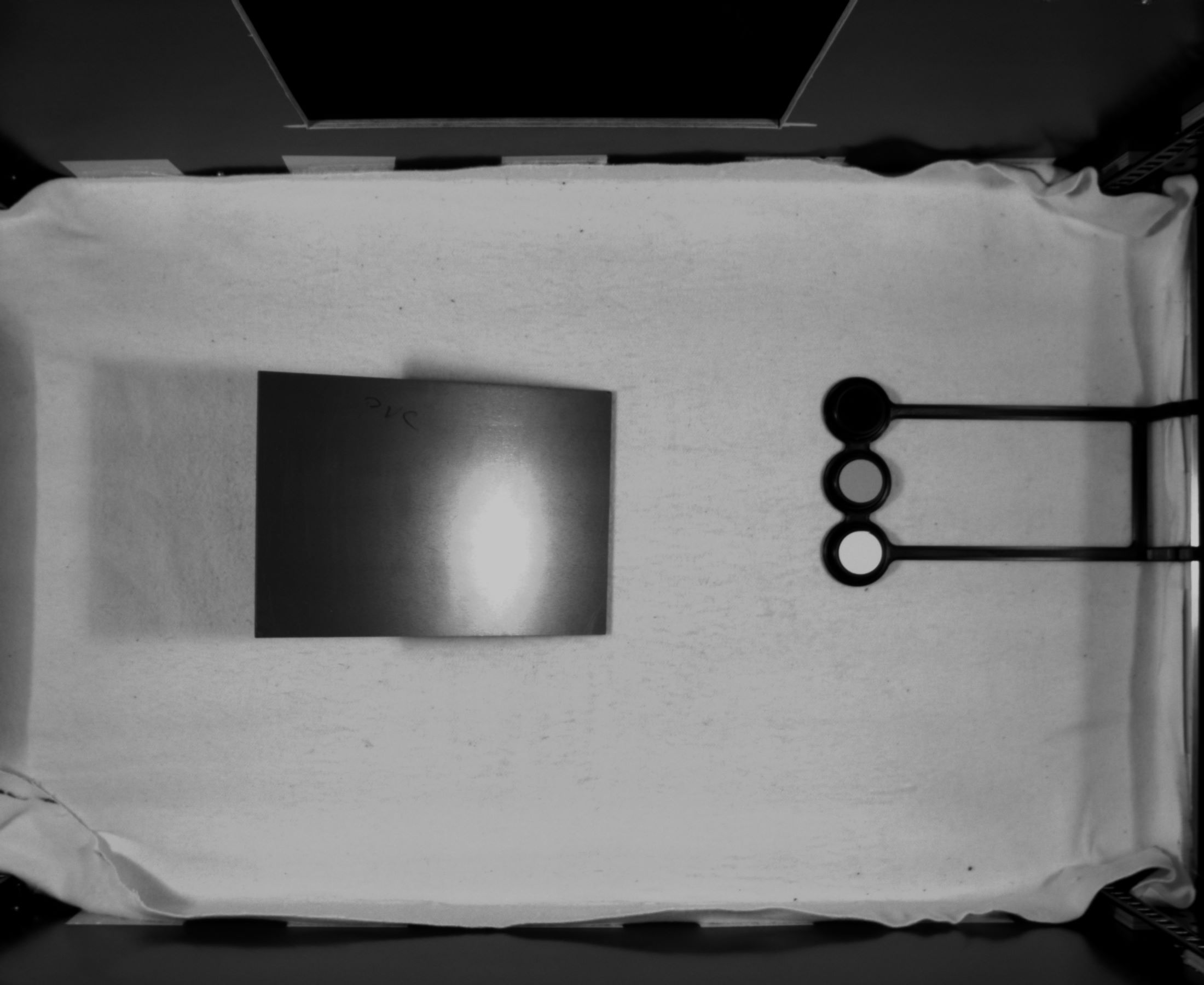}};
		\node(M11)[right of=Or1, xshift=0.8cm]{\includegraphics[width=0.09\textwidth]{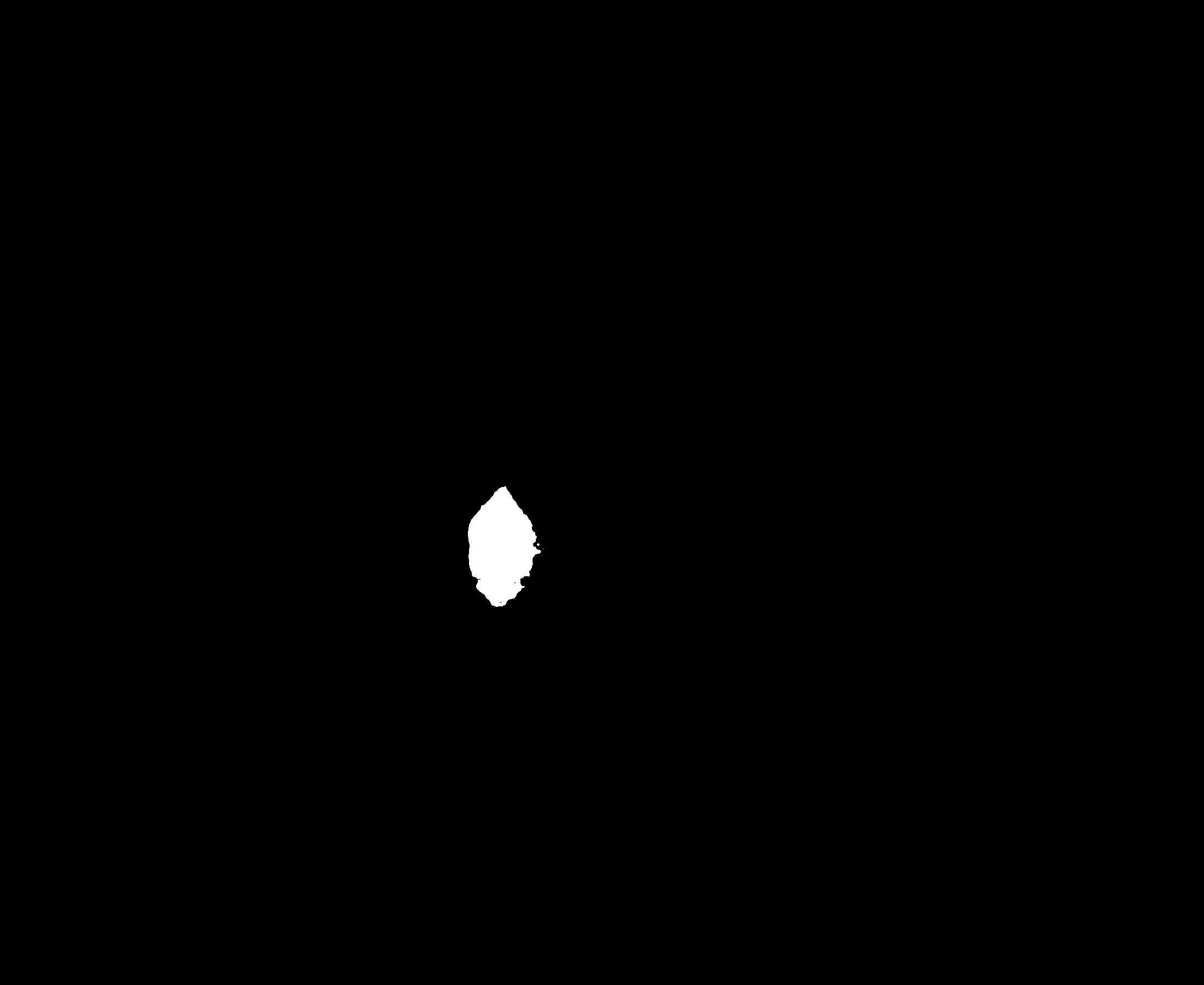}};
		\node(M21)[right of=M11,xshift=0.7cm]{\includegraphics[width=0.09\textwidth]{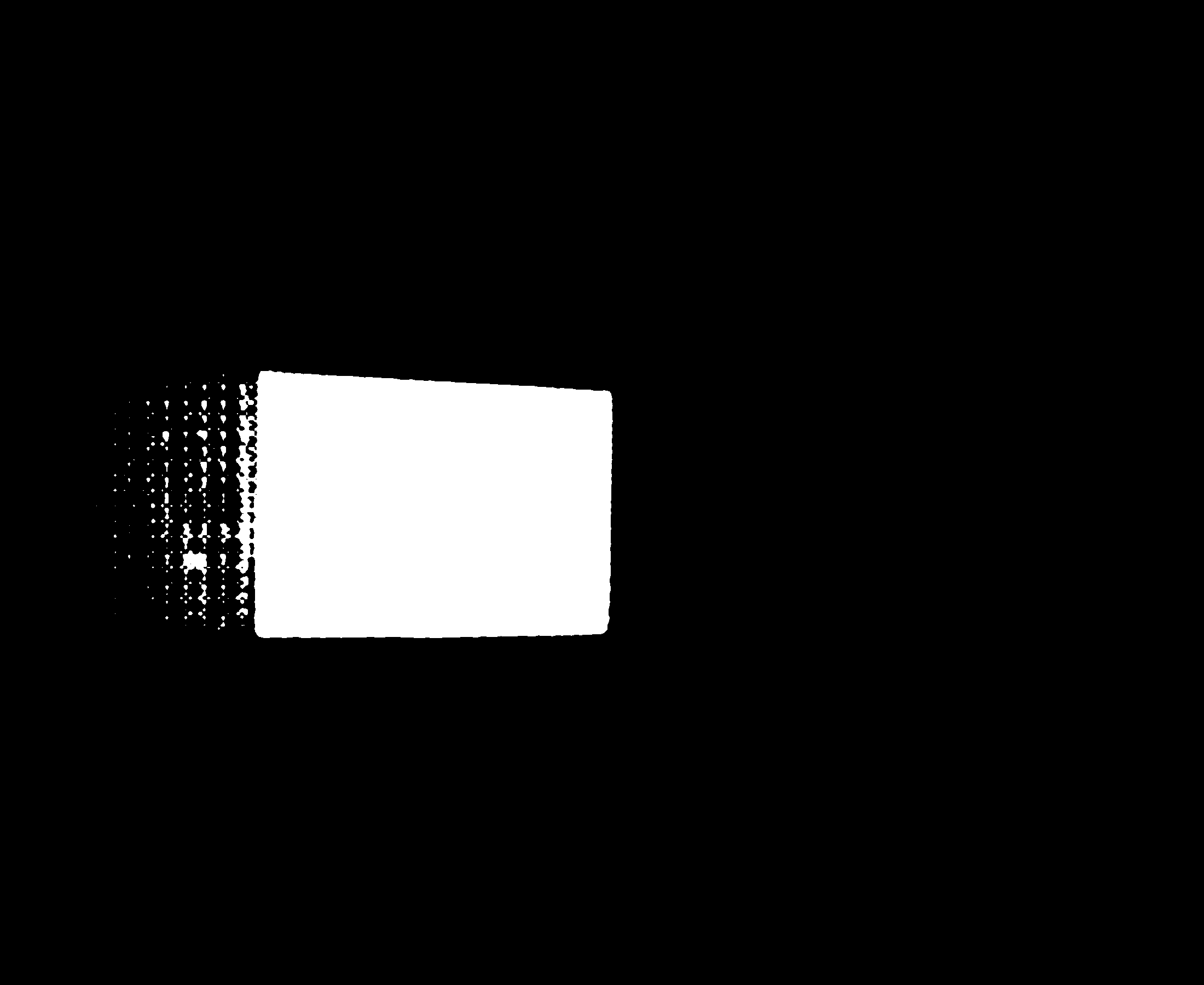}};
		\node(M31)[right of=M21,xshift=0.7cm]{\includegraphics[width=0.09\textwidth]{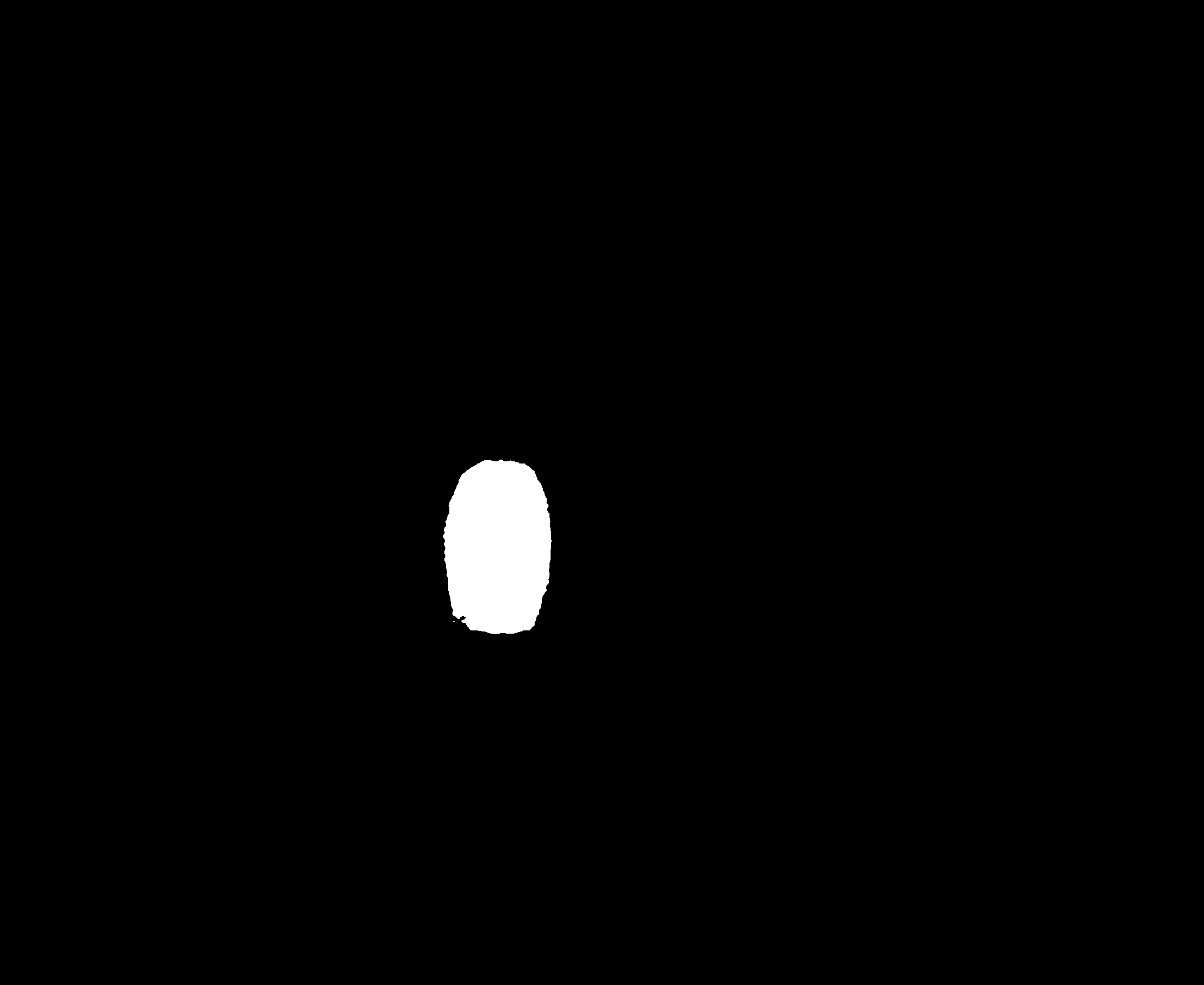}};
		\node(MF1)[right of=M31,xshift=1cm]{\includegraphics[width=0.09\textwidth]{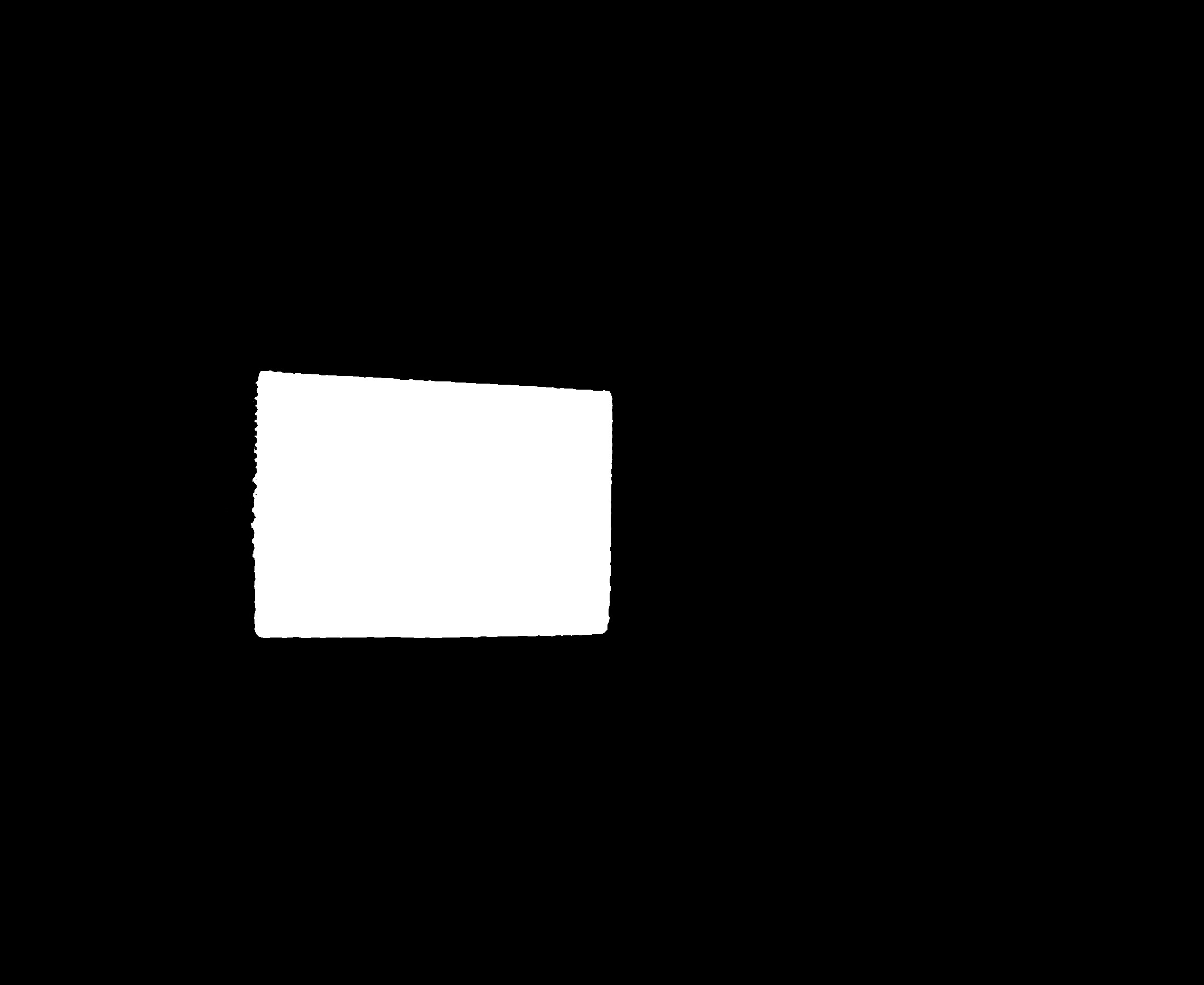}};
		
		\node(Or2)[below of=Or1, yshift=-0.4cm]{\includegraphics[width=0.09\textwidth]{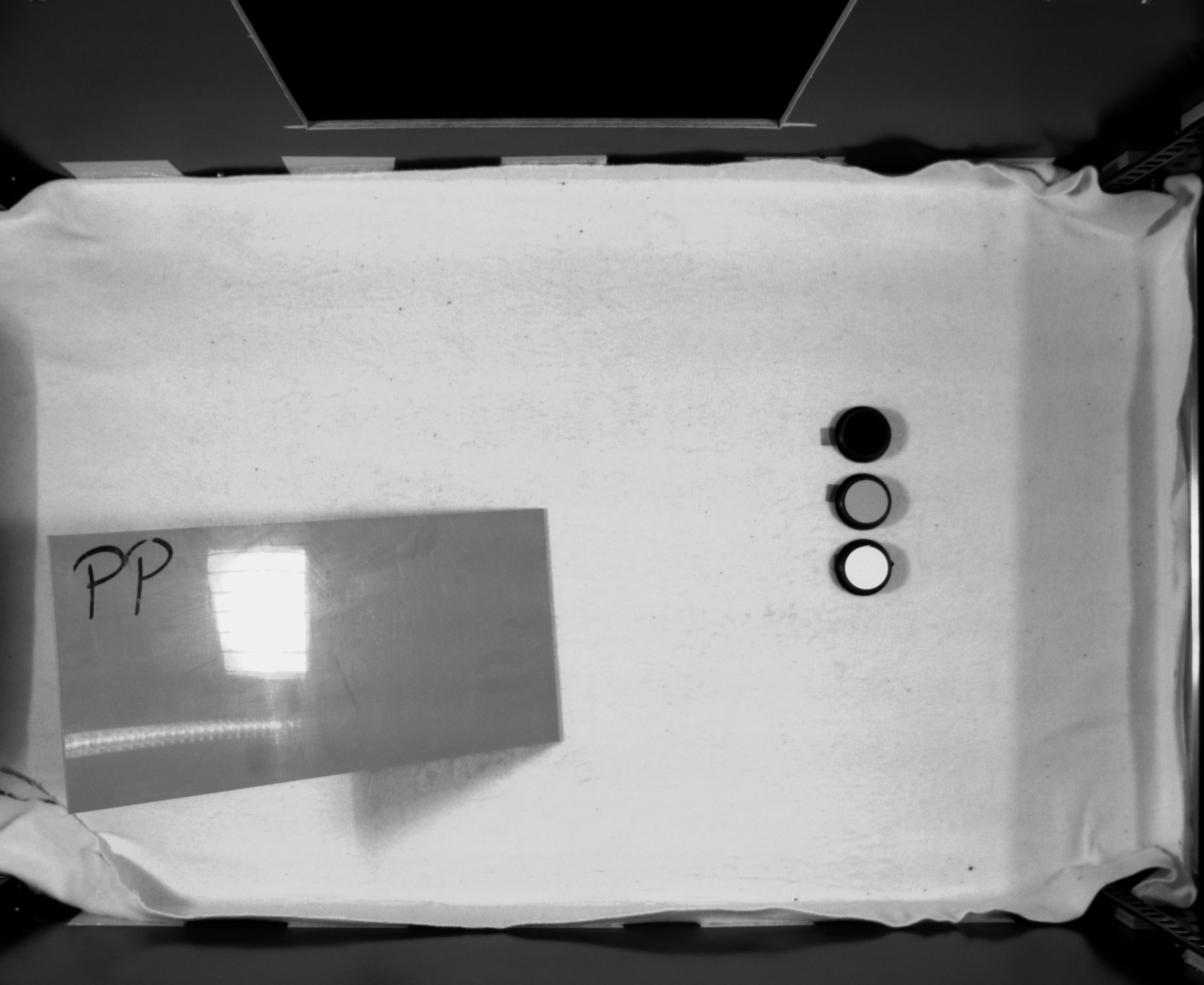}};
		\node(M12)[below of=M11, yshift=-0.4cm]{\includegraphics[width=0.09\textwidth]{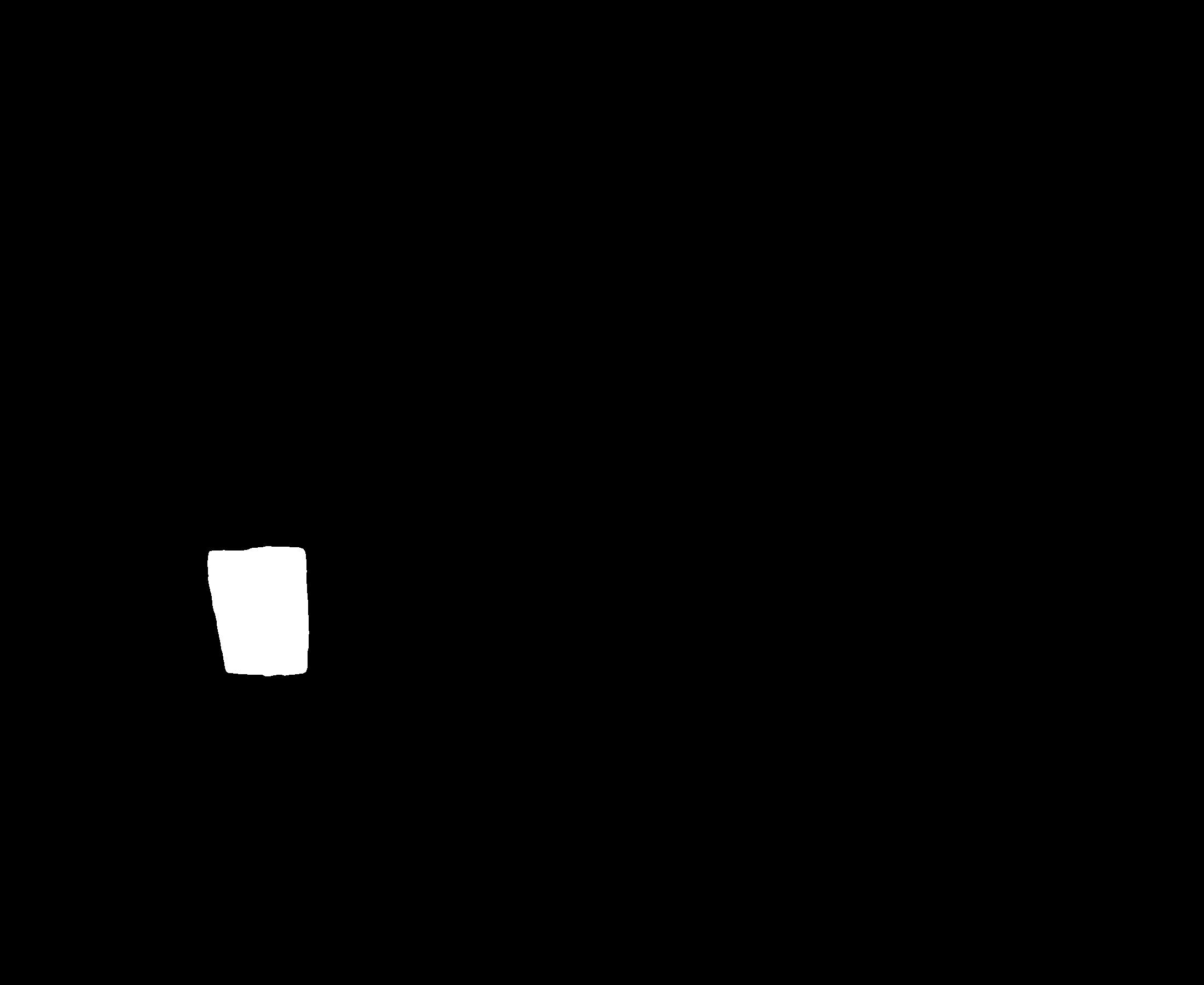}};
		\node(M22)[below of=M21, yshift=-0.4cm]{\includegraphics[width=0.09\textwidth]{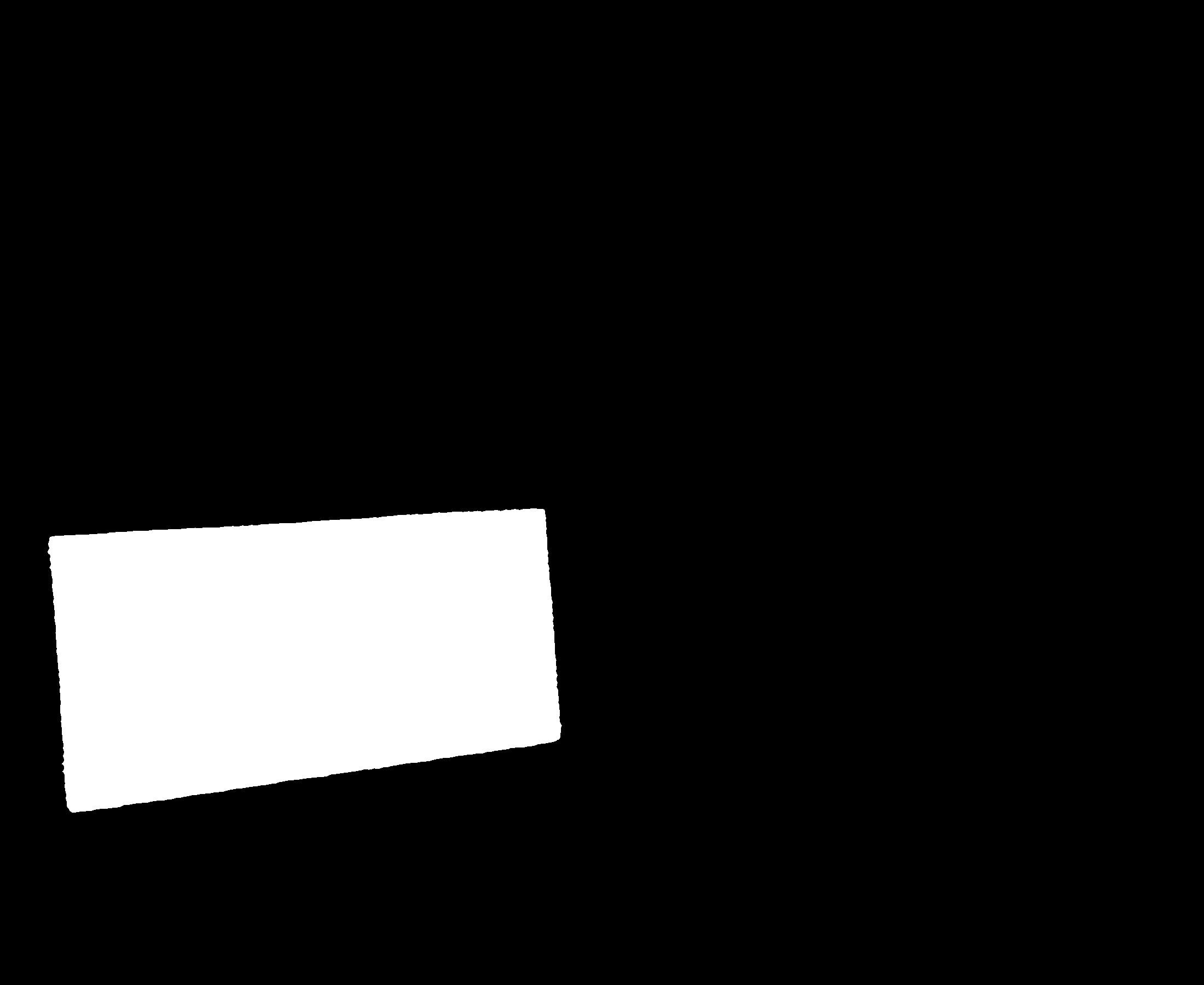}};
		\node(M32)[below of=M31, yshift=-0.4cm]{\includegraphics[width=0.09\textwidth]{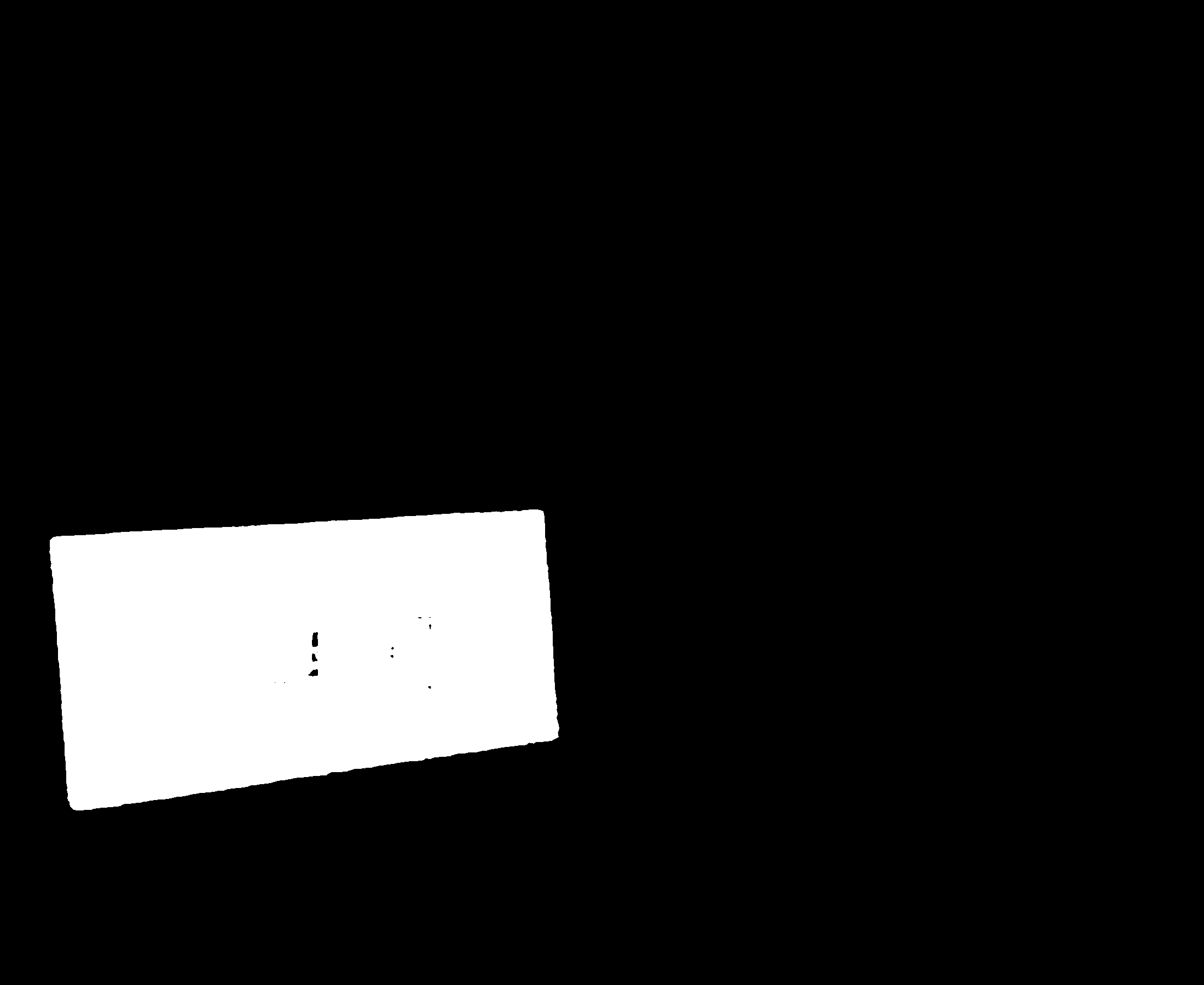}};
		\node(MF2)[below of=MF1, yshift=-0.4cm]{\includegraphics[width=0.09\textwidth]{Images/scene5/output_mask_1.jpg}};
		
		\node(Or3)[below of=Or2, yshift=-0.4cm]{\includegraphics[width=0.09\textwidth]{Images/scene4/cam-4.jpg}};
		\node(M13)[below of=M12, yshift=-0.4cm]{\includegraphics[width=0.09\textwidth]{Images/scene4/output_mask_0.png}};
		\node(M23)[below of=M22, yshift=-0.4cm]{\includegraphics[width=0.09\textwidth]{Images/scene4/output_mask_1.png}};
		\node(M33)[below of=M32, yshift=-0.4cm]{\includegraphics[width=0.09\textwidth]{Images/scene4/output_mask_2.png}};
		\node(MF3)[below of=MF2, yshift=-0.4cm]{\includegraphics[width=0.09\textwidth]{Images/scene4/output_mask_2_1.jpg}};
		
		\node(Or4)[below of=Or3, yshift=-0.4cm]{\includegraphics[width=0.09\textwidth]{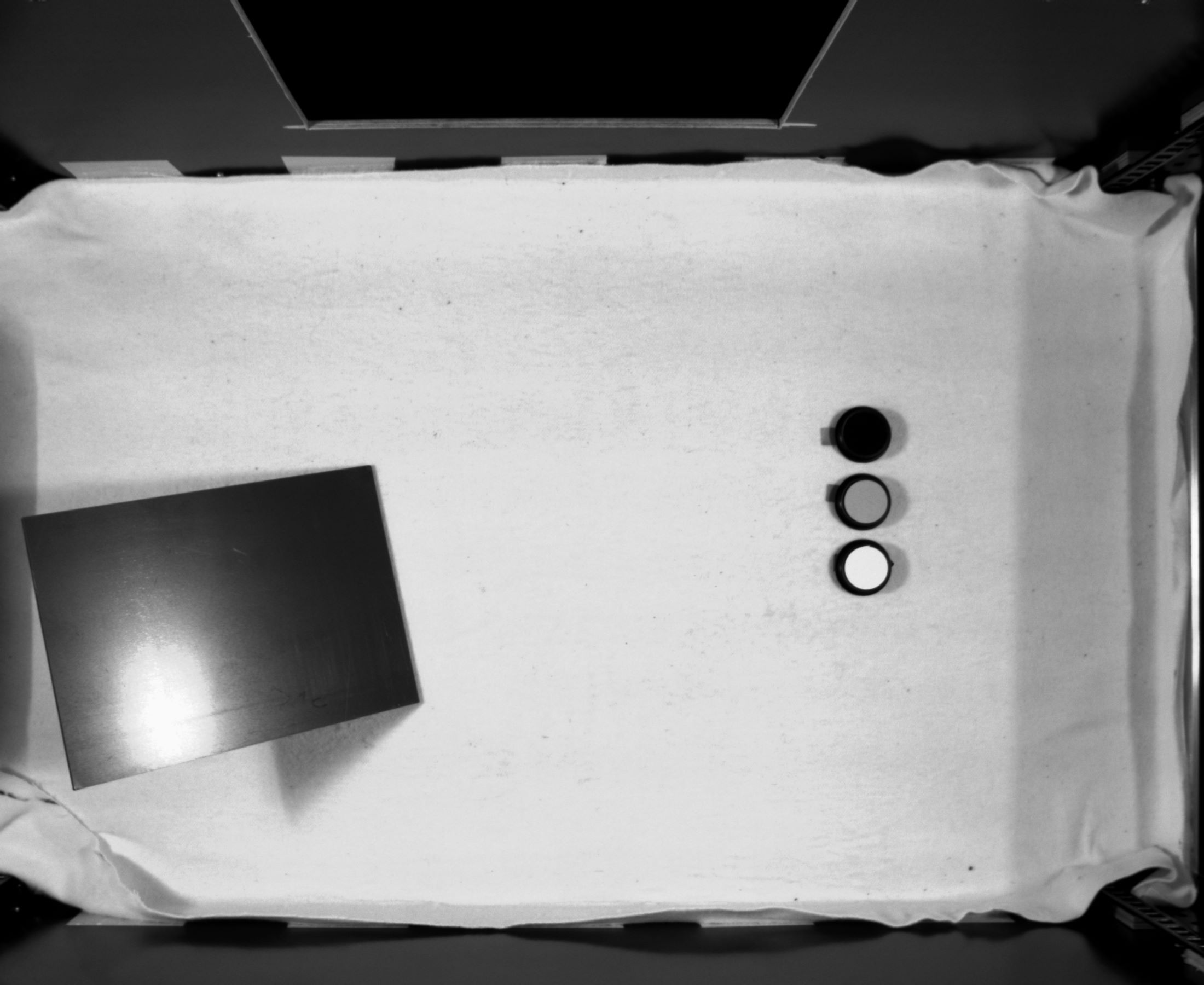}};
		\node(M14)[below of=M13, yshift=-0.4cm]{\includegraphics[width=0.09\textwidth]{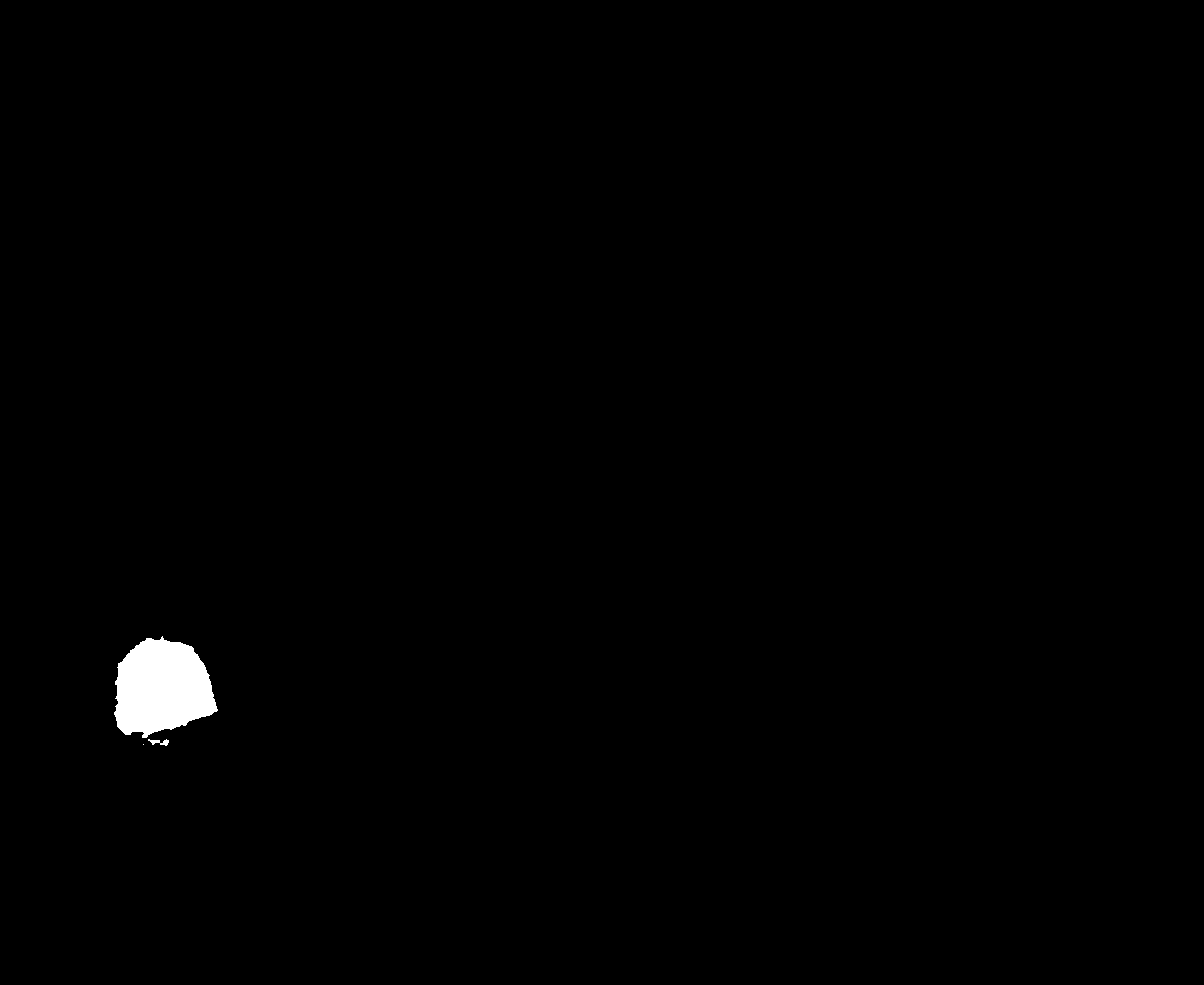}};
		\node(M24)[below of=M23, yshift=-0.4cm]{\includegraphics[width=0.09\textwidth]{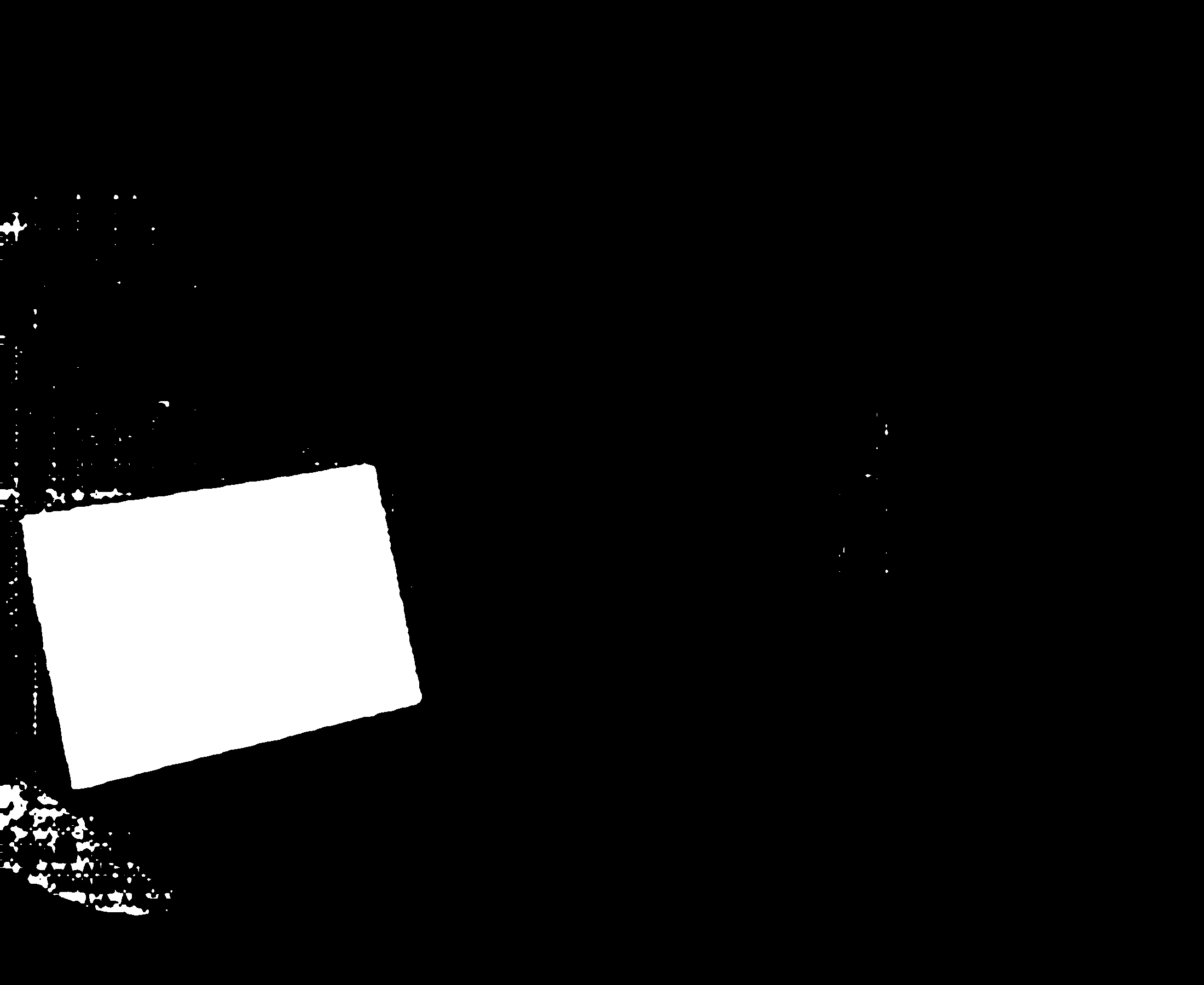}};
		\node(M34)[below of=M33, yshift=-0.4cm]{\includegraphics[width=0.09\textwidth]{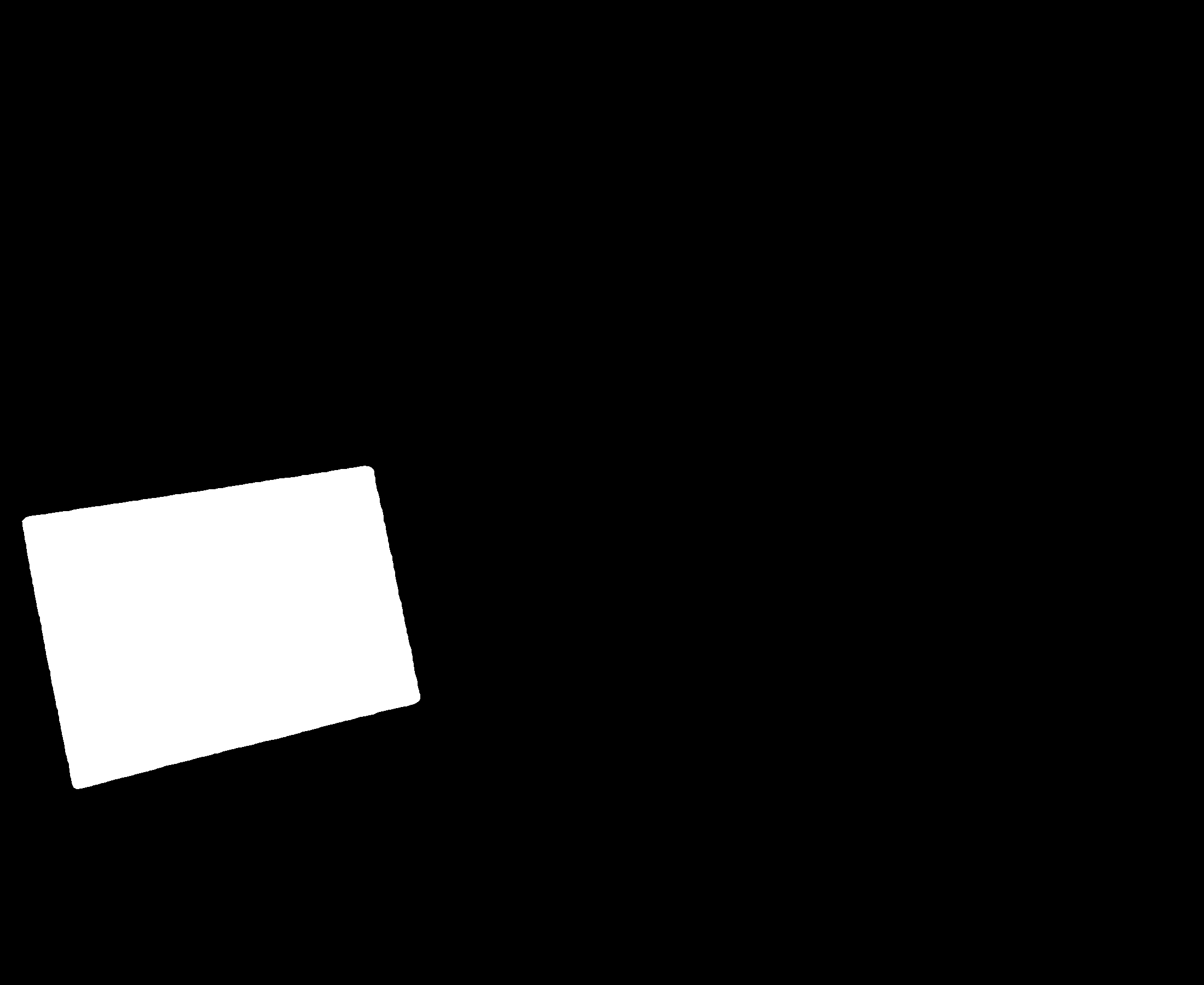}};
		\node(MF4)[below of=MF3, yshift=-0.4cm]{\includegraphics[width=0.09\textwidth]{Images/scene6/output_mask_2.jpg}};
		
		\begin{pgfonlayer}{background}
			\node[draw=blue, thick, dashed, fill=blue!10, inner sep=0.001cm, fit=(M11)(M21)(M31)(M12)(M22)(M32)(M13)(M23)(M33)(M14)(M24)(M34)] {};
			\node[draw=red, thick, dashed, fill=red!10, inner sep=0.001cm, text width=1cm, fit=(MF1)(MF2)(MF3)(MF4)] {};
		\end{pgfonlayer}
		
		\node(masksel)[below of=M24, yshift=-0.1cm]{\textbf{\color{blue}Mask selector}};
		\node(postproc)[below of=MF4, text width=1.5cm, yshift=-0.3cm, align=center]{\textbf{\color{red}Post\\processing}};
		\node(ori)[above of=Or1]{Original};
		\node(m1)[above of=M11]{$\mathcal{M}_1$};
		\node(m2)[above of=M21]{$\mathcal{M}_2$};
		\node(m3)[above of=M31]{$\mathcal{M}_3$};
		\node(pp)[above of=MF1]{$\mathcal{M}$};
	\end{tikzpicture}
	\vspace{-0.8cm}
	\caption{Original image (left), masks $\mathcal{M}_1$ - $\mathcal{M}_3$ generated by SAM2 for a given input specular reflection center $[c_x, c_y]$ (middle) and the final mask $\mathcal{M}$ after post processing (right). The best candidate mask of $\mathcal{M}_1$ - $\mathcal{M}_3$ is selected automatically by the mask selector.}
	\label{fig:ExampleMasks}
\end{figure}

After the specular reflection center of mass $C$ has been identified, its coordinates $[c_x,c_y]$ are passed into SAM2. The model is configured in multimask-mode, allowing it to generate up to three potential segmentation results $\mathcal{M}_1, \mathcal{M}_2$ and $\mathcal{M}_3$ for the object to which the reflection center $C$ most likely belongs. This setup enables SAM2 to reflect inherent ambiguity in boundary interpretation, especially in scenes with complex reflections. Thus, these masks show different plausible interpretations of the object surrounding the given point. For example, one mask may represent only the specular reflection, another may include the surface containing the reflection, and the third may capture a broader area, sometimes including background elements. Examples for $\mathcal{M}_1, \mathcal{M}_2$ and $\mathcal{M}_3$ are shown in Fig. \ref{fig:ExampleMasks} in the middle. 

As the segmentation output is ambiguous by design, a subsequent mask selector is applied to determine the most likely mask representing the reflective surface. This selection process is based on two complementary principles. First, the pixel ratio-based selection prioritizes the mask with the highest proportion of white (foreground) pixels, under the assumption that the reflective surface covers a larger area than the specular highlight. This can be expressed as
\begin{equation}
	 \mathcal{M}_{\text{pre-selected}} = \underset{\mathcal{M}_i}{\arg\max} \; R_i  ~~~,
\end{equation}
where $R_i$ is the white pixel ratio of mask $\mathcal{M}_i$. Second, to avoid selecting masks that include unrelated background regions especially in complex scenes, a maximum area threshold $R_{\text{max}}$ is introduced. Any mask exceeding this upper bound is excluded from the selection:
\begin{equation}
	\mathcal{M}_{\text{valid}} = \left\{ \mathcal{M}_i \mid R_i \leq R_{\text{max}} \right\} ~~~.
\end{equation}
Typical values for $R_{\text{max}}$ range between 0.3 and 0.6, depending on the scene complexity. The final selected mask is determined by 
\begin{equation}
	\mathcal{M}_\text{selected} = \underset{\mathcal{M}_i \in \mathcal{M}_{\text{valid}}}{\arg\max} \; R_i ~~~.
\end{equation}
This two-stage selection ensures that the mask corresponds to the relevant surface, excluding smaller specular highlights and overly large background regions.

After selecting the most suitable mask, a post processing step is applied to clean the binary output and remove small noise artifacts such as isolated white or black dots. These can result from lighting effects, reflections, or shadows. The goal is to produce a coherent mask that accurately captures the reflective surface. 

In the first step, foreground noise outside the main object is removed. This is achieved using connected component analysis (CCA) \cite{CCA}, which groups all connected foreground pixels in $\mathcal{M}_\text{selected}$ into disjoint sets $A_k$, also called components. Each component $A_k$ contains a set of connected pixel positions $p = (x,y)$, and no pixel belongs to more than one component:
\begin{equation}
	\bigcup_{k=1}^{K} A_k = \{p \in \Omega | \mathcal{M}_\text{selected}(p) = 1\} ~,~ A_i \cap A_j = \emptyset ~\text{for}~ i \neq j.
\end{equation}
Here, $K$ is the number of connected components, and $i,j$ are indices over these components. Then, from all detected regions, only the largest component $A_\text{max}$ is retained. Thus, isolated white specks that may lie outside the main reflective surface are successfully removed.

However, the selected mask may still contain small black areas within the main white region, which are not removed in the first step. These appear as background within the object and reduce the mask's completeness. An inversion of the mask transforms black holes within the object into small white regions on a black background. Another CCA is subsequently performed to remove these small regions. Finally, the result is inverted back to restore the correct polarity. The resulting mask represents a clean, hole-free segmentation of the reflective surface, preserving only the largest continuous region and eliminating both white speckles and internal black voids:
\begin{equation}
	\mathcal{M} = 1 - \begin{cases}
		1 & \text{if}~~ p \in A_{\text{max}} \\
		0  & \text{otherwise  .}
	\end{cases}
\end{equation}
Examples of $\mathcal{M}$ after post-processing are illustrated in Fig. \ref{fig:ExampleMasks} on the right.

\section{Experimental Results}
\begin{figure*}[t!]
	\centering
	\begin{tikzpicture}
		\node(I1)[]{\includegraphics[width=0.118\textwidth]{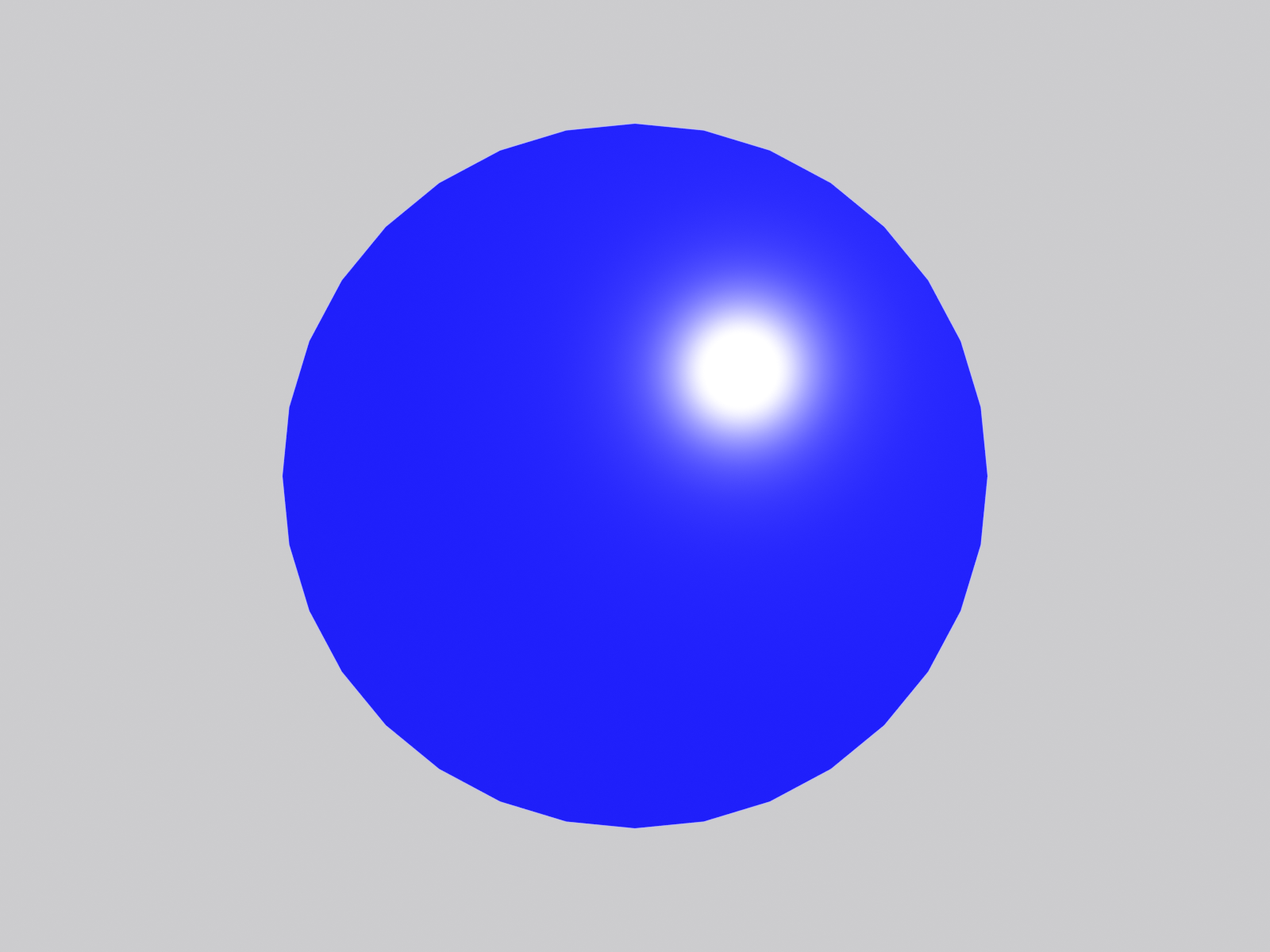}};
		\node(M1GT)[below of=I1, yshift=-0.67cm]{\includegraphics[width=0.118\textwidth]{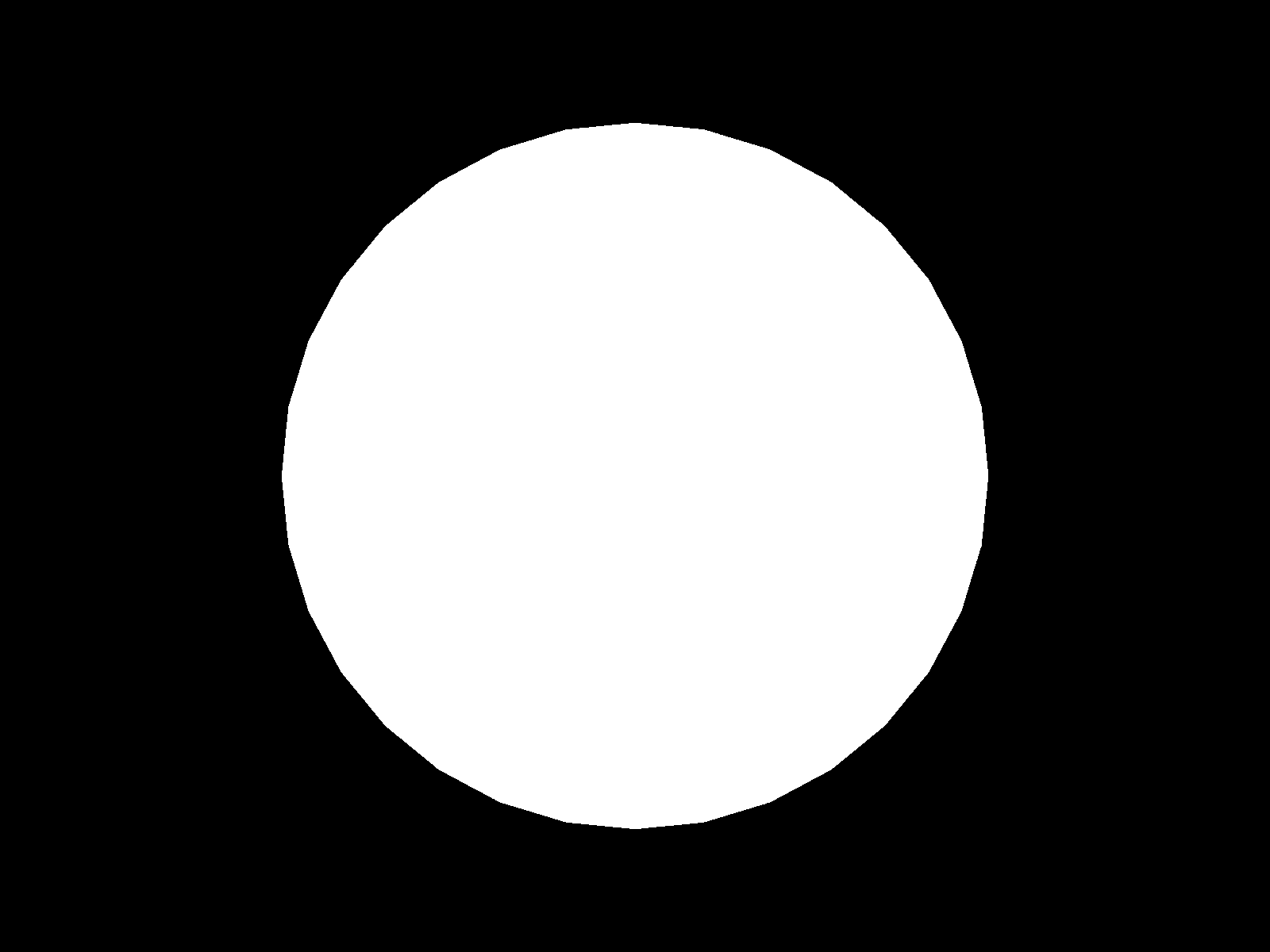}};
		\node(M1O)[below of=M1GT, yshift=-0.67cm]{\includegraphics[width=0.118\textwidth]{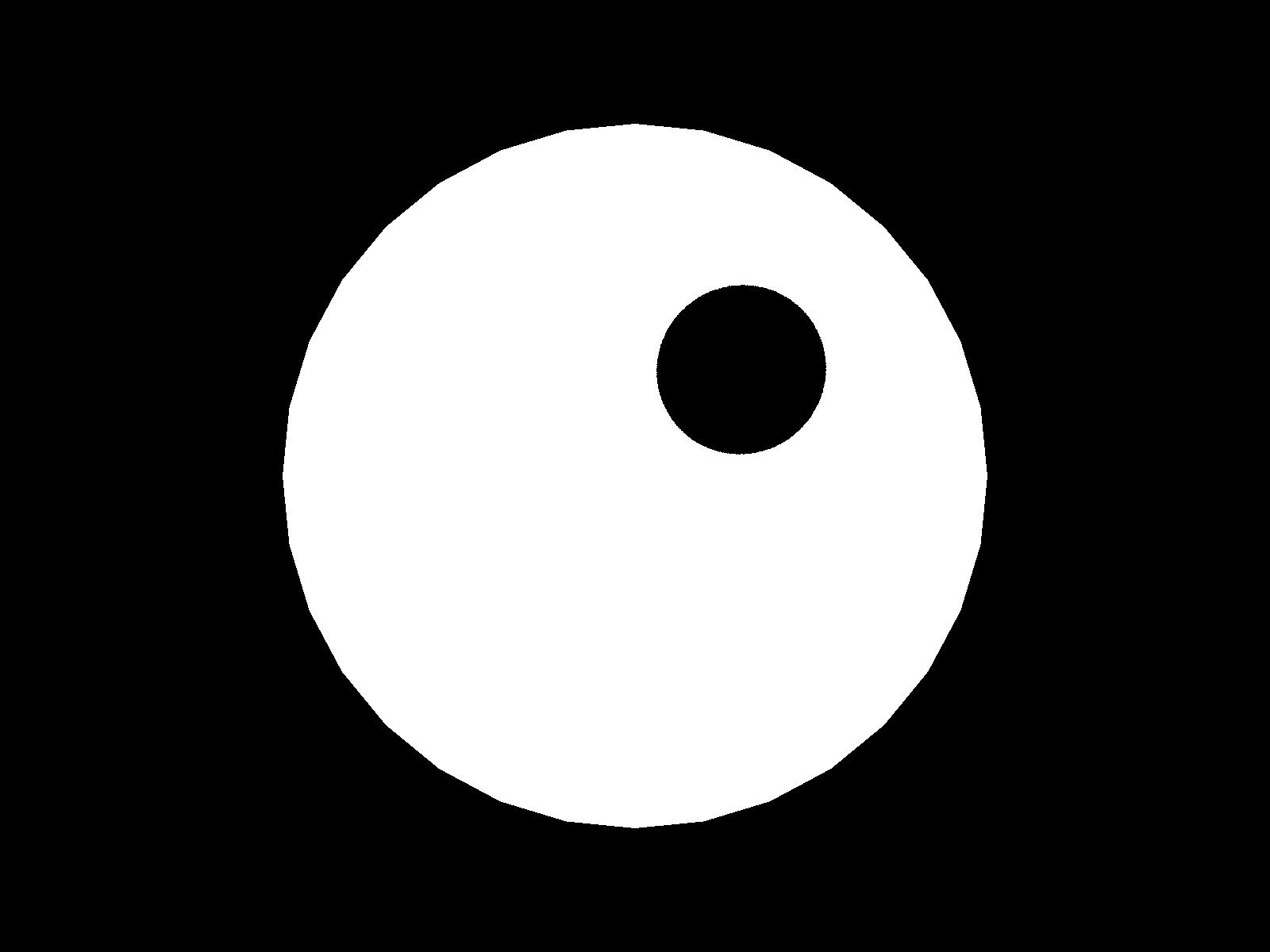}};
		\node(M1Y)[below of=M1O, yshift=-0.67cm]{\includegraphics[width=0.118\textwidth]{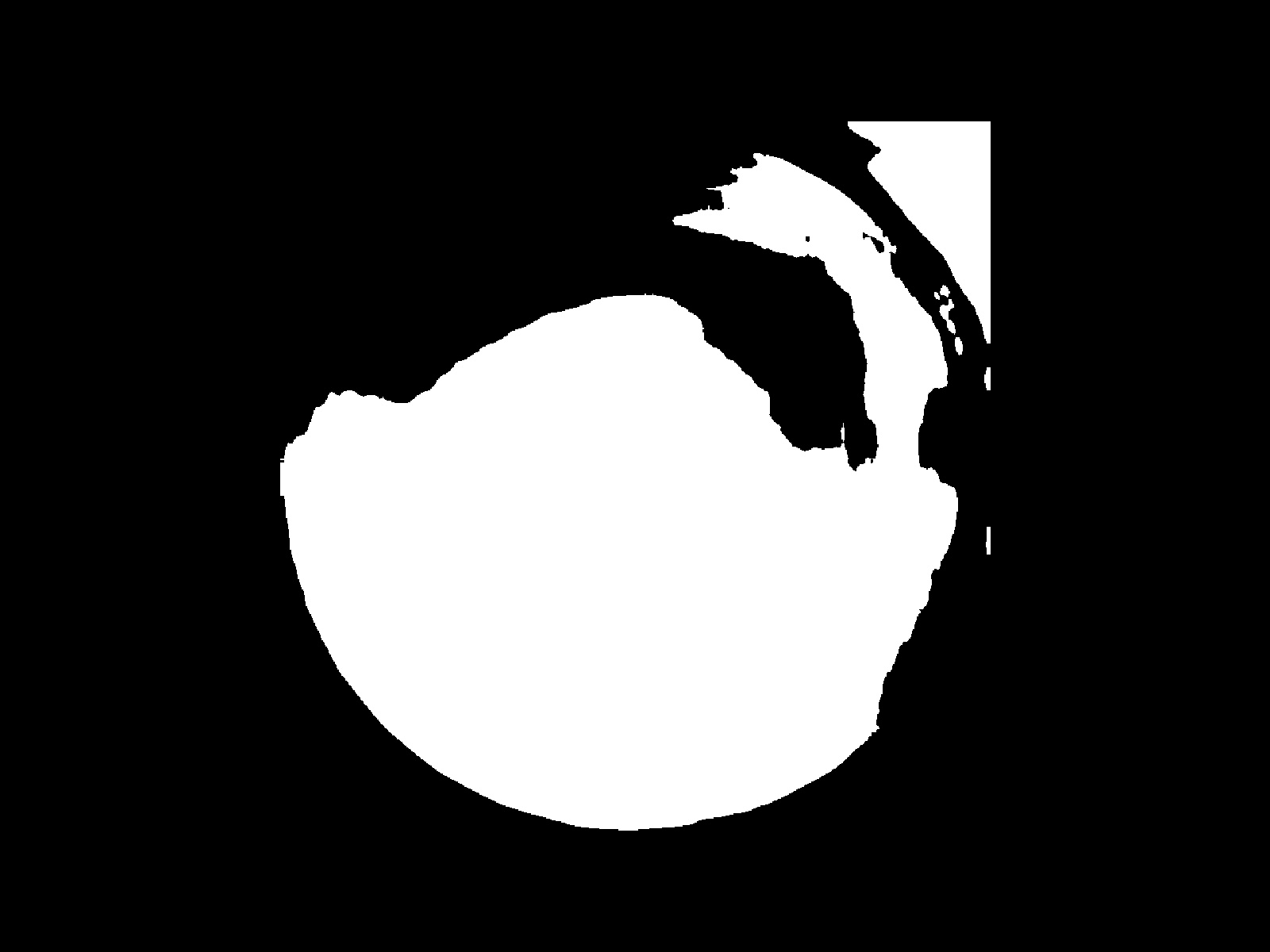}};
		\node(M1S)[below of=M1Y, yshift=-0.67cm]{\includegraphics[width=0.118\textwidth]{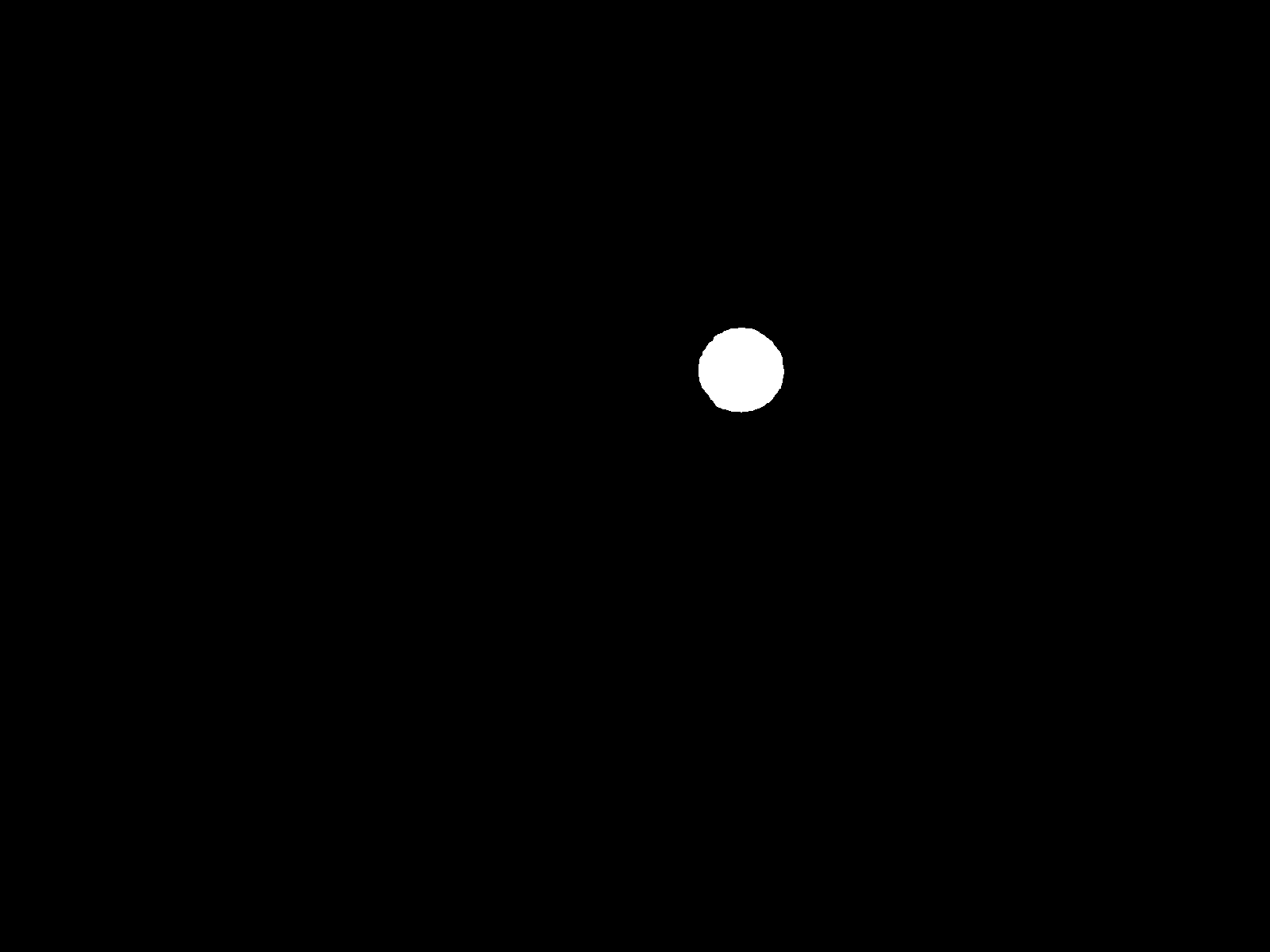}};
		\node(M1W)[below of=M1S, yshift=-0.67cm]{\includegraphics[width=0.118\textwidth]{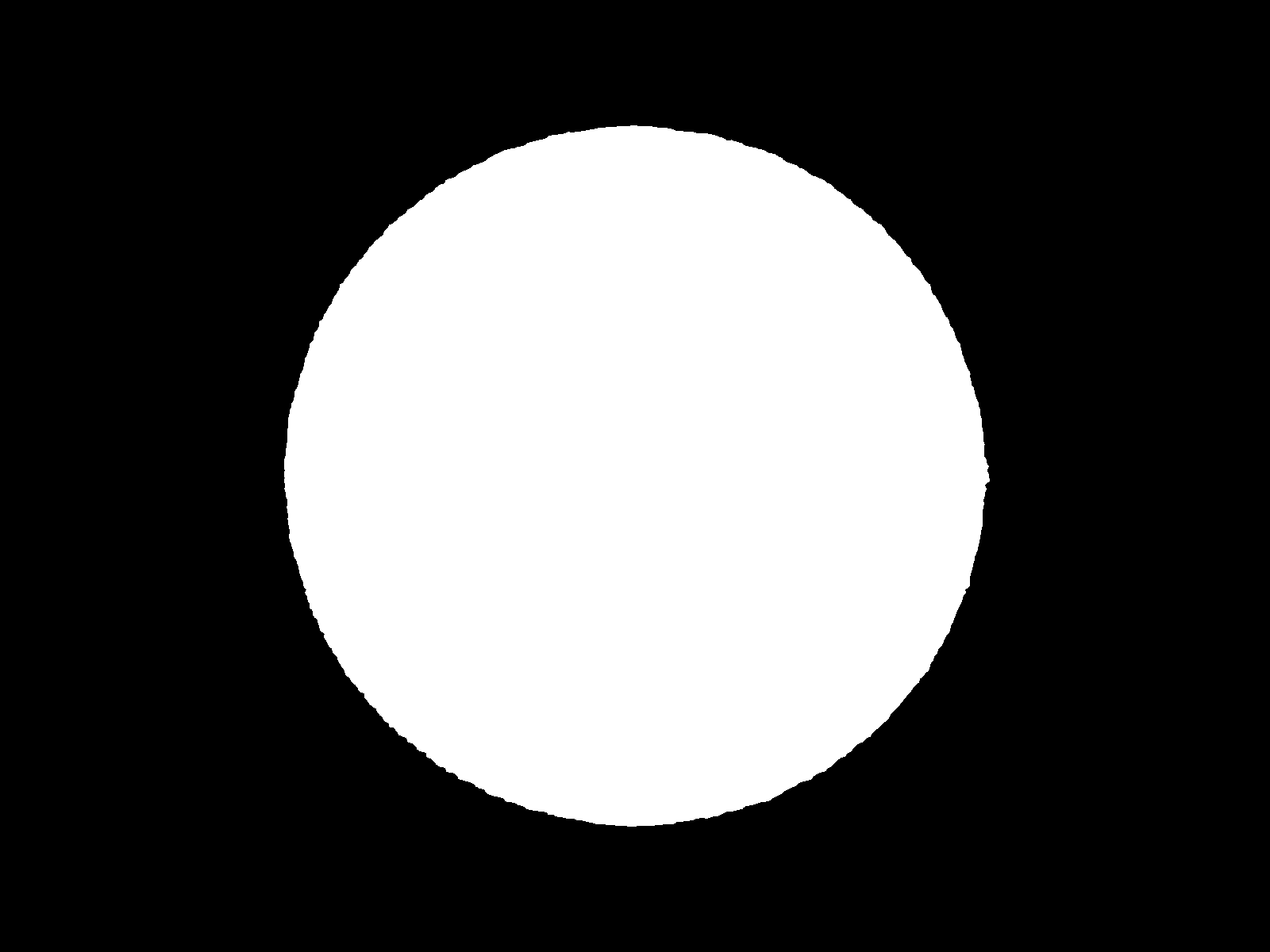}};
		
		\node(i1)[left of=I1, xshift=-0.7cm, align=center]{Original};
		\node(m1O)[left of=M1GT, text width=1cm, , xshift=-0.7cm, align=center]{Ground truth};
		\node(m1GT)[left of=M1O, xshift=-0.7cm, text width=1cm, align=center]{Otsu \cite{Otsu}};
		\node(m1Y)[left of=M1Y, xshift=-0.7cm, text width=1cm, align=center]{YOLO \cite{YOLO}};
		\node(m1S)[left of=M1S, xshift=-0.7cm, text width=1cm, align=center]{SAM2 \cite{SAM2}};
		\node(m1W)[left of=M1W, xshift=-0.7cm, text width=1cm, align=center]{Ours};
		
		\node(I2)[right of=I1, xshift=1.2cm]{\includegraphics[width=0.118\textwidth]{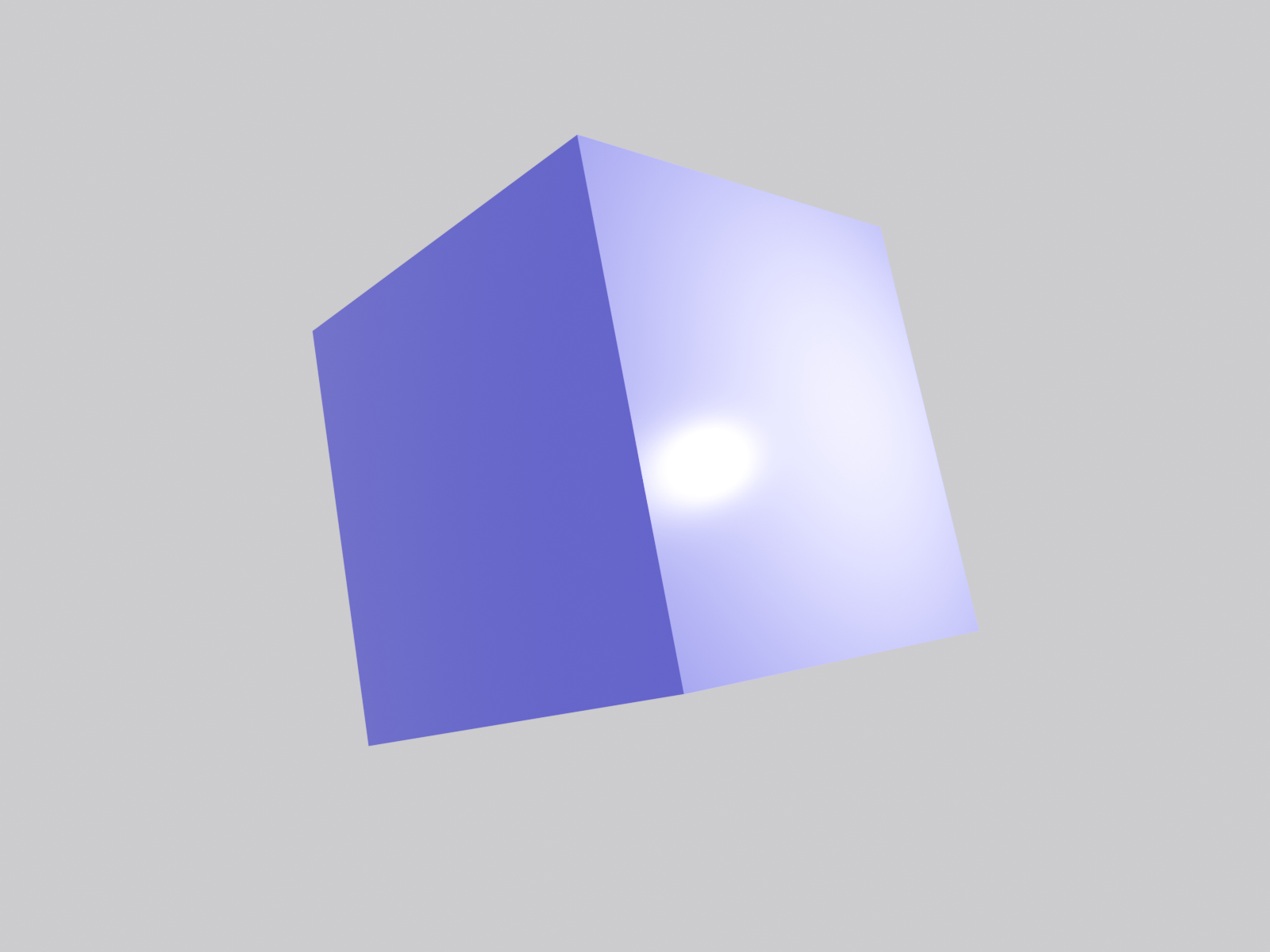}};
		\node(M2GT)[below of=I2, yshift=-0.67cm]{\includegraphics[width=0.118\textwidth]{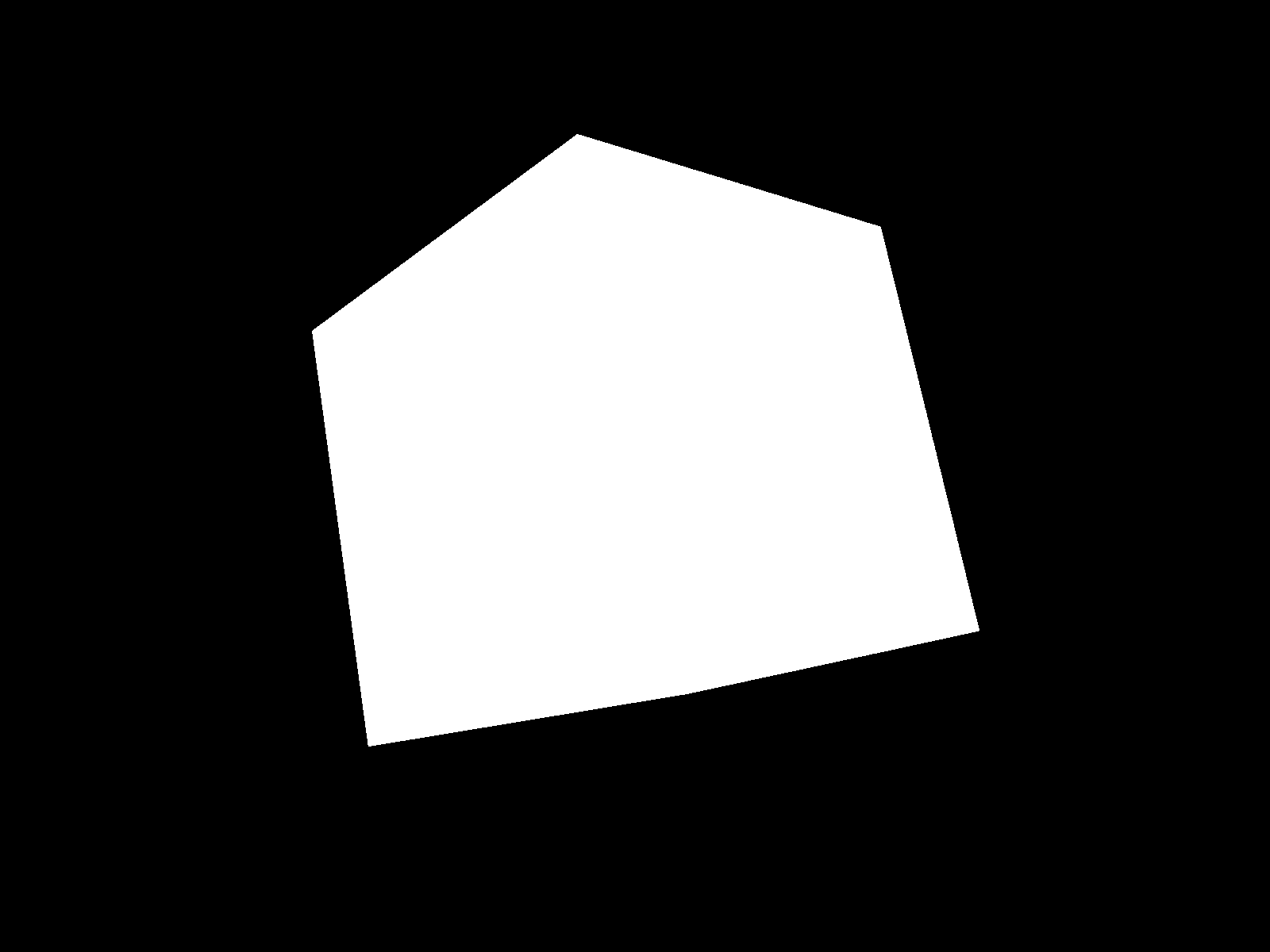}};
		\node(M2O)[below of=M2GT, yshift=-0.67cm]{\includegraphics[width=0.118\textwidth]{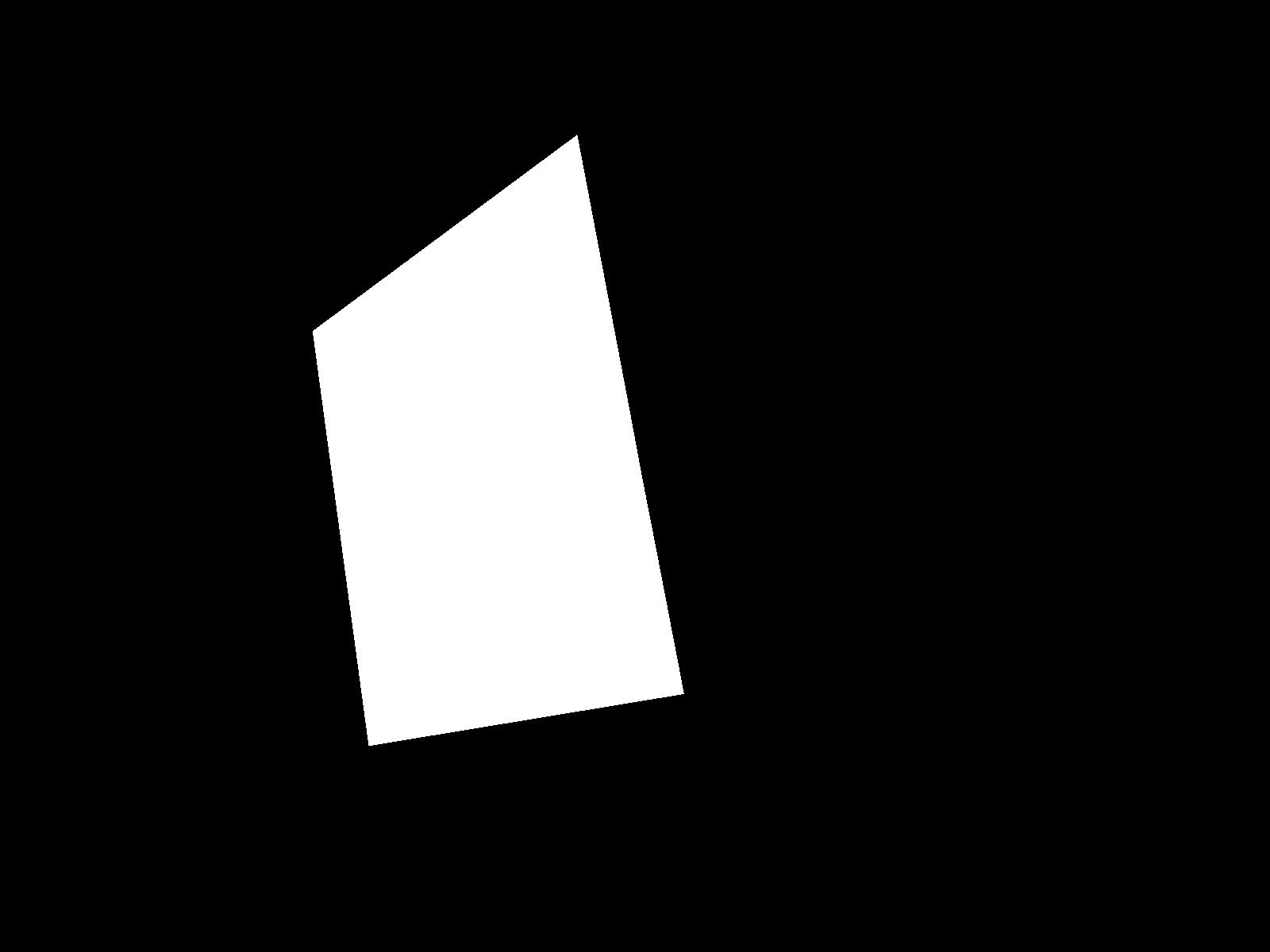}};
		\node(M2Y)[below of=M2O, yshift=-0.67cm]{\includegraphics[width=0.118\textwidth]{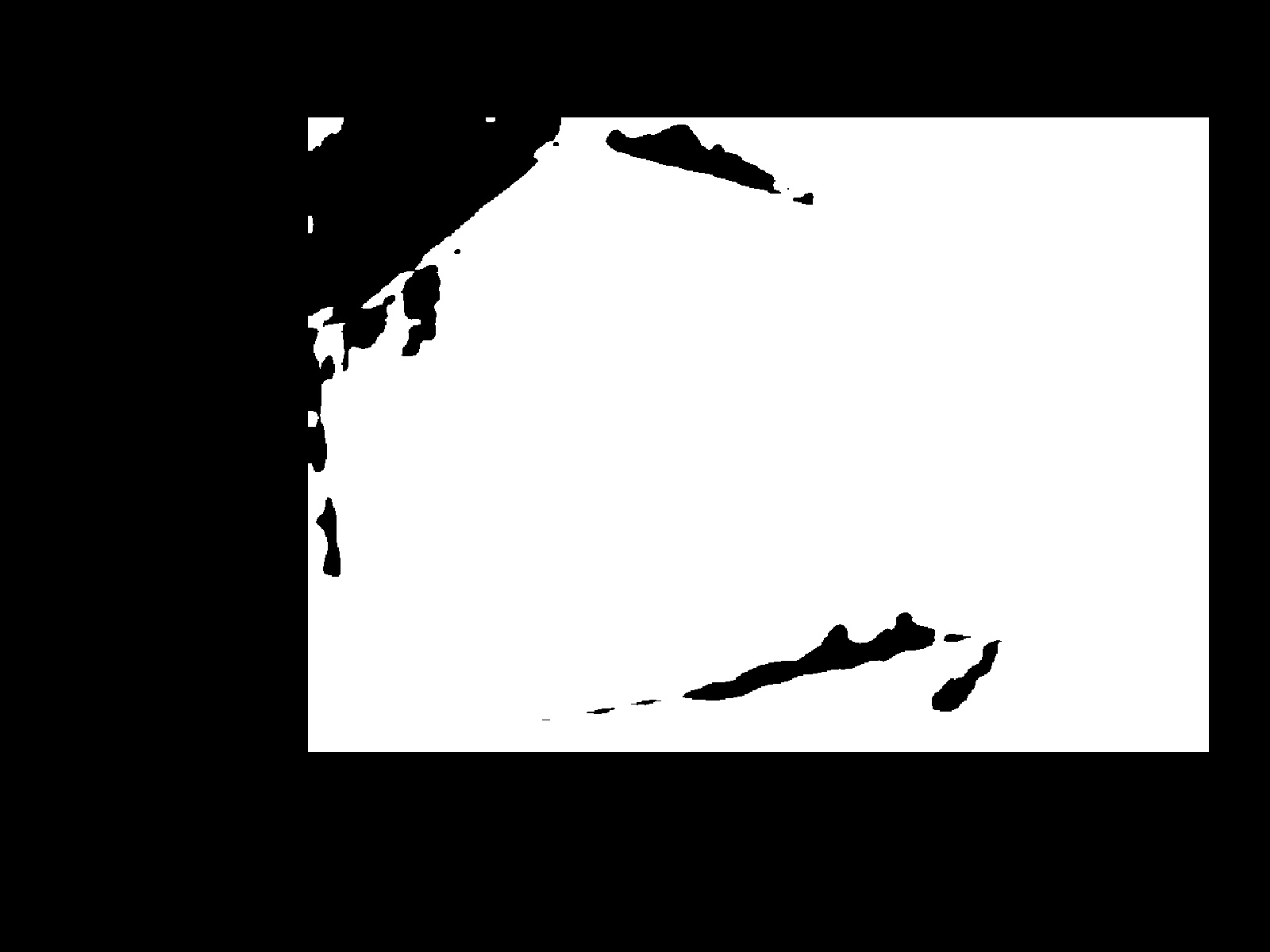}};
		\node(M2S)[below of=M2Y, yshift=-0.67cm]{\includegraphics[width=0.118\textwidth]{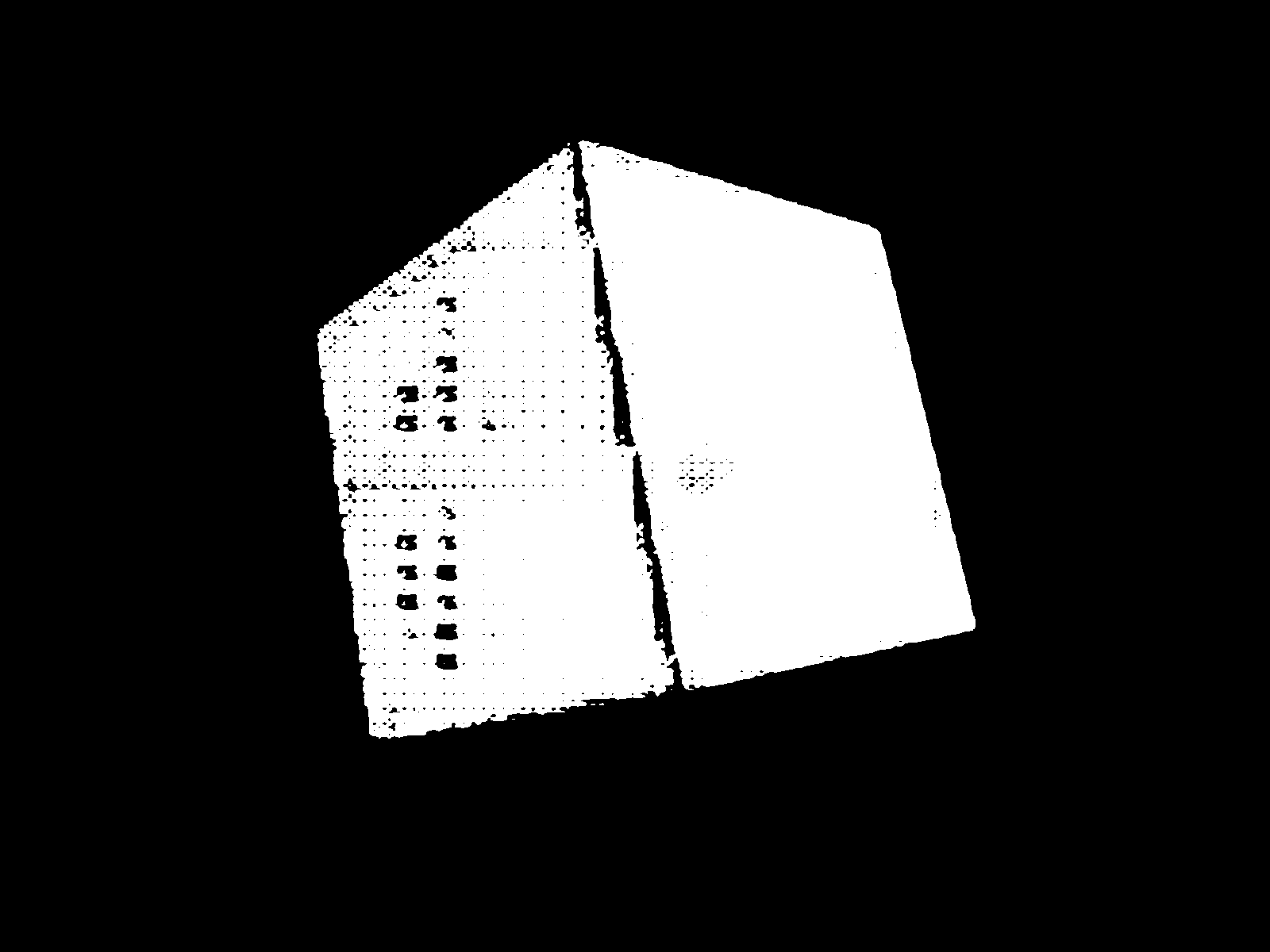}};
		\node(M2W)[below of=M2S, yshift=-0.67cm]{\includegraphics[width=0.118\textwidth]{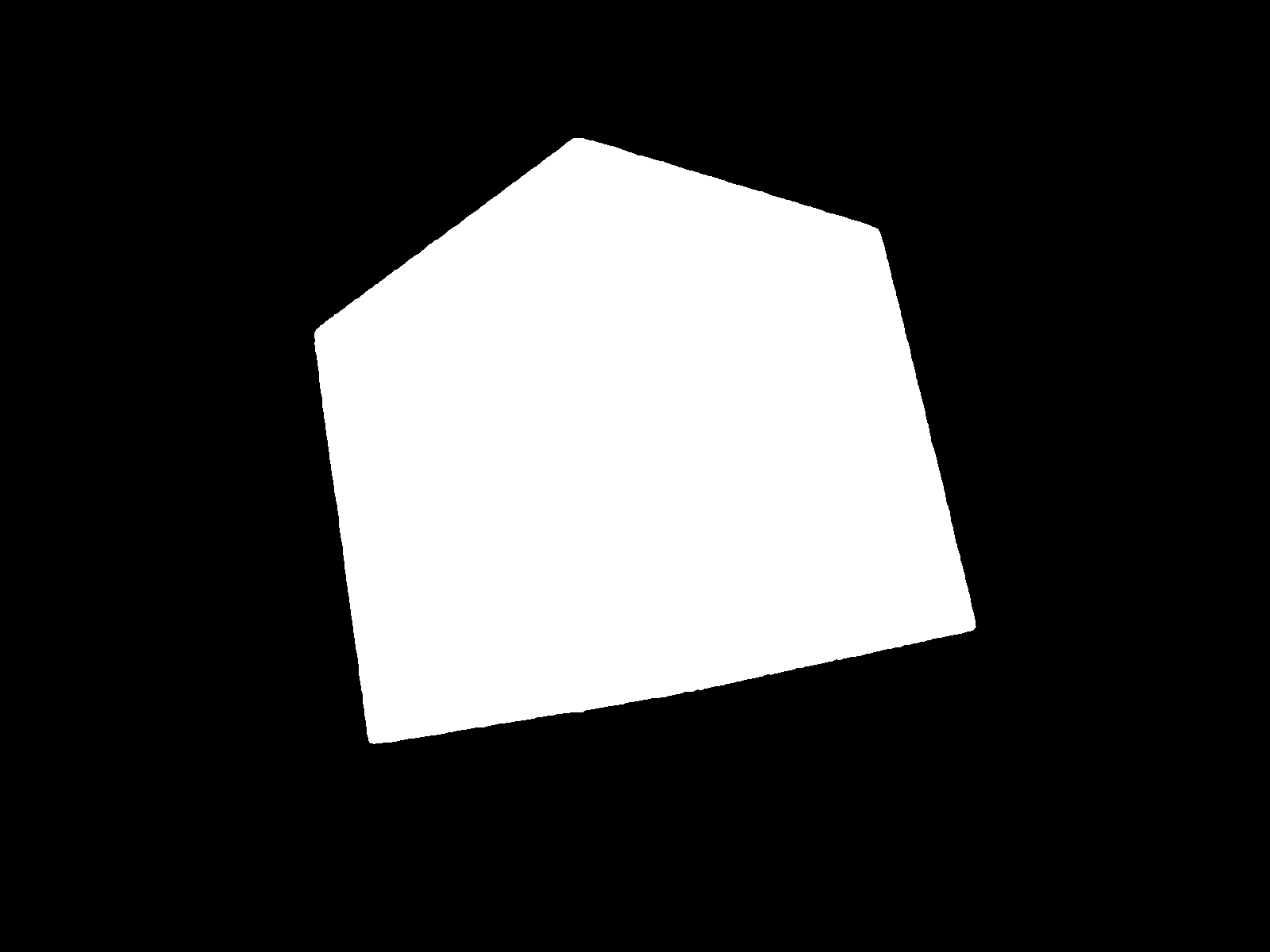}};
		
		\node(I3)[right of=I2, xshift=1.2cm]{\includegraphics[width=0.118\textwidth]{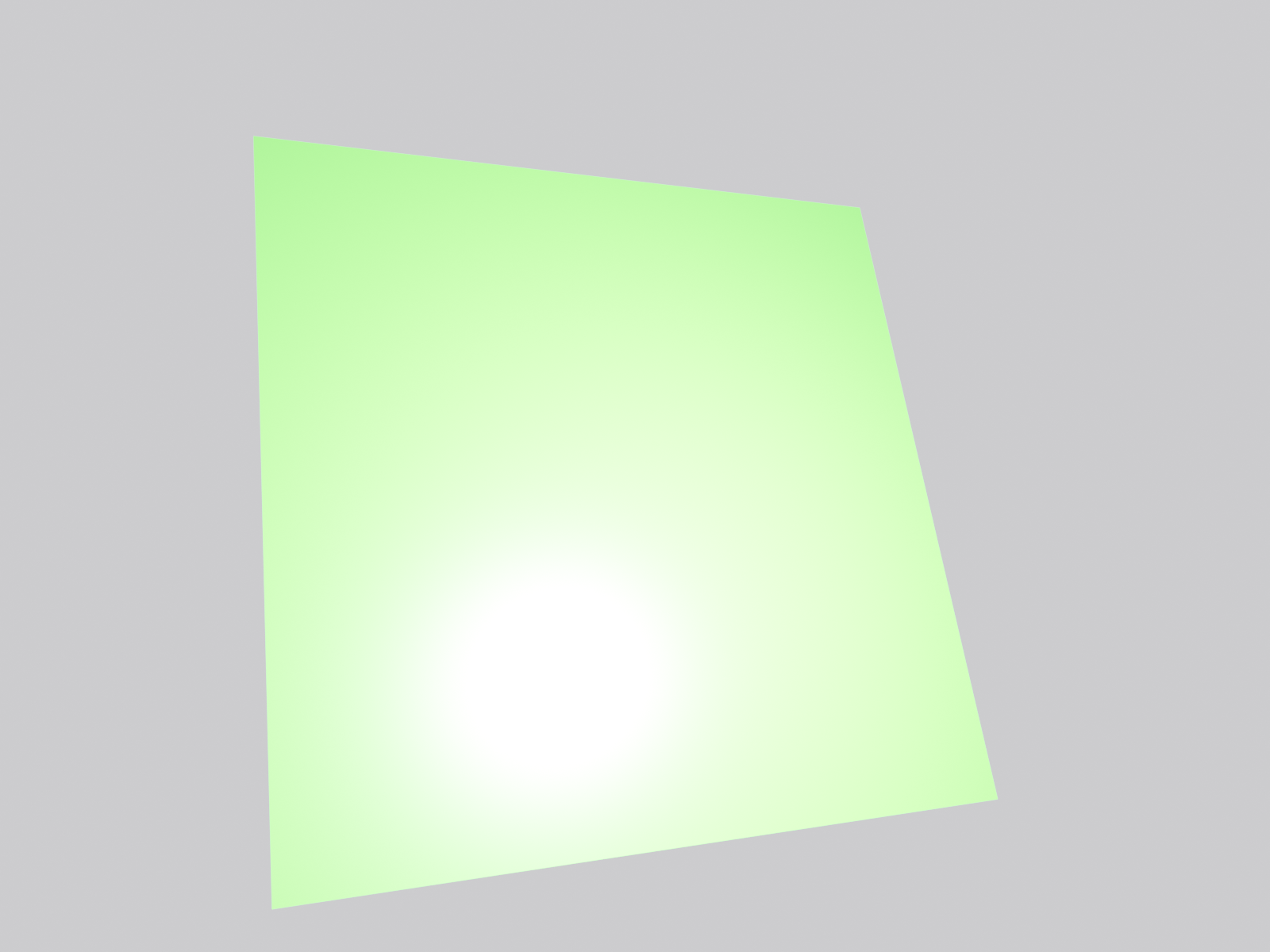}};	
		\node(M3GT)[below of=I3, yshift=-0.67cm]{\includegraphics[width=0.118\textwidth]{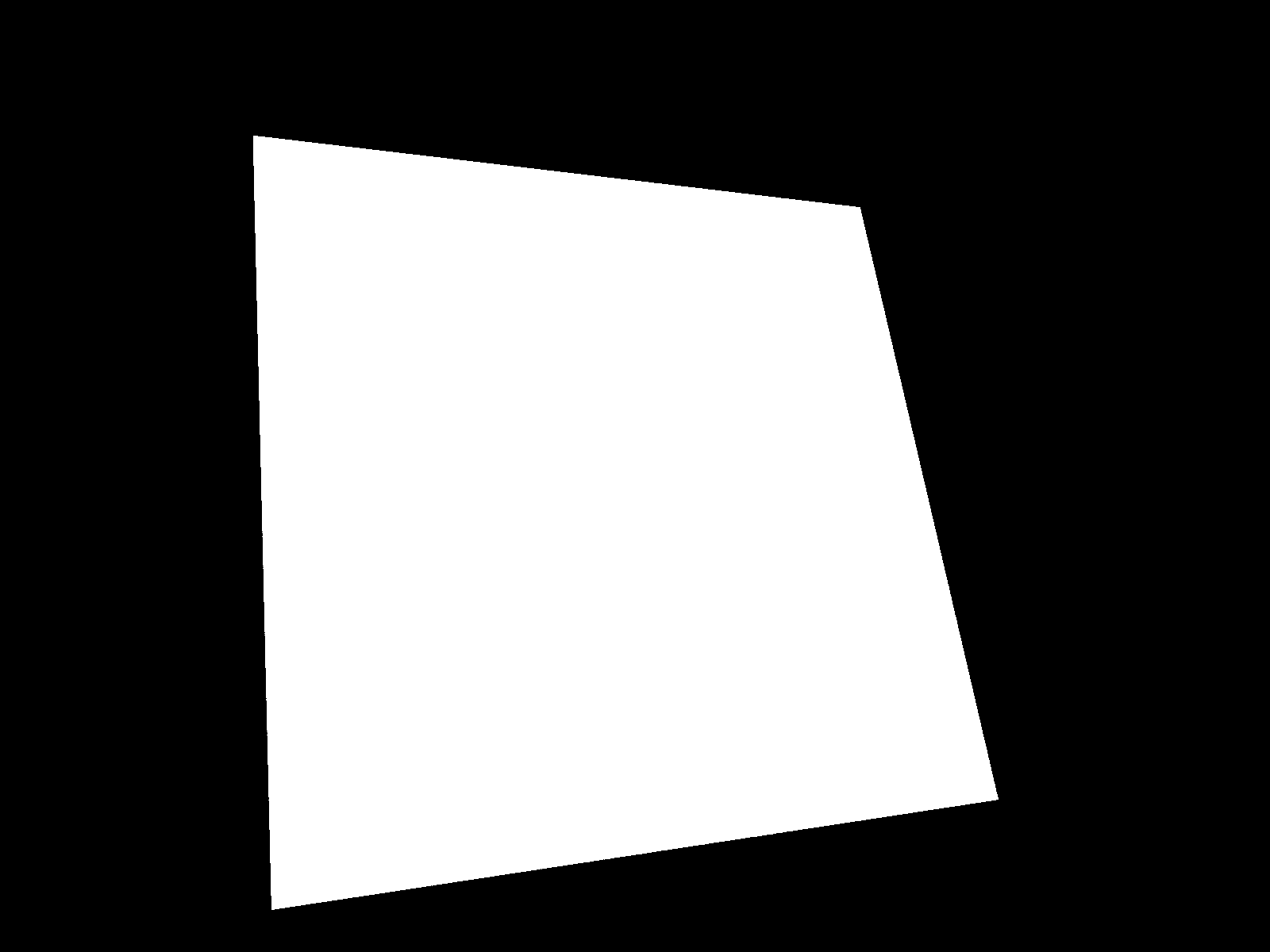}};
		\node(M3O)[below of=M3GT, yshift=-0.67cm]{\includegraphics[width=0.118\textwidth]{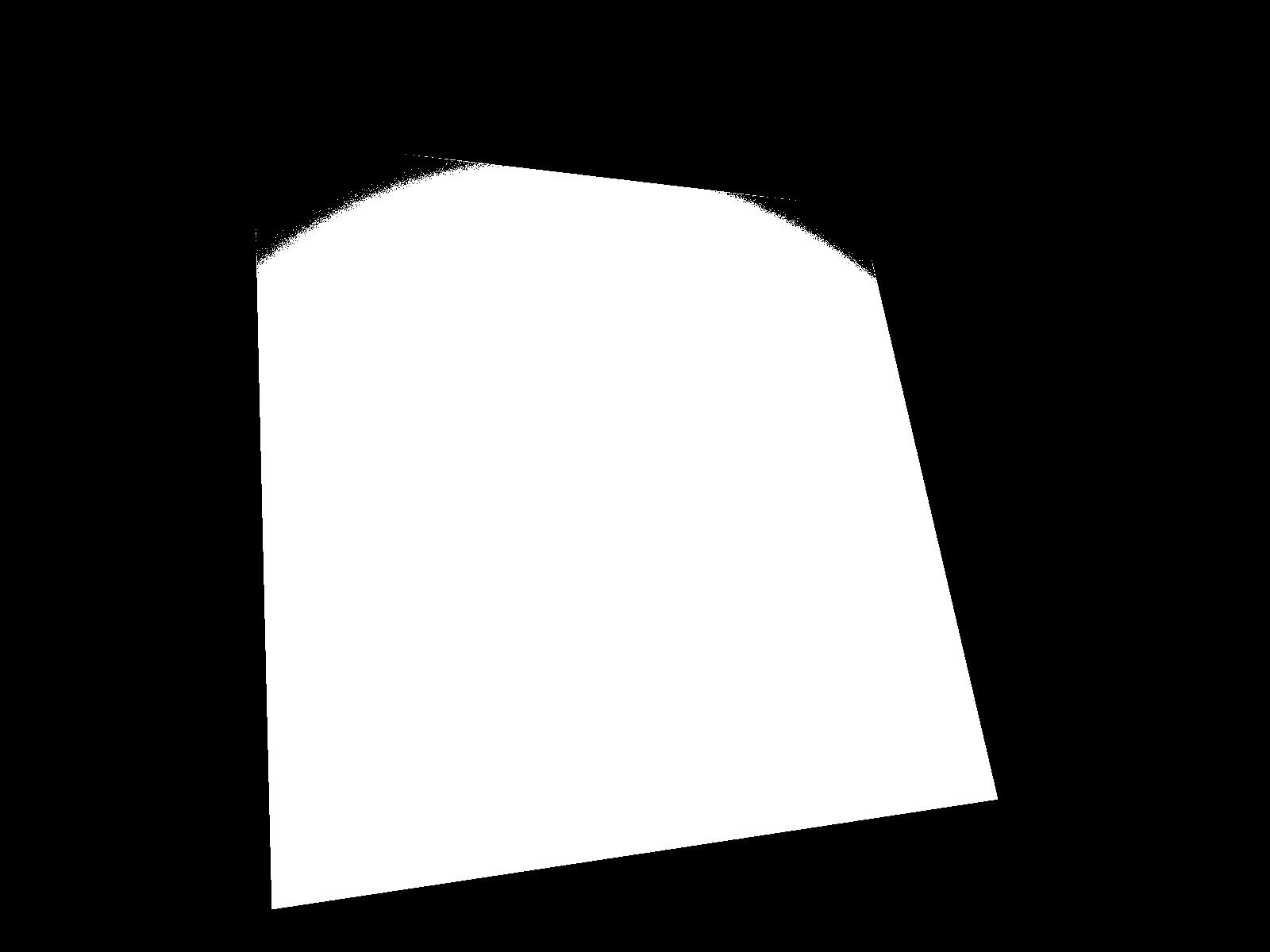}};
		\node(M3Y)[below of=M3O, yshift=-0.67cm]{\includegraphics[width=0.118\textwidth]{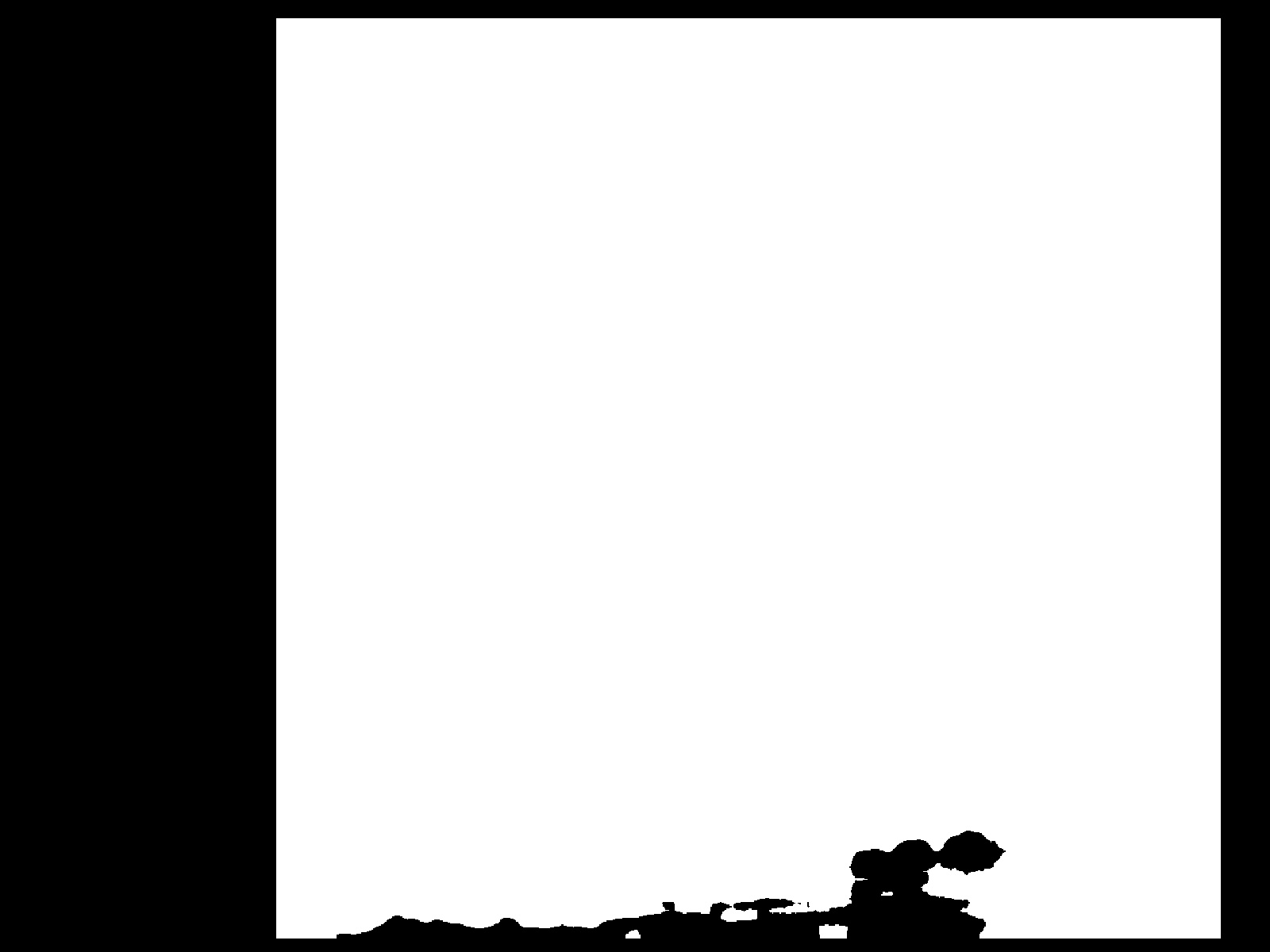}};
		\node(M3S)[below of=M3Y, yshift=-0.67cm]{\includegraphics[width=0.118\textwidth]{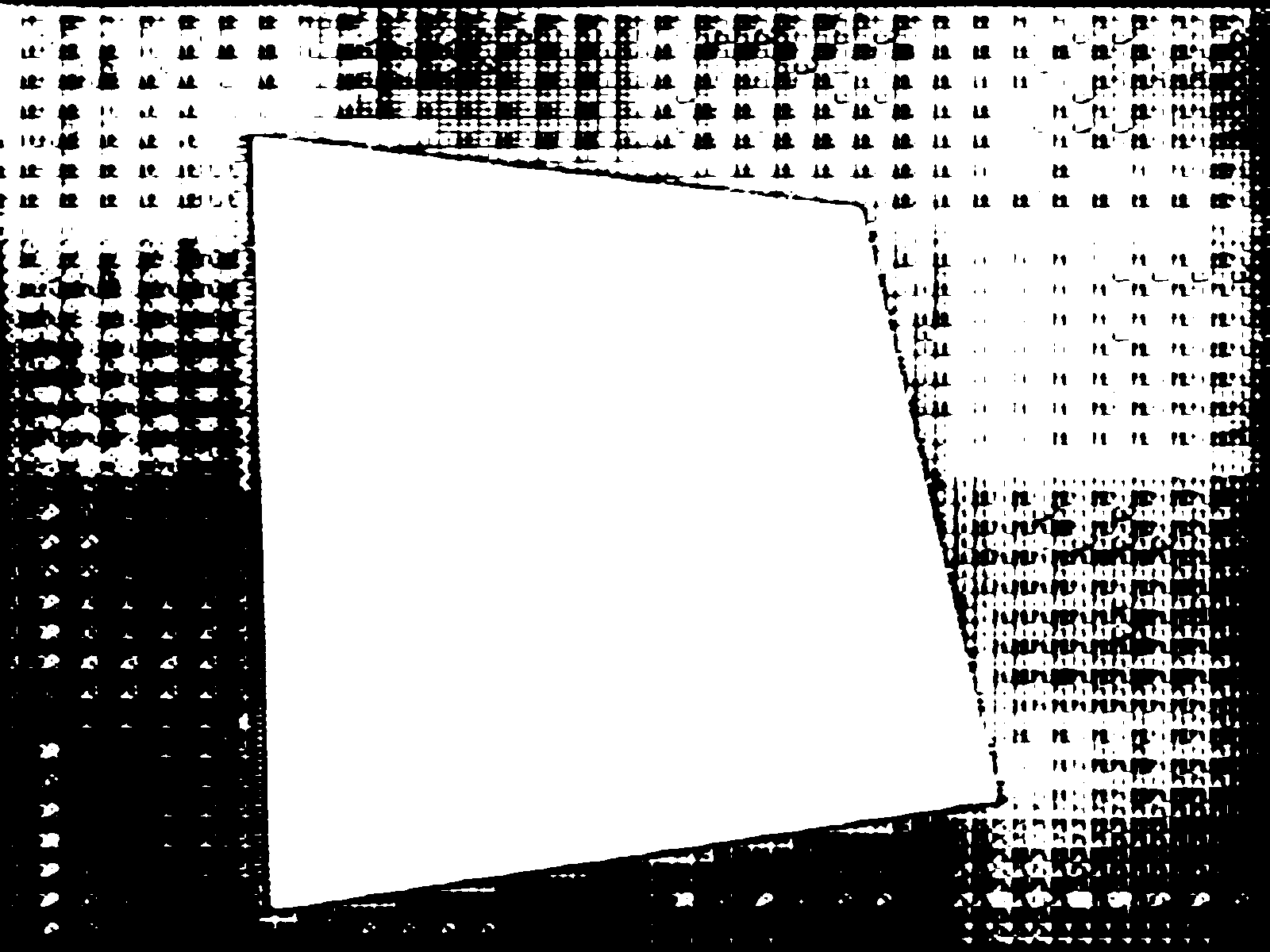}};
		\node(M3W)[below of=M3S, yshift=-0.67cm]{\includegraphics[width=0.118\textwidth]{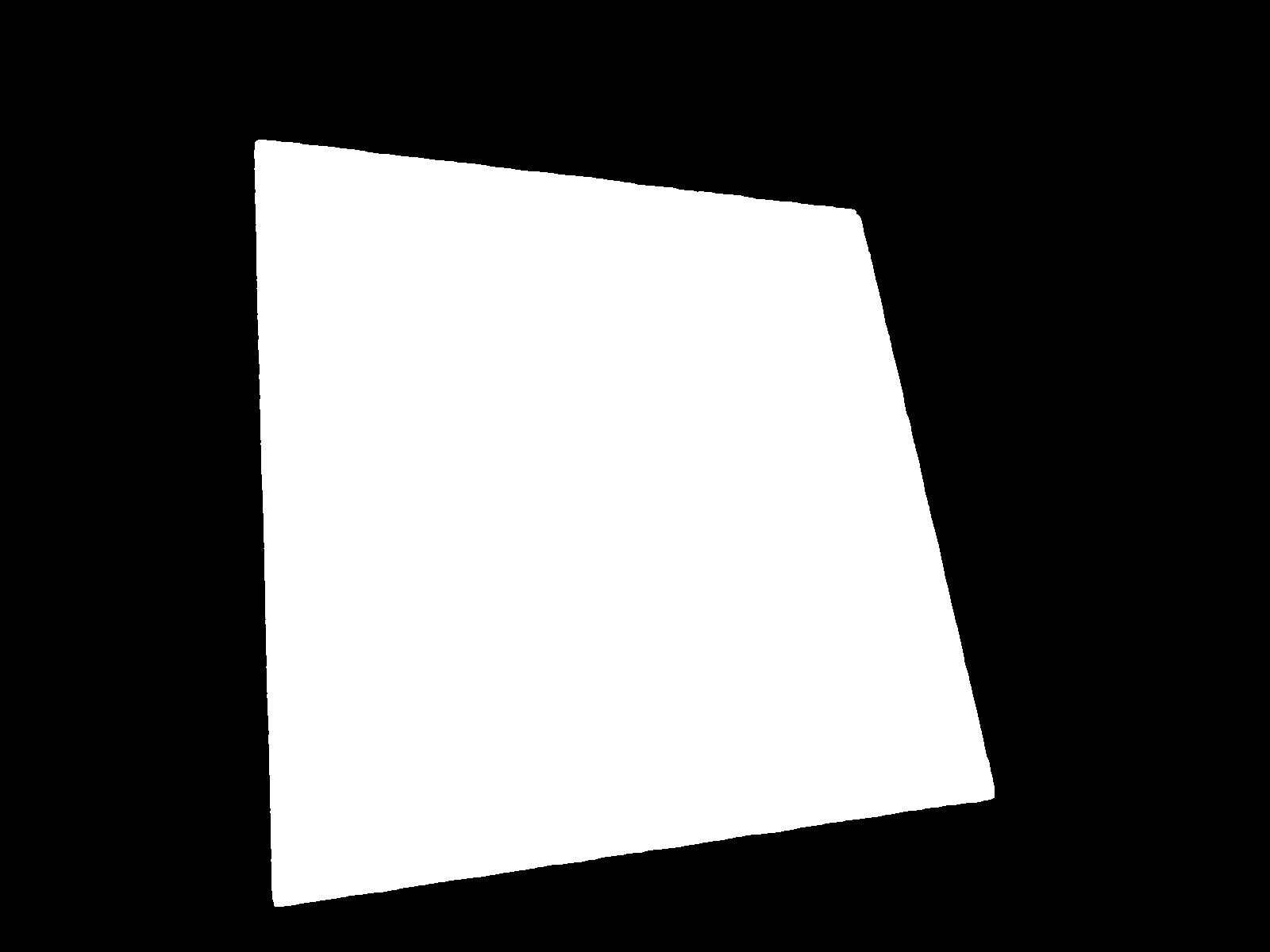}};
		
		\node(I4)[right of=I3, xshift=1.21cm]{\includegraphics[width=0.118\textwidth]{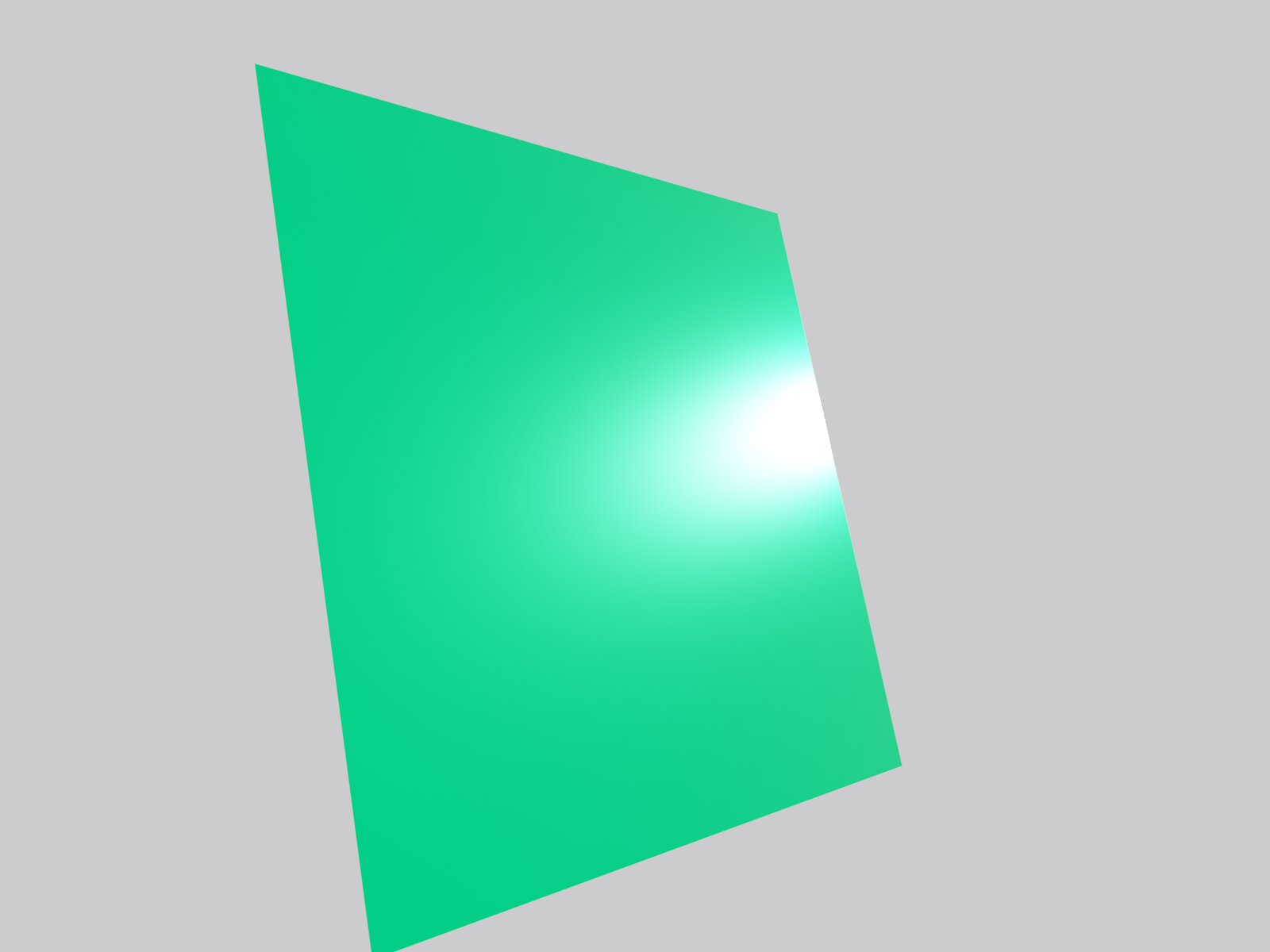}};	
		\node(M4GT)[below of=I4, yshift=-0.67cm]{\includegraphics[width=0.118\textwidth]{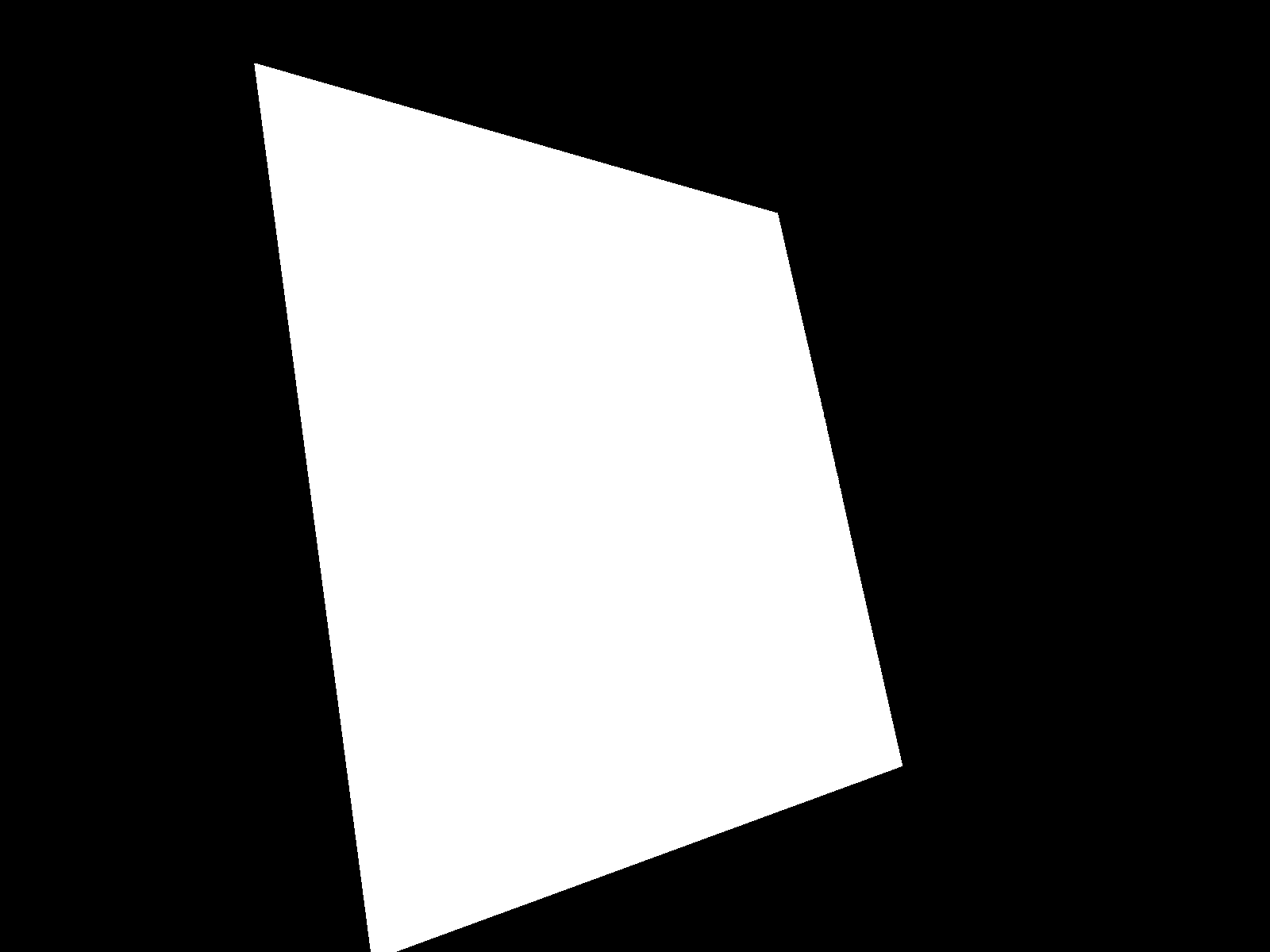}};
		\node(M4O)[below of=M4GT, yshift=-0.67cm]{\includegraphics[width=0.118\textwidth]{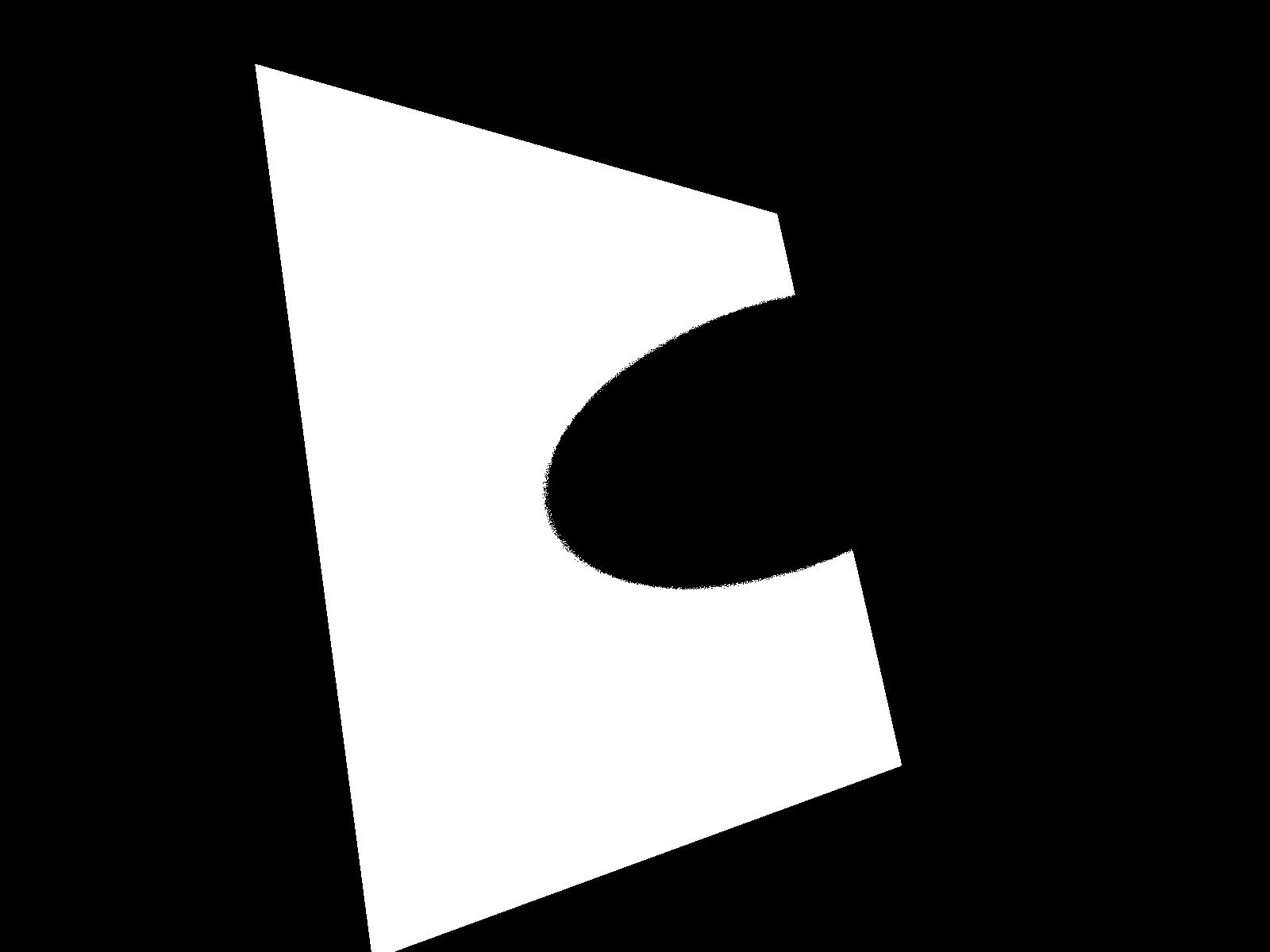}};
		\node(M4Y)[below of=M4O, yshift=-0.67cm]{\includegraphics[width=0.118\textwidth]{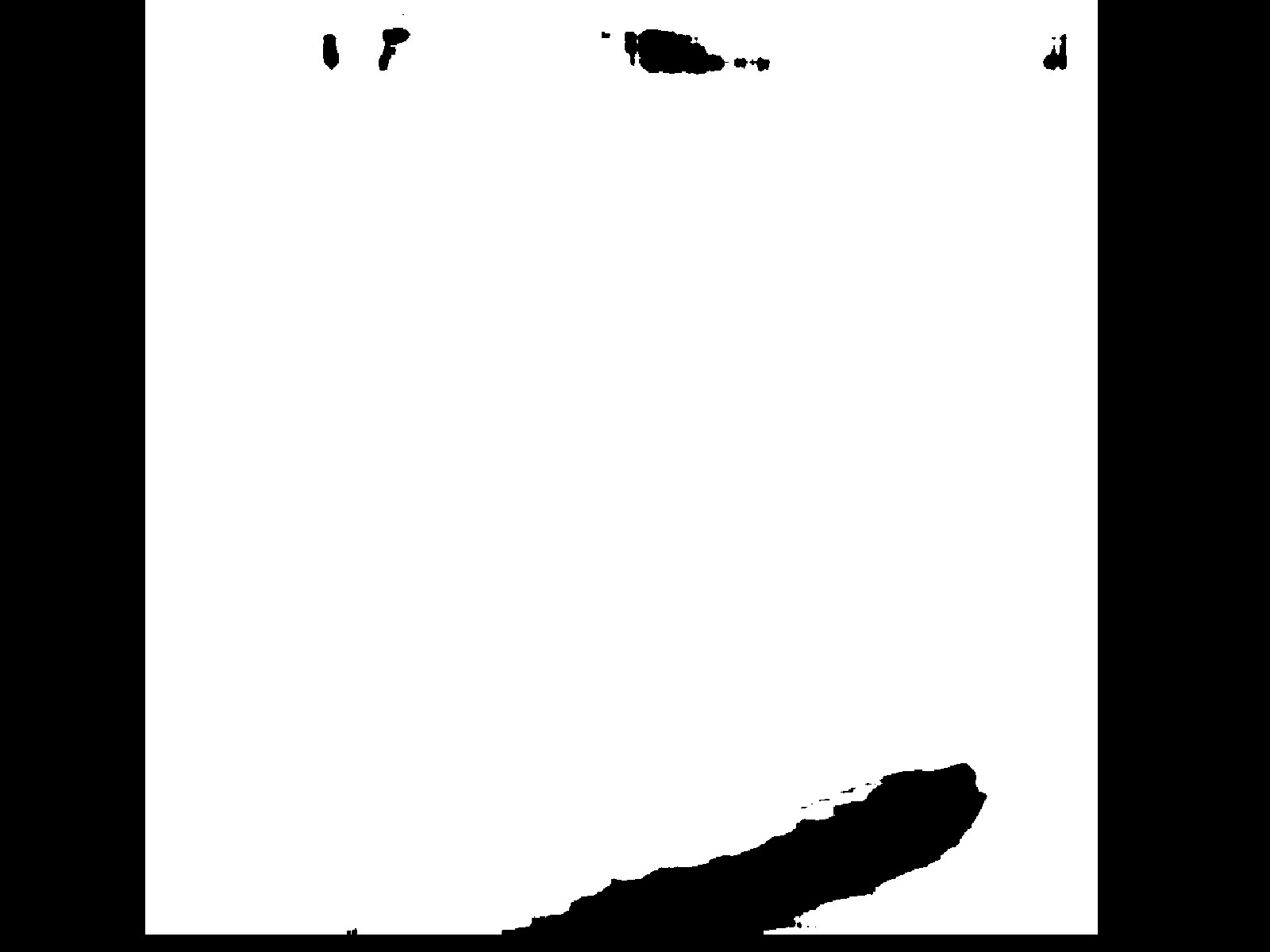}};
		\node(M4S)[below of=M4Y, yshift=-0.67cm]{\includegraphics[width=0.118\textwidth]{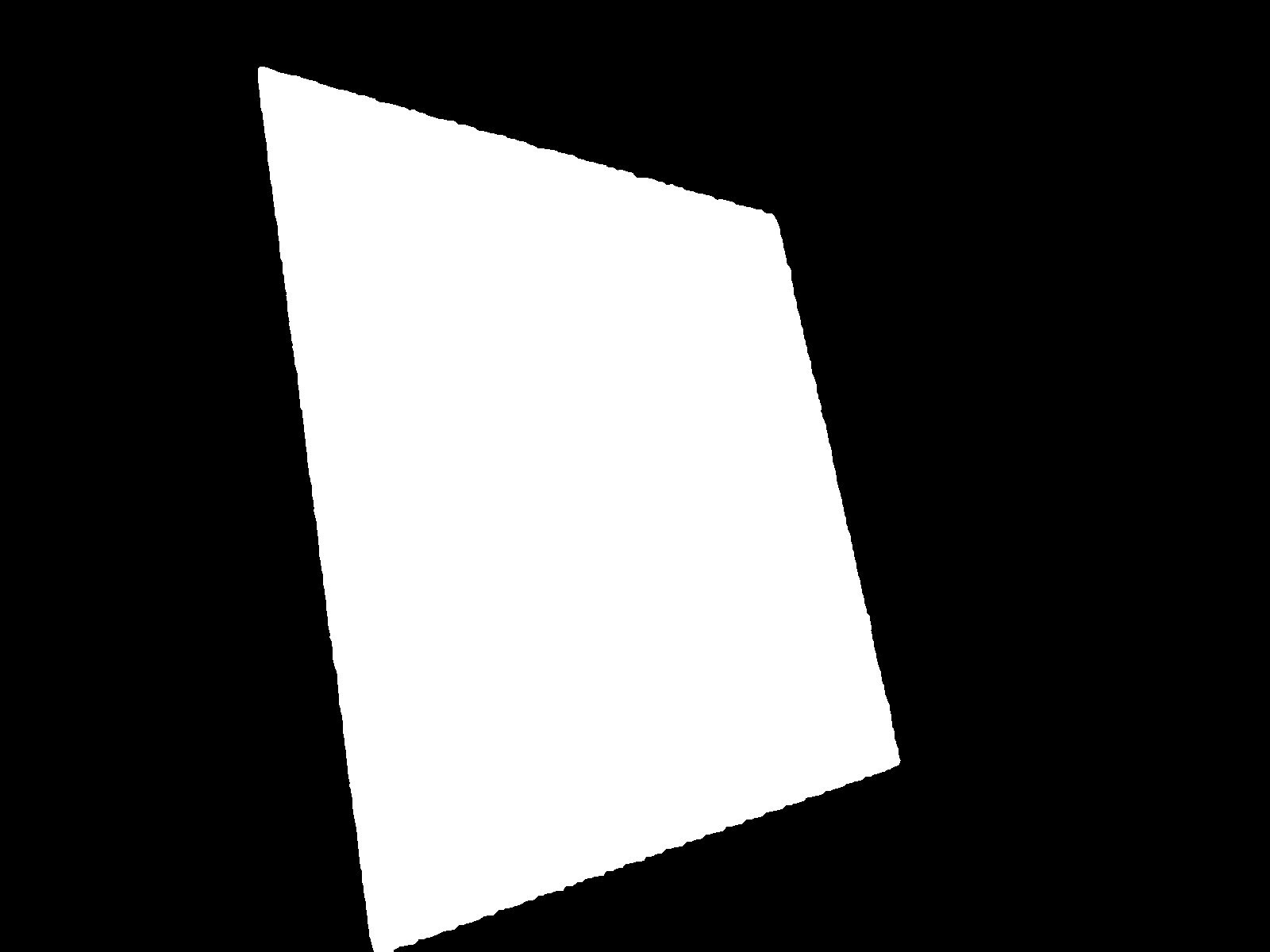}};
		\node(M4W)[below of=M4S, yshift=-0.67cm]{\includegraphics[width=0.118\textwidth]{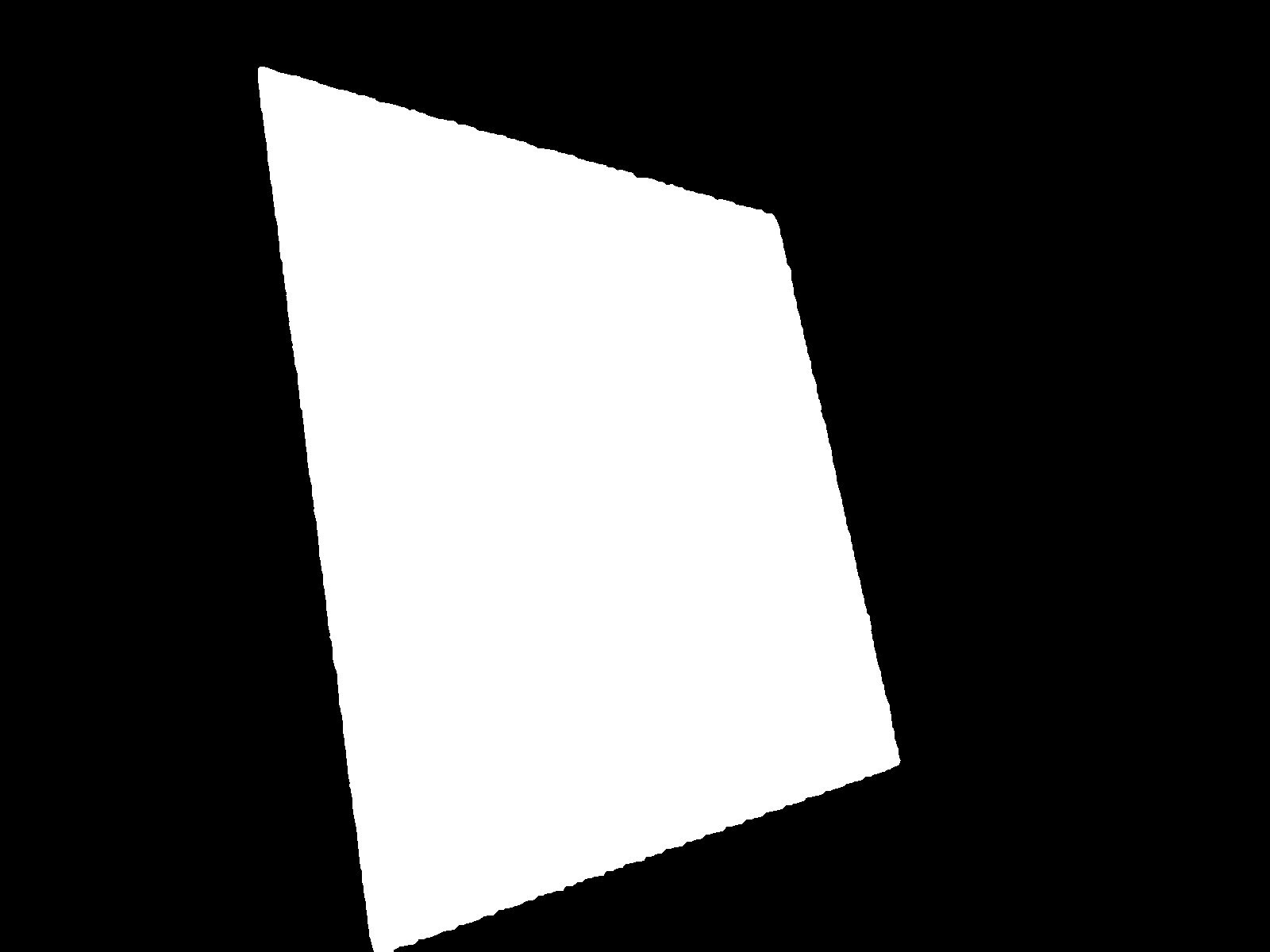}};
		
		
		
		\node(I5)[right of=I4, xshift=1.22cm]{\includegraphics[width=0.118\textwidth]{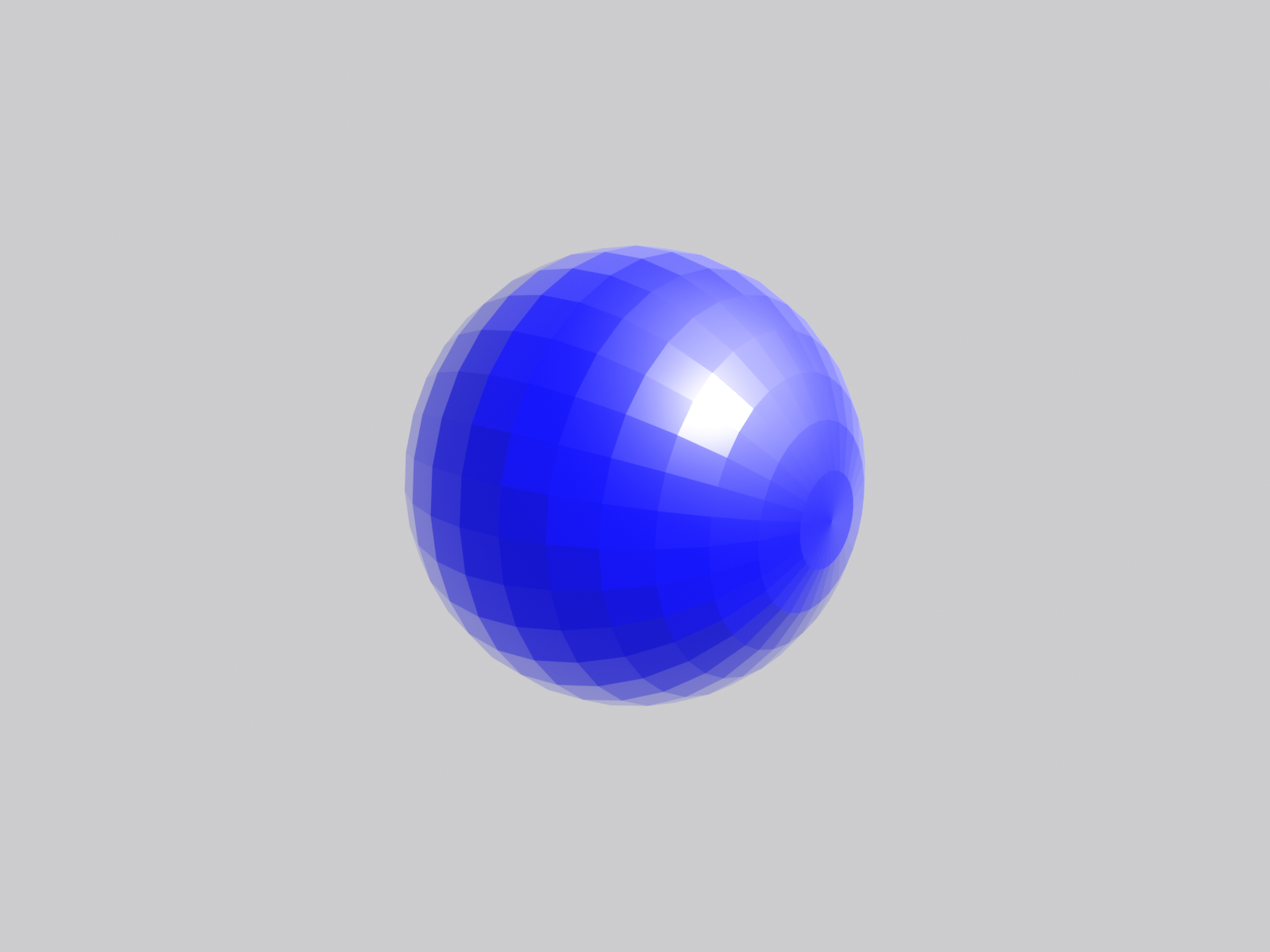}};	
		\node(M5GT)[below of=I5, yshift=-0.67cm]{\includegraphics[width=0.118\textwidth]{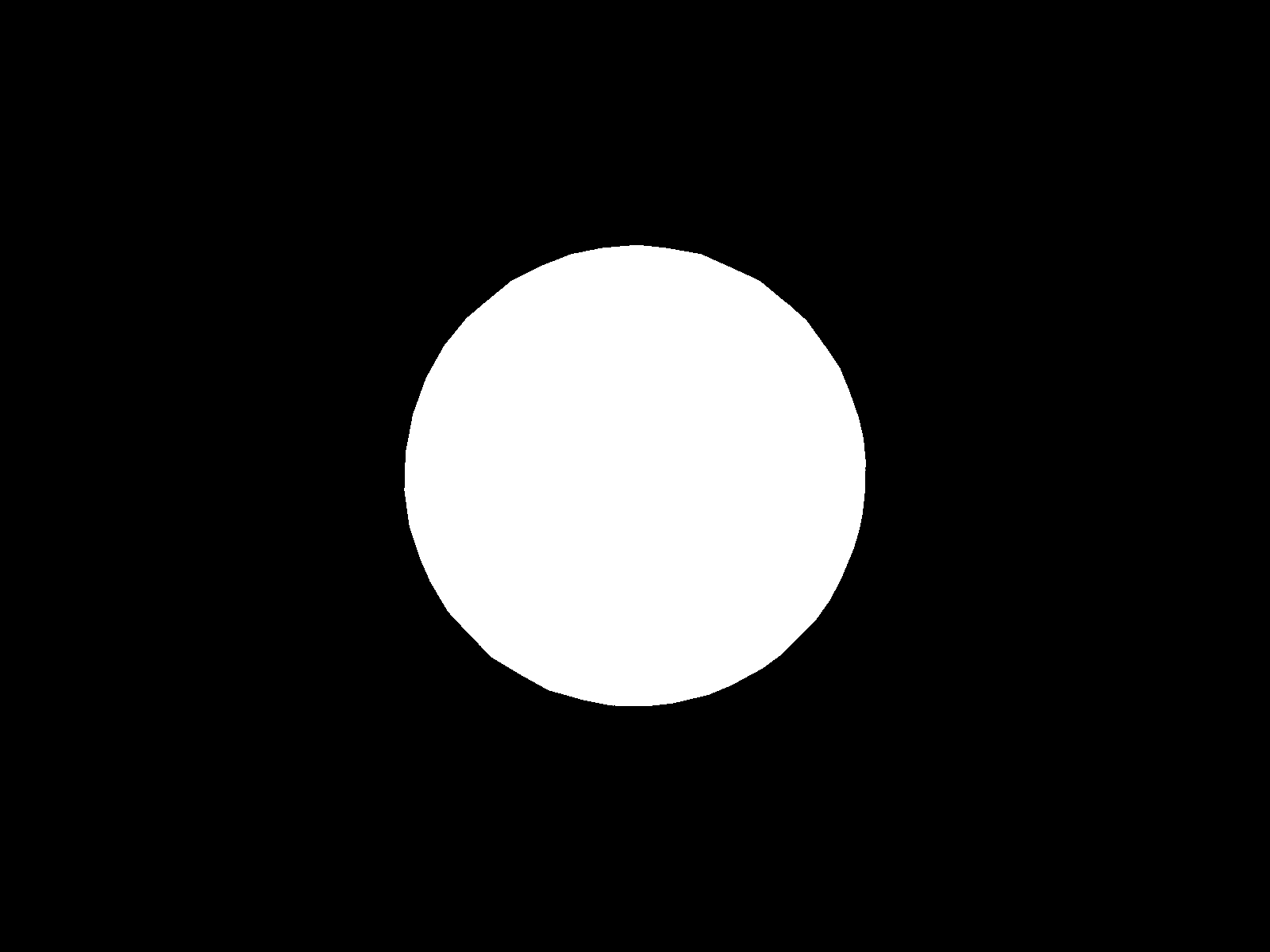}};
		\node(M5O)[below of=M5GT, yshift=-0.67cm]{\includegraphics[width=0.118\textwidth]{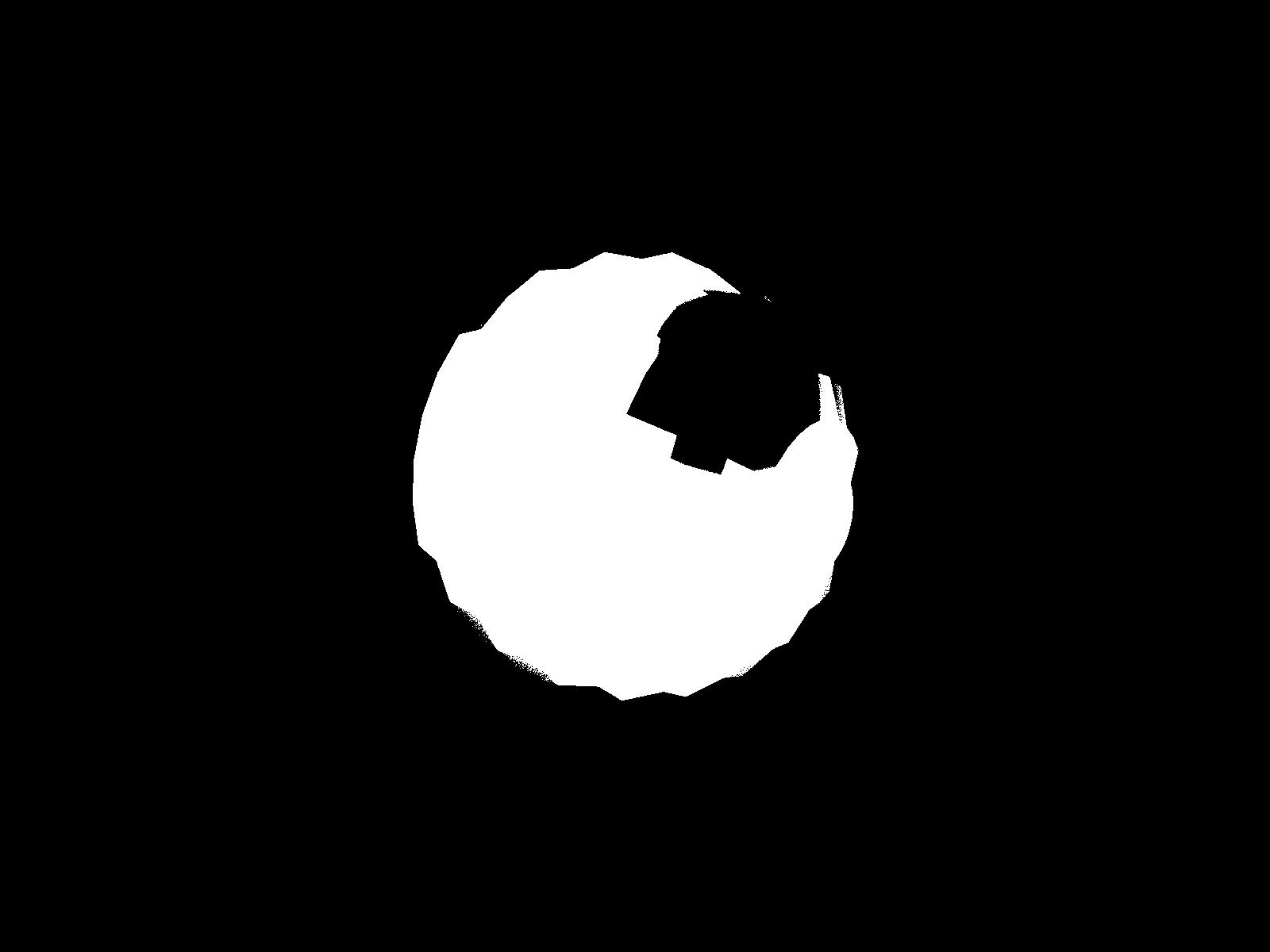}};
		\node(M5Y)[below of=M5O, yshift=-0.67cm]{\includegraphics[width=0.118\textwidth]{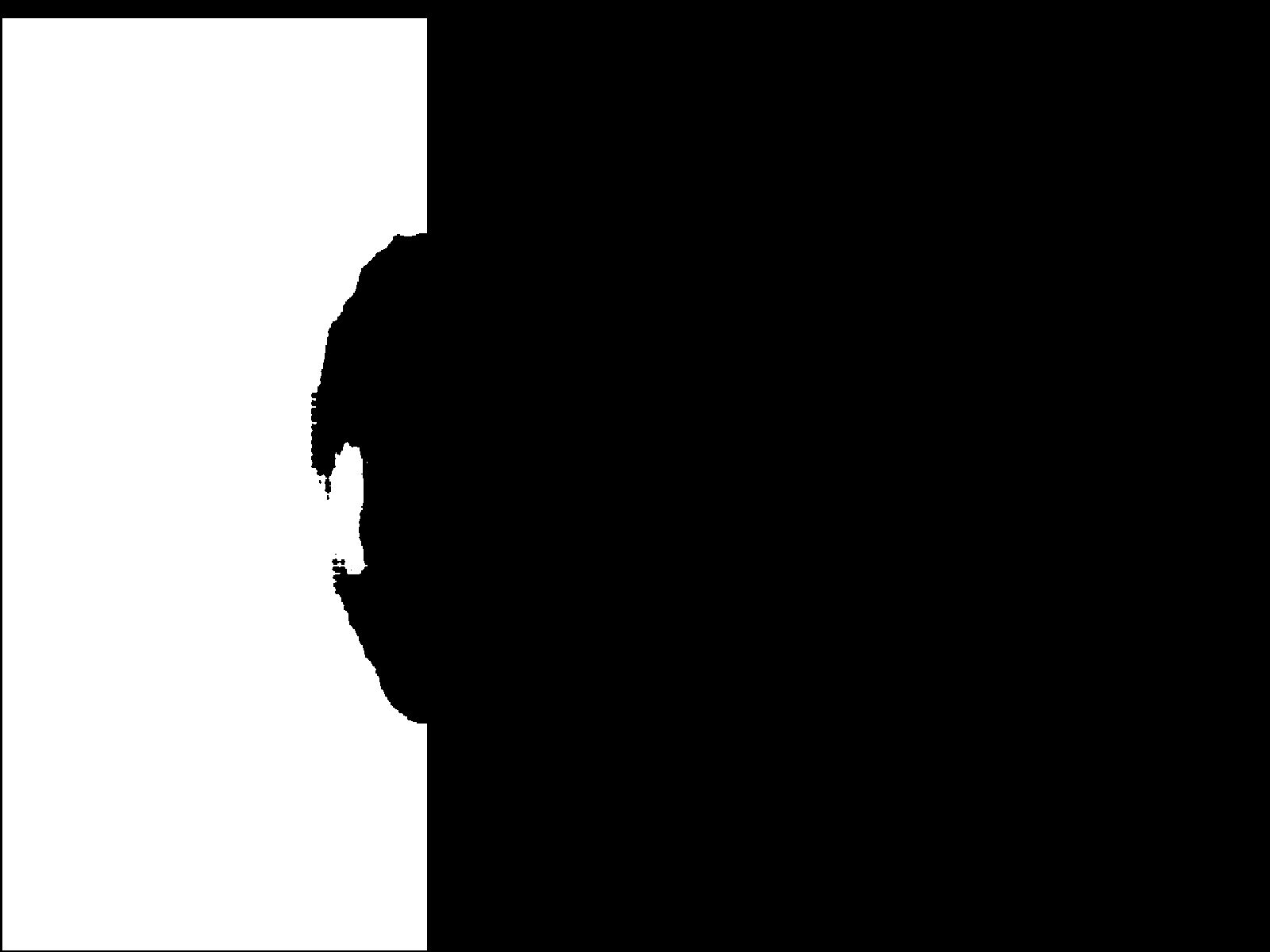}};
		\node(M5S)[below of=M5Y, yshift=-0.67cm]{\includegraphics[width=0.118\textwidth]{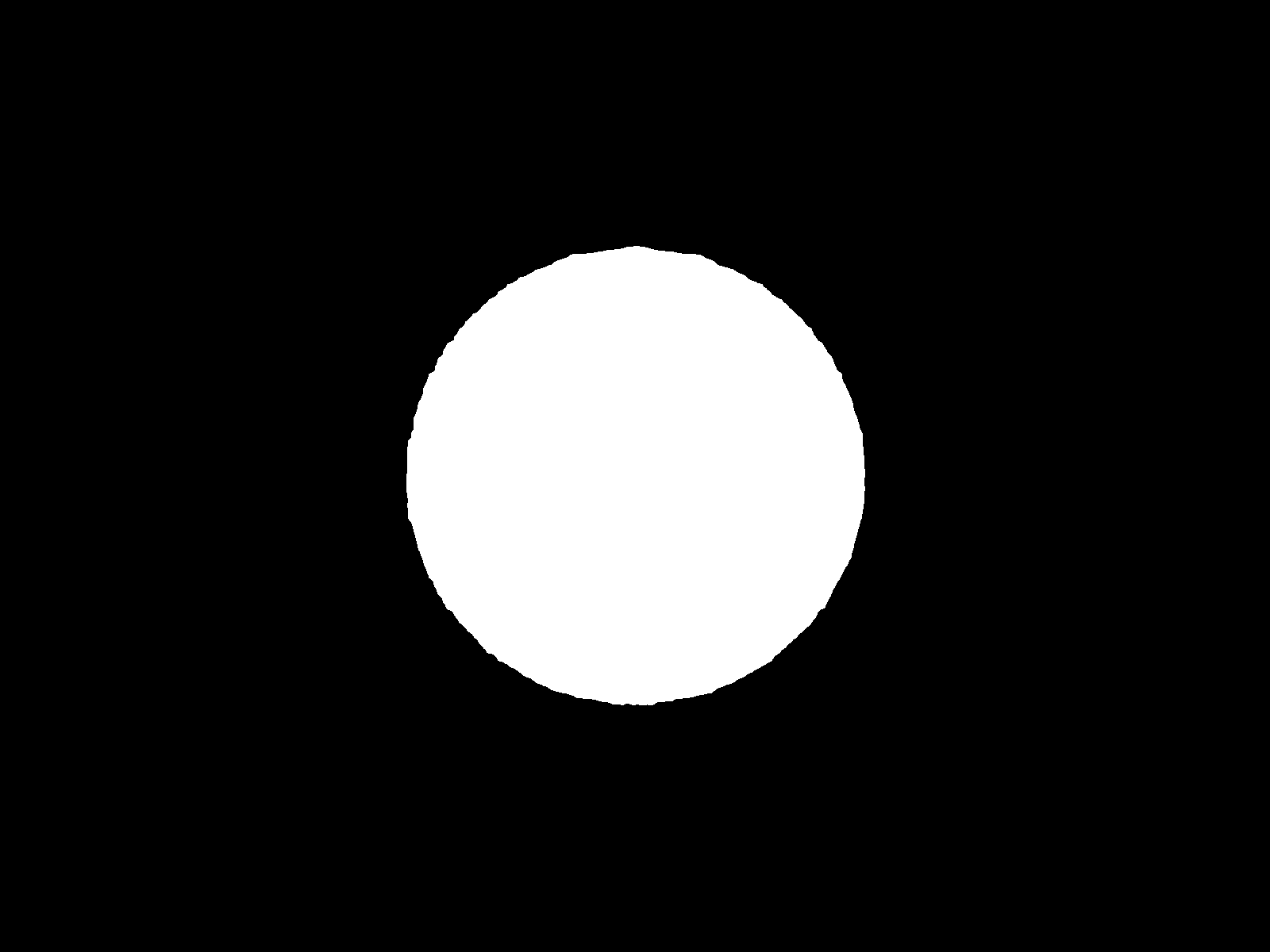}};
		\node(M5W)[below of=M5S, yshift=-0.67cm]{\includegraphics[width=0.118\textwidth]{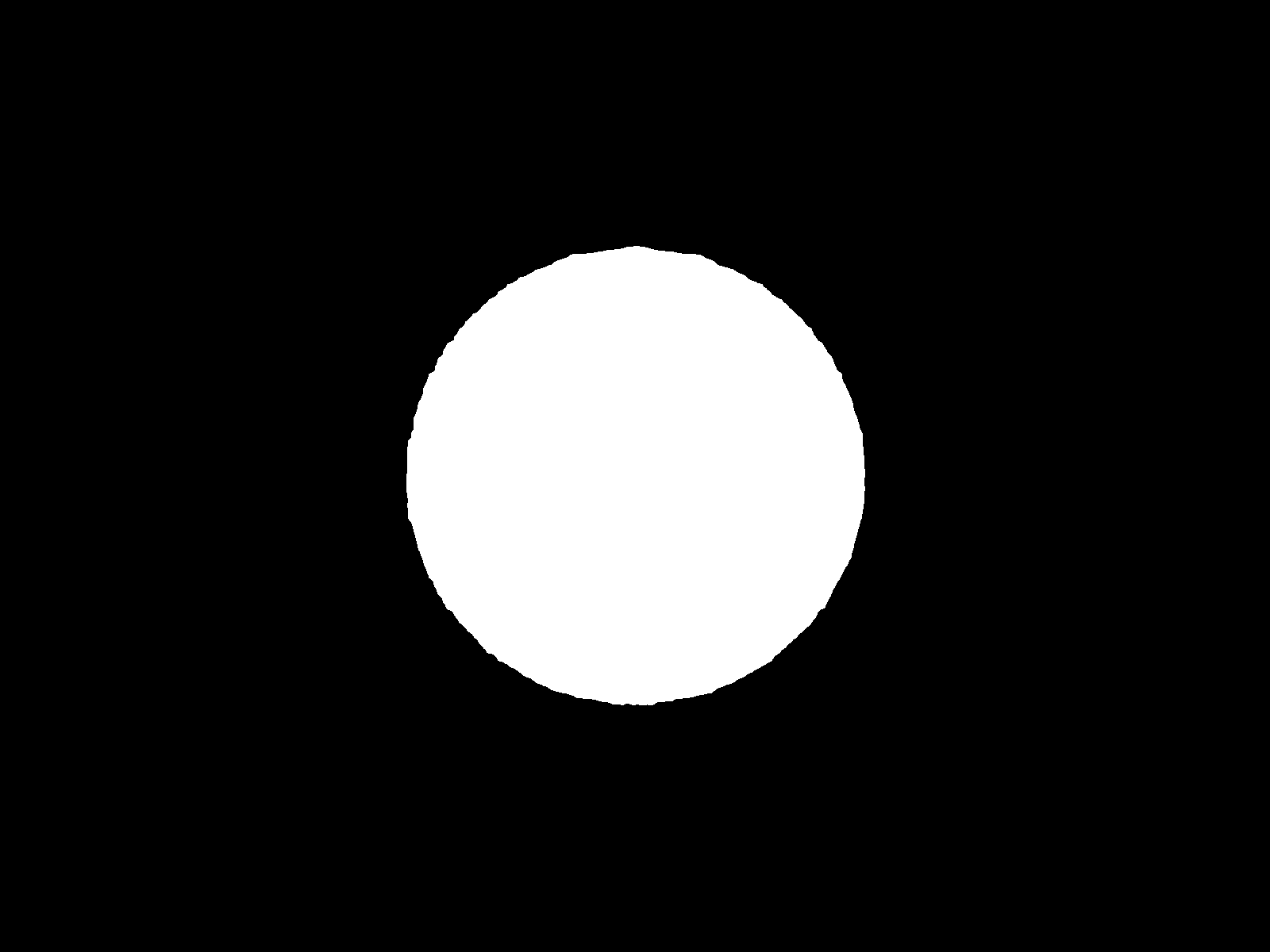}};	
		
		\node(I6)[right of=I5, xshift=1.11cm]{\includegraphics[width=0.108\textwidth]{Images/scene3/cam-4.jpg}};
		\node(M6GT)[below of=I6, yshift=-0.67cm]{\includegraphics[width=0.108\textwidth]{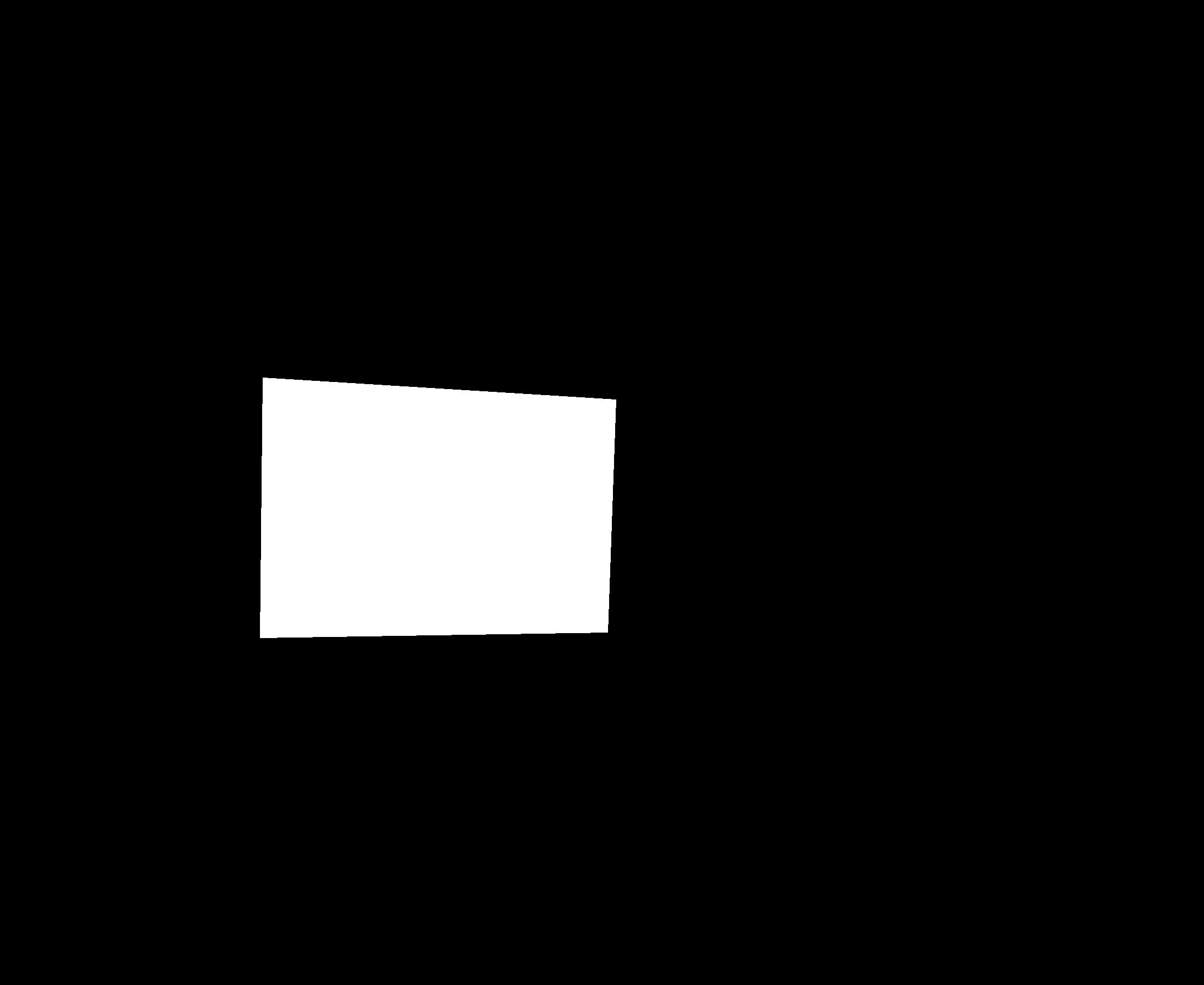}};	
		\node(M6O)[below of=M6GT, yshift=-0.67cm]{\includegraphics[width=0.108\textwidth]{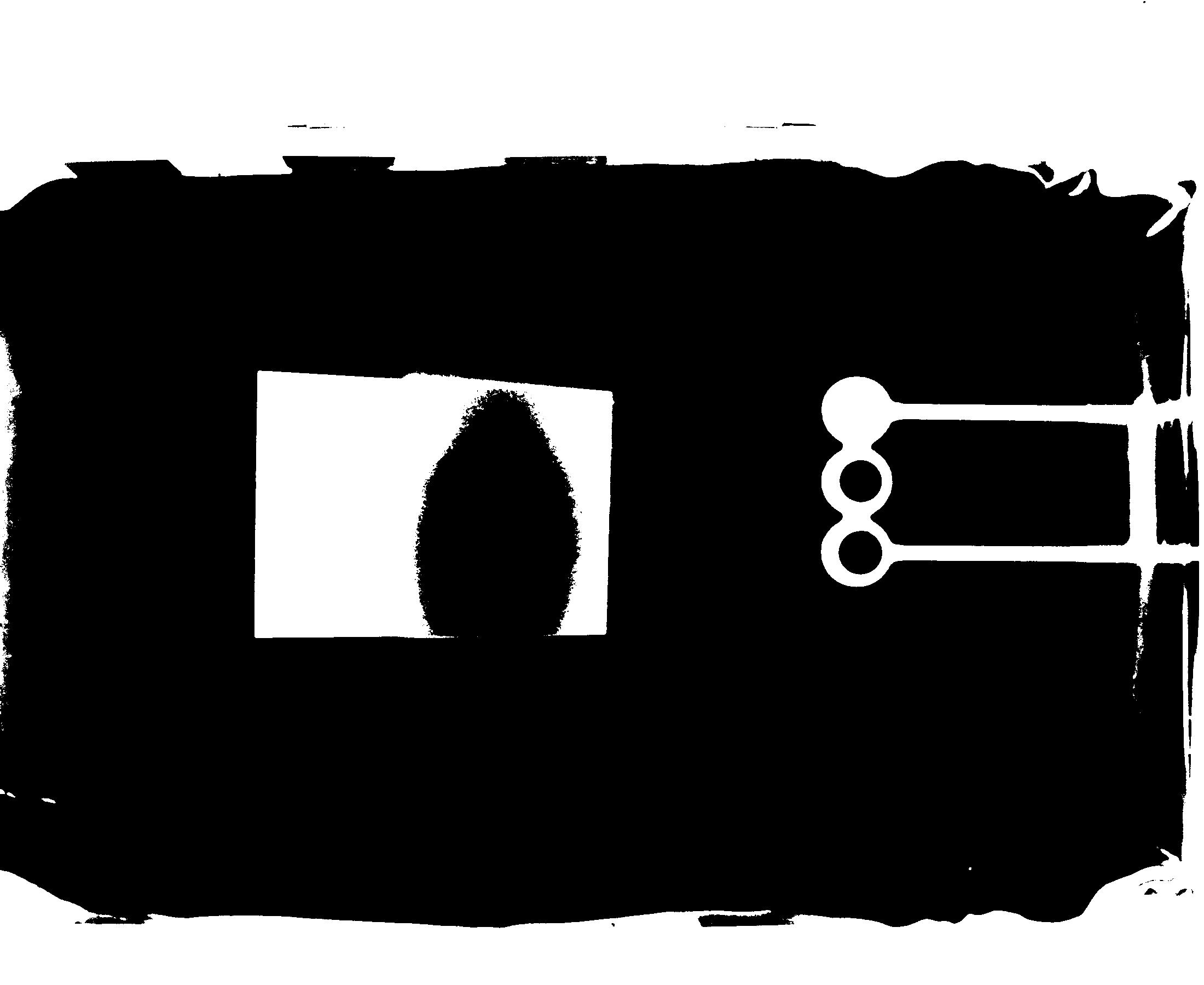}};
		\node(M6Y)[below of=M6O, yshift=-0.67cm]{\includegraphics[width=0.108\textwidth]{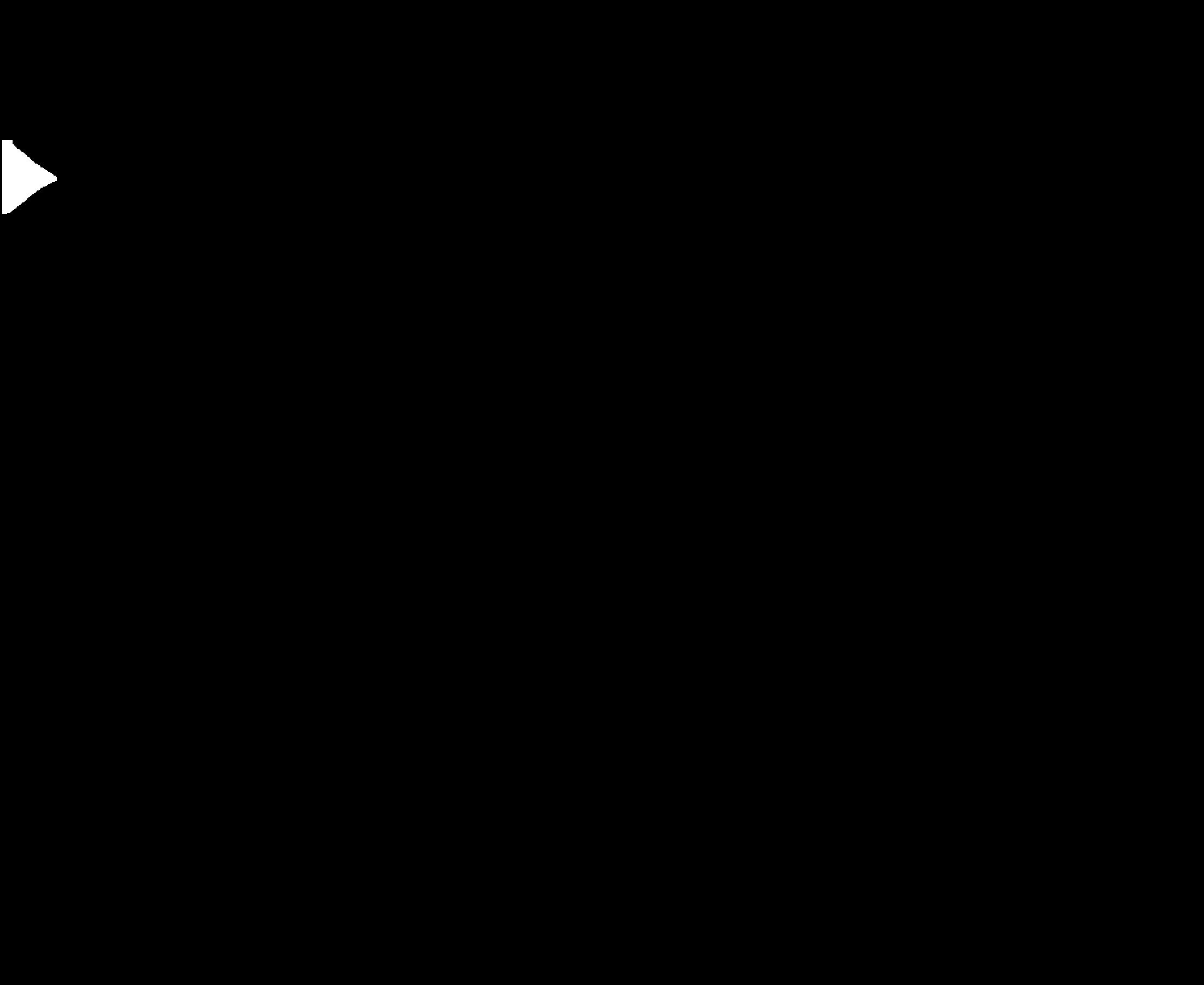}};
		\node(M6S)[below of=M6Y, yshift=-0.67cm]{\includegraphics[width=0.108\textwidth]{Images/scene3/output_mask_1.png}};
		\node(M6W)[below of=M6S, yshift=-0.67cm]{\includegraphics[width=0.108\textwidth]{Images/scene3/output_mask_2_1.jpg}};	
		
		\node(I7)[right of=I6, xshift=1.03cm]{\includegraphics[width=0.108\textwidth]{Images/scene5/cam-4.jpg}};
		\node(M7GT)[below of=I7, yshift=-0.67cm]{\includegraphics[width=0.108\textwidth]{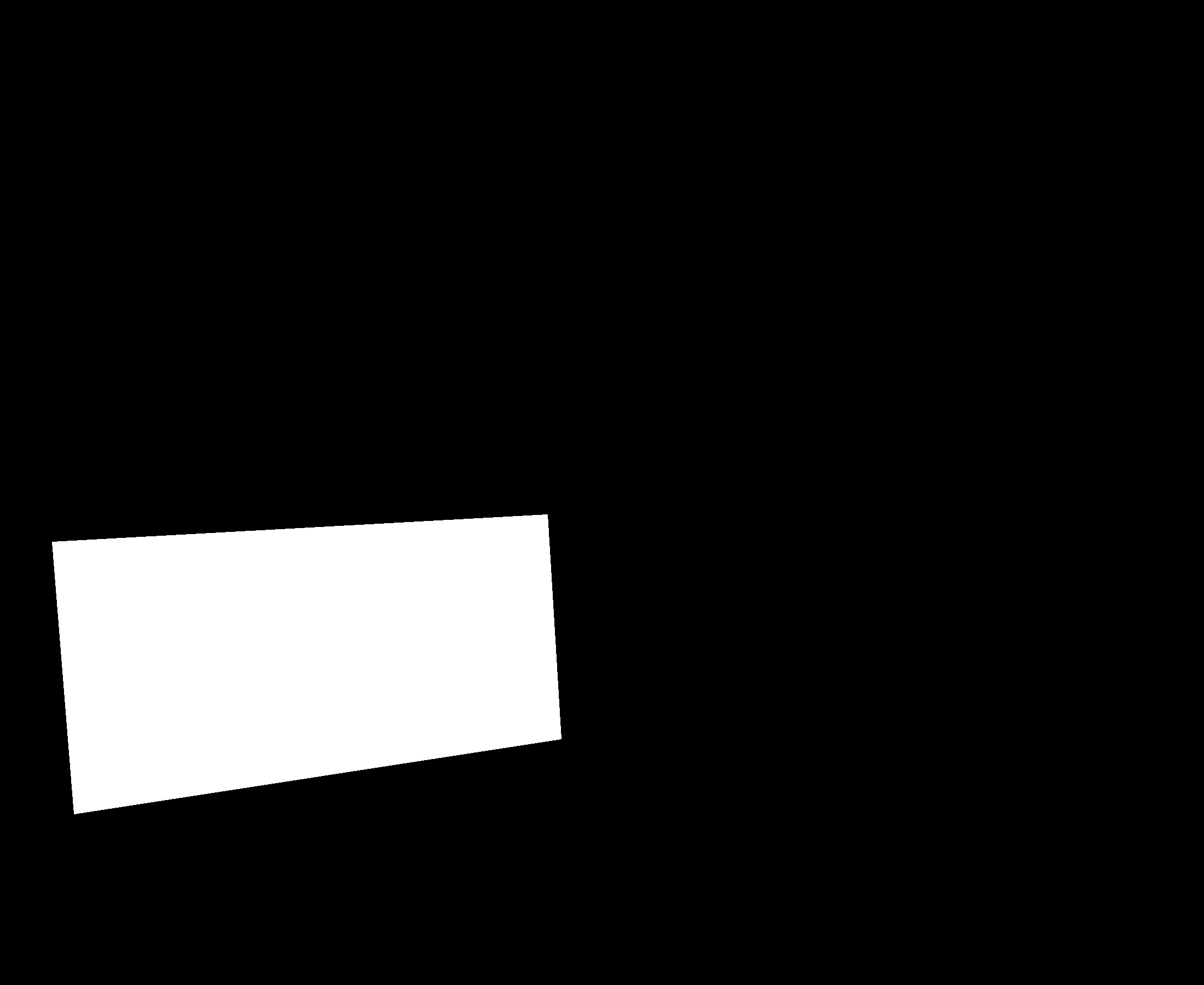}};		
		\node(M7O)[below of=M7GT, yshift=-0.67cm]{\includegraphics[width=0.108\textwidth]{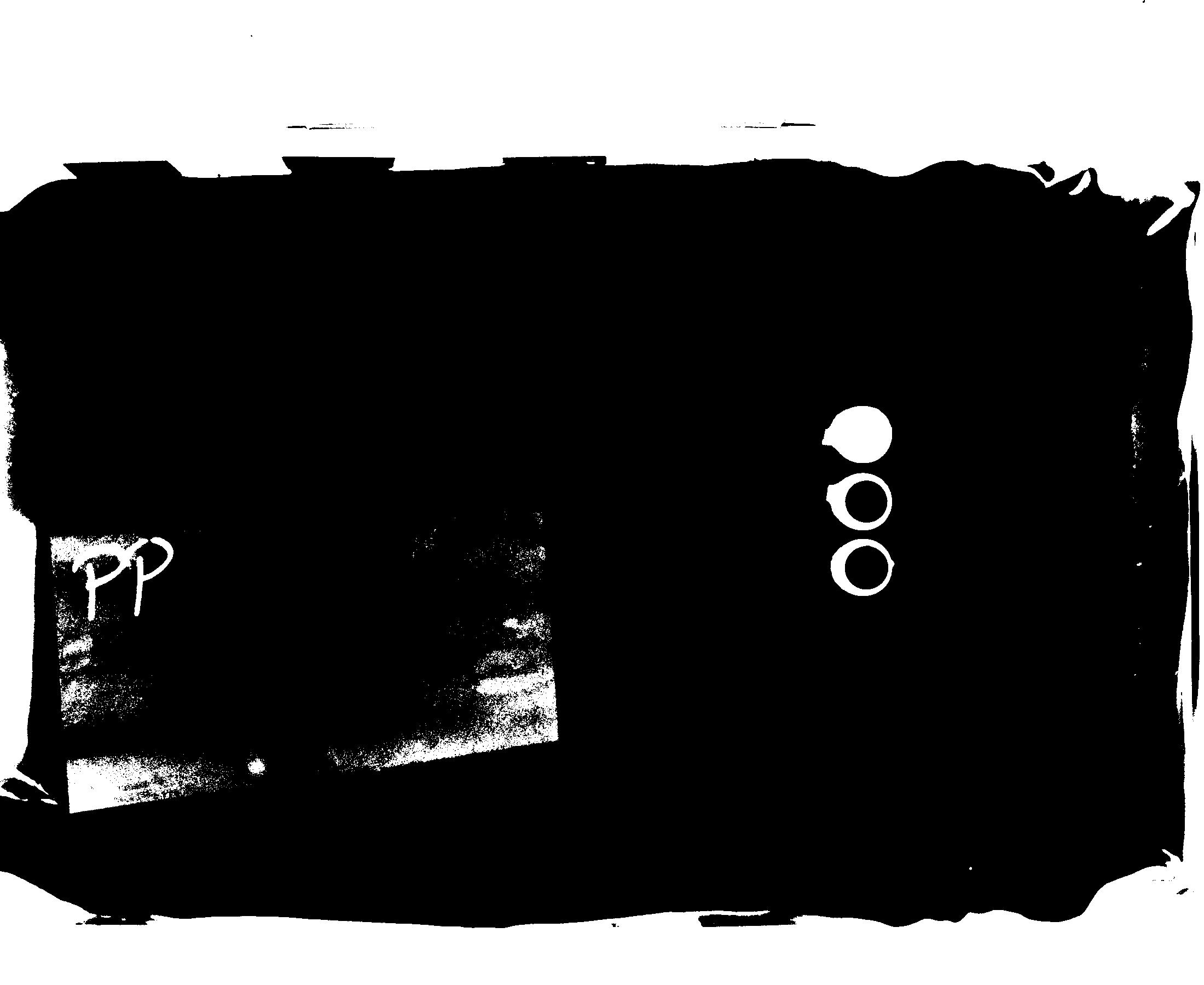}};
		\node(M7Y)[below of=M7O, yshift=-0.67cm]{\includegraphics[width=0.108\textwidth]{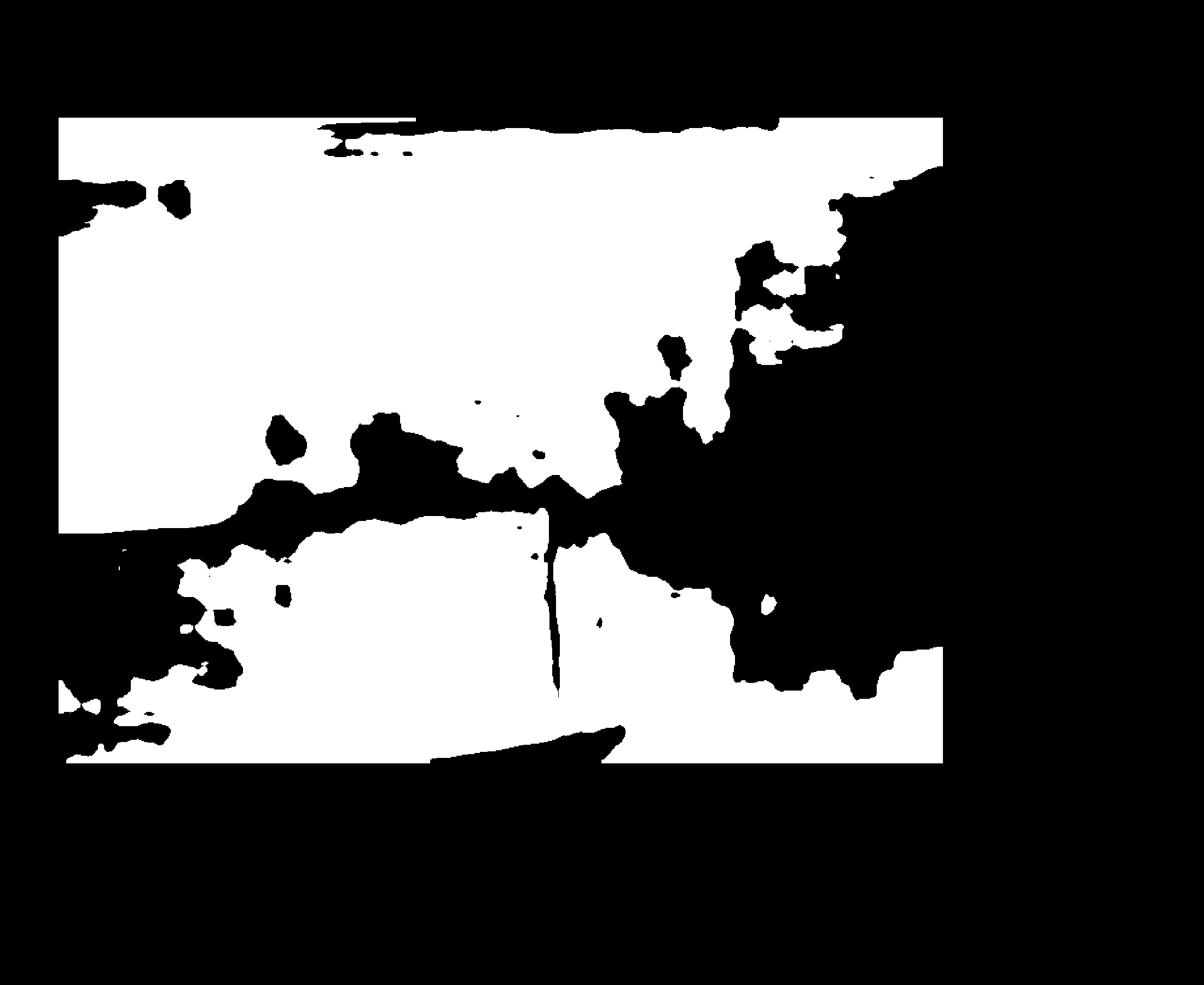}};
		\node(M7S)[below of=M7Y, yshift=-0.67cm]{\includegraphics[width=0.108\textwidth]{Images/scene5/output_mask_2.jpg}};
		\node(M7W)[below of=M7S, yshift=-0.67cm]{\includegraphics[width=0.108\textwidth]{Images/scene5/output_mask_1.jpg}};

		\node(I8)[right of=I7, xshift=1.03cm]{\includegraphics[width=0.108\textwidth]{Images/scene4/cam-4.jpg}};
		\node(M8GT)[below of=I8, yshift=-0.67cm]{\includegraphics[width=0.108\textwidth]{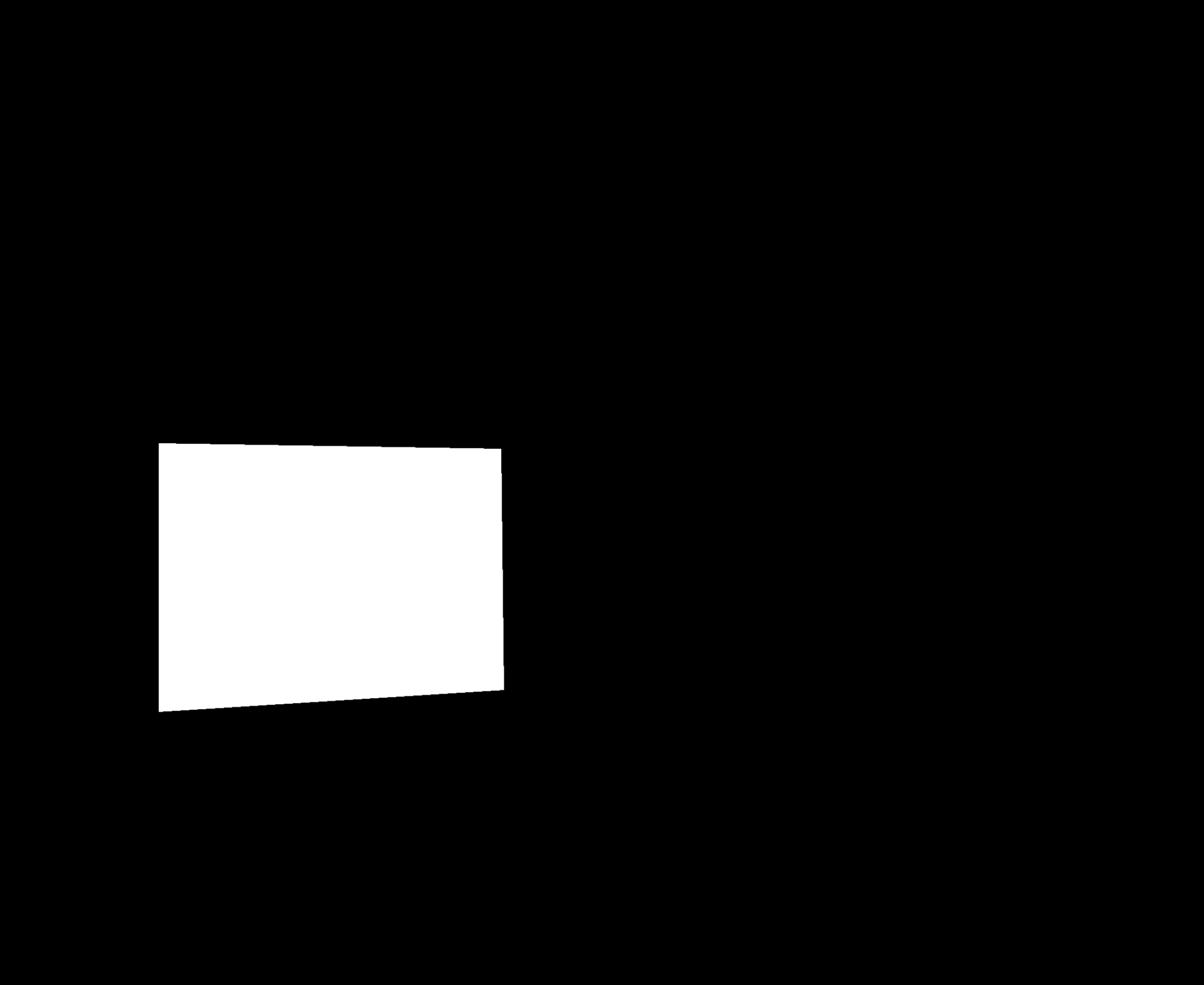}};		
		\node(M8O)[below of=M8GT, yshift=-0.67cm]{\includegraphics[width=0.108\textwidth]{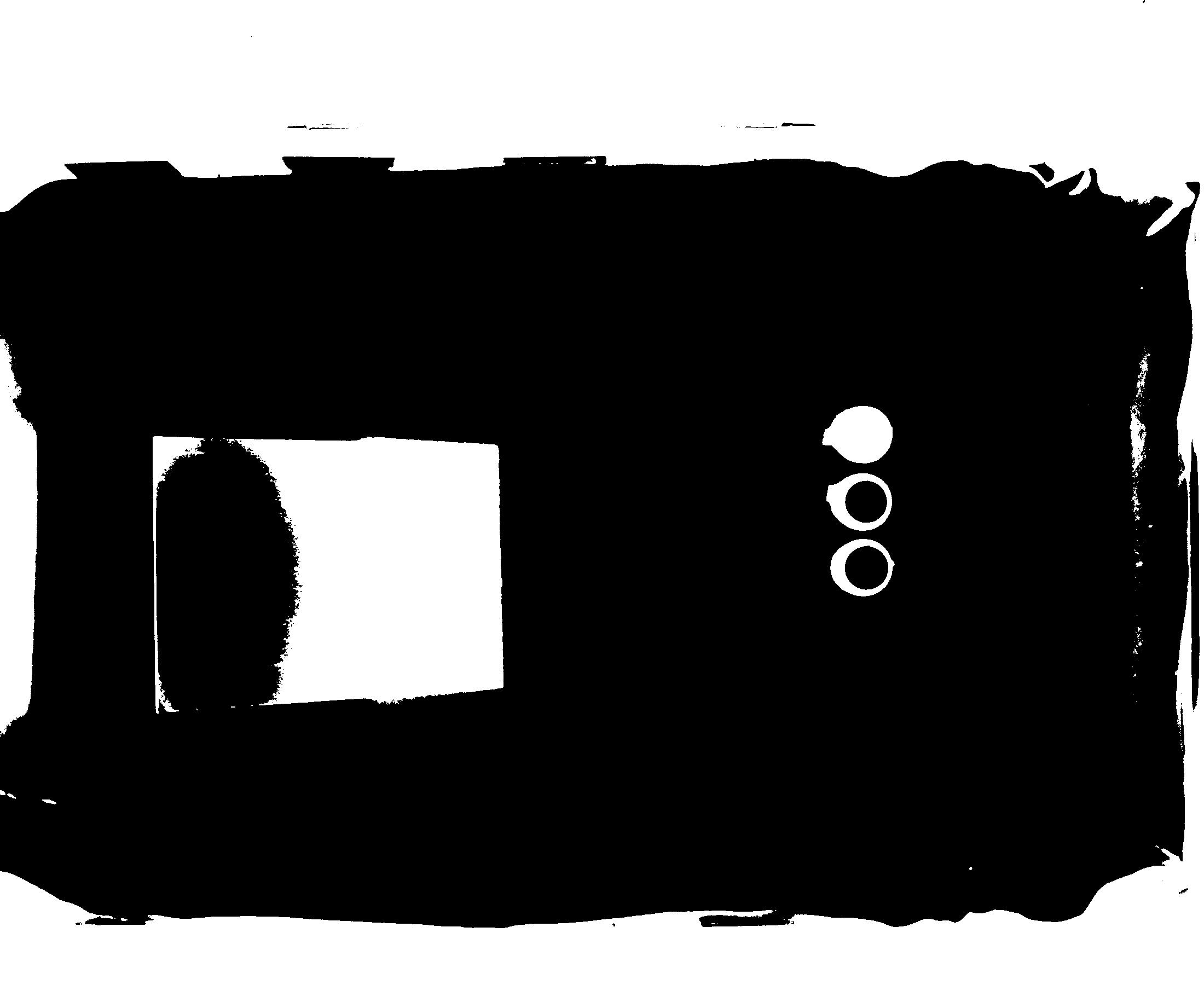}};
		\node(M8Y)[below of=M8O, yshift=-0.67cm]{\includegraphics[width=0.108\textwidth]{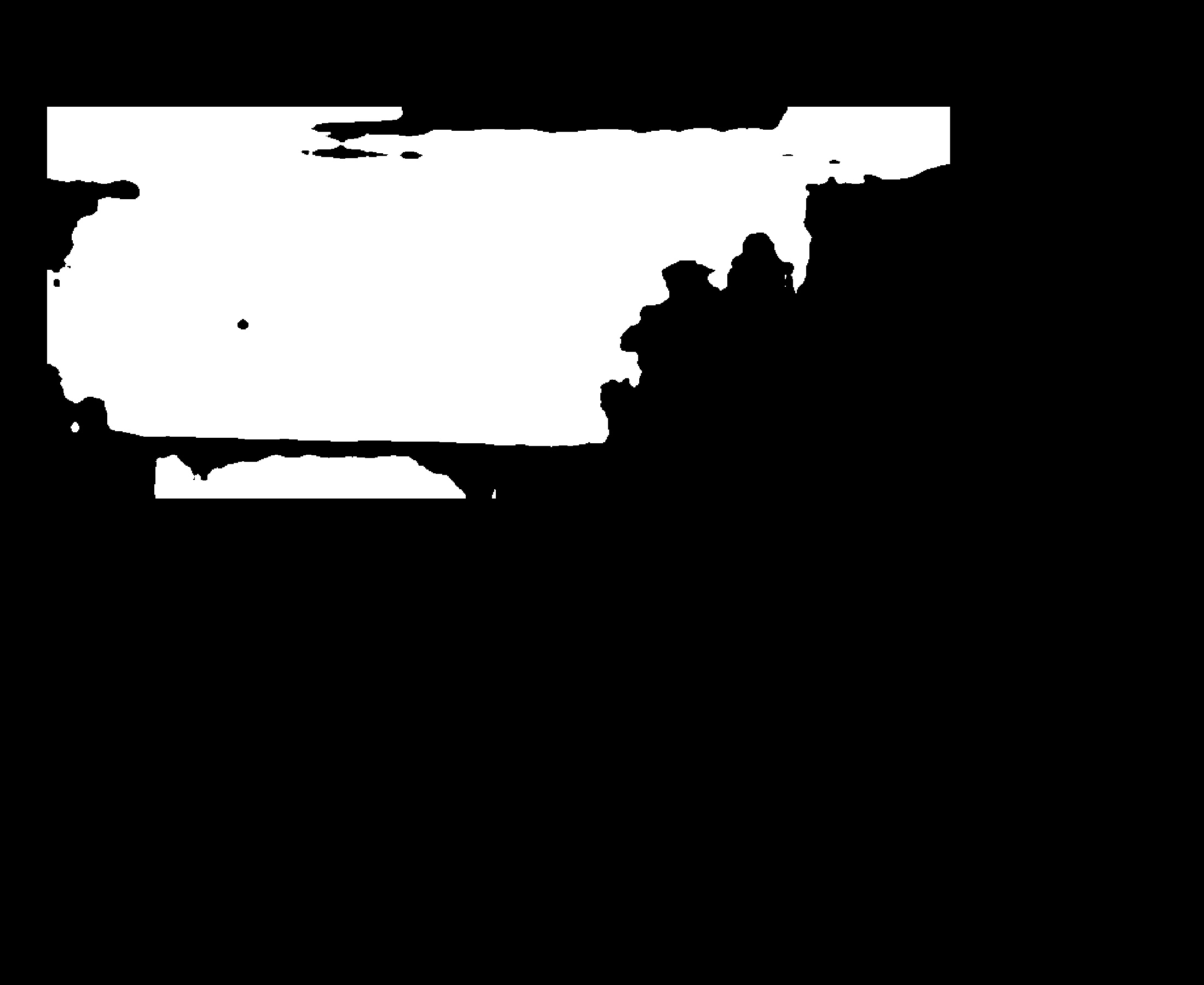}};
		\node(M8S)[below of=M8Y, yshift=-0.67cm]{\includegraphics[width=0.108\textwidth]{Images/scene4/output_mask_2.png}};
		\node(M8W)[below of=M8S, yshift=-0.67cm]{\includegraphics[width=0.108\textwidth]{Images/scene4/output_mask_2_1.jpg}};	
	\end{tikzpicture}
	\vspace{-0.7cm}
	\caption{Qualitative evaluation of the segmentation masks on synthetic and real world images using Otsu \cite{Otsu}, YOLO \cite{YOLO}, SAM2 \cite{SAM2} and our method RePoSeg. The examples shown here are representative, the complete setup was conducted on a substantially larger dataset.}
	\label{fig:QualEval}
	\vspace{-0.5cm}
\end{figure*} 


\begin{table}[t]
	\centering
	\caption{Quantitative evaluation of Otsu \cite{Otsu}, YOLO \cite{YOLO}, SAM2 \cite{SAM2} and our proposed method RePoSeg on synthetic and real-world data. The best results are marked in \textbf{bold}.}
\begin{minipage}{0.45\linewidth}
	\centering
	\textbf{Synthetic data}
	\begin{tabular}{l|c}
		Metric & Value [\%] \\
		\hline
		IoU (Otsu)     & 69.98 \\
		IoU (YOLO)     & 36.96 \\
		IoU (SAM2)     & 78.03 \\
		IoU (RePoSeg)     & \textbf{98.86} \\
		\hline
		DSC (Otsu)     & 79.92 \\
		DSC (YOLO)     & 48.40 \\
		DSC (SAM2)     & 81.29 \\
		DSC (RePoSeg)     & \textbf{99.43} \\
		\hline
		Pixel Acc. (Otsu) & 98.77 \\
		Pixel Acc. (YOLO) & 67.69 \\
		Pixel Acc. (SAM2) & 90.87 \\
		Pixel Acc. (RePoSeg) & \textbf{99.68} \\
	\end{tabular}
\end{minipage}
\hfill
\begin{minipage}{0.45\linewidth}
	\centering
	\textbf{Real-world data}
	\begin{tabular}{l|c}
		Metric & Value [\%] \\
		\hline
		IoU (Otsu)     & 9.45 \\
		IoU (YOLO)     & 7.78 \\
		IoU (SAM2)     & 88.76 \\
		IoU (RePoSeg)     & \textbf{94.22} \\
		\hline
		DSC (Otsu)     & 16.83 \\
		DSC (YOLO)     & 13.30 \\
		DSC (SAM2)     & 93.93 \\
		DSC (RePoSeg)     & \textbf{97.02} \\
		\hline
		Pixel Acc. (Otsu) & 68.22 \\
		Pixel Acc. (YOLO) & 77.46 \\
		Pixel Acc. (SAM2) & 99.10 \\
		Pixel Acc. (RePoSeg) & \textbf{99.52} \\
	\end{tabular}
\end{minipage}
\label{tab:QuantitativeResults}
\vspace{-0.5cm}
\end{table}


RePoSeg is evaluated quantitatively and qualitatively on synthetic images and real world data, captured with a camera array \cite{CAMSI}. The synthetic images were rendered in Blender \cite{Blender}, allowing the generation of accurate ground truth masks for objective evaluation. The masks for the real world images were manually annotated. Representative examples for both datasets are shown in Fig. \ref{fig:QualEval}. 

To quantify segmentation performance, we use three commonly used metrics. Intersection over Union (IoU) \cite{IoU} measures the overlap between prediction and ground truth, the Dice Similarity Coefficient (DSC) \cite{Dice} emphasizes correctly predicted foreground pixels, and pixel accuracy (pix. acc.), which reflects the proportion of correctly classified pixels in the entire image, including both foreground and background.

Quantitative results for both datasets are shown in Table \ref{tab:QuantitativeResults}. RePoSeg is evaluated alongside the three established baselines of Otsu \cite{Otsu}, the YOLO model \cite{YOLO}, and SAM2 \cite{SAM2}. Our approach achieves superior performance across all metrics. Specifically, for the synthetic images we obtain an IoU of 98.86\%, a DSC of 99.43\%, and a pixel accuracy of 99.68\%. Compared to the strongest baseline SAM2, RePoSeg improves the IoU by 26.7\%, the DSC by 22.3\% and pixel accuracy by 9.7\%. Against the classical Otsu method, improvements are 41.3\% in IoU, 24.4\% in DSC and 0.9\% in pixel accuracy. In comparison to the masks obtained with YOLO, even larger relative gains of 176.5\% in IoU, 105.4\% in DSC and 47.2\% in pixel accuracy are achieved. Similarly, on the real-world data, we observe consistently strong results with an IoU of 94.22\%, a DSC of 97.02\% and a pixel accuracy of 99.52\%. These results clearly demonstrate that the proposed method remains highly effective even when applied to a segmentation model that was not specifically trained to handle specular reflections. The approach yields a high robustness and precision, particularly in capturing specular regions that often challenge conventional and learning-based segmentation methods.

Fig. \ref{fig:QualEval} presents qualitative examples of segmentation outputs for both synthetic and real images. For synthetic cases, the ground truth mask is shown alongside the outputs of Otsu \cite{Otsu}, YOLO \cite{YOLO}, SAM2 \cite{SAM2} and RePoSeg. The visual results confirm that our method produces cleaner, more precise boundaries and preserves object regions better than the reference approaches, even in the presence of complex highlights and specular reflections. Importantly, it performs consistently well on both grayscale and color images, showing no noticeable limitations with respect to input modality.

\vspace{-0.1cm}
\section{Conclusion}
\vspace{-0.1cm}
In this paper, a novel approach for obtaining object segmentation masks in the presence of specular reflections was proposed. RePoSeg leverages the physical constraint that specular highlights must lie on the object surface. By identifying the largest connected region containing such highlights, the object can be segmented reliably, even in cases where conventional methods or deep neural networks struggle due to the misleading visual cues introduced by reflections. Our strategy enables the use of generic segmentation networks, that were not explicitly trained to handle specular reflections, improving their applicability to a wider range of real world scenarios. The effectiveness of the proposed method is supported by strong quantitative results on synthetic data as well as consistent qualitative performance on synthetic and real world imagery, demonstrating both robustness and generalizability independent of the input modality. 

\balance

\bibliographystyle{IEEEbib}
\bibliography{refs}

\end{document}